\documentclass[10pt,journal,compsoc]{IEEEtran}

\ifCLASSOPTIONcompsoc
  \usepackage[nocompress]{cite}
\else
  \usepackage{cite}
\fi

\usepackage{graphicx,subfigure}
\usepackage{amsmath,amsthm,amssymb}
\usepackage{ulem}

\usepackage{epstopdf}
\usepackage{algorithm,algorithmic}

\def\T{\mathcal{T}}

\newcommand{\vv}[1]{\boldsymbol{#1}} 
\newcommand{\mean}[2]{{\mu}_{{#1}{#2}}} 
\usepackage{xcolor}

\newcommand{\edit}[1]{{#1}}

\title{Quick \& Plenty: Achieving Low Delay \& High Rate in 802.11ac Edge Networks}
\author{Hamid Hassani, Francesco Gringoli, Douglas J. Leith
\IEEEcompsocitemizethanks{\IEEEcompsocthanksitem F. Gringoli is with University of Brescia, Italy.\IEEEcompsocthanksitem H.Hassani and D. J. Leith are with Trinity College Dublin, Ireland. HH was supported by Science Foundation Ireland under Grant No. 13/RC/2077.}
}

\begin{document}

\global\csname @topnum\endcsname 0
\global\csname @botnum\endcsname 0

\IEEEtitleabstractindextext{
\begin{abstract}
We consider transport layer approaches for achieving high rate, low delay communication over edge paths where the bottleneck is an 802.11ac WLAN.   We first show that by regulating send rate so as to maintain a target aggregation level it is possible to realise high rate, low delay communication over 802.11ac WLANs.  We then address two important practical issues arising in production networks, namely that (i) many client devices are non-rooted mobile handsets/tablets and (ii) the bottleneck may lie in the backhaul rather than the WLAN, or indeed vary between the two over time.  We show that both these issues can be resolved by use of simple and robust machine learning techniques.  We present a prototype transport layer implementation of our low delay rate allocation approach and use this to evaluate performance under real radio conditions.
\end{abstract}
}
\maketitle

\section{Introduction}
While much attention in 5G has been focussed on the physical and link layers, it is increasingly being realised that a wider redesign of network protocols is also needed in order to meet 5G requirements.   Transport protocols are of particular relevance for end-to-end performance, including end-to-end latency.   For example, ETSI has recently set up a working group to study next generation protocols for 5G \cite{etsi}.  The requirement for major upgrades to current transport protocols is also reflected in initiatives such as Google QUIC \cite{quic}, Coded TCP \cite{ctcp14} and the Open Fast Path Alliance \cite{ofp}.   

\begin{figure}
\centering
\subfigure[]{
\raisebox{0.25in}{\includegraphics[width=0.47\columnwidth]{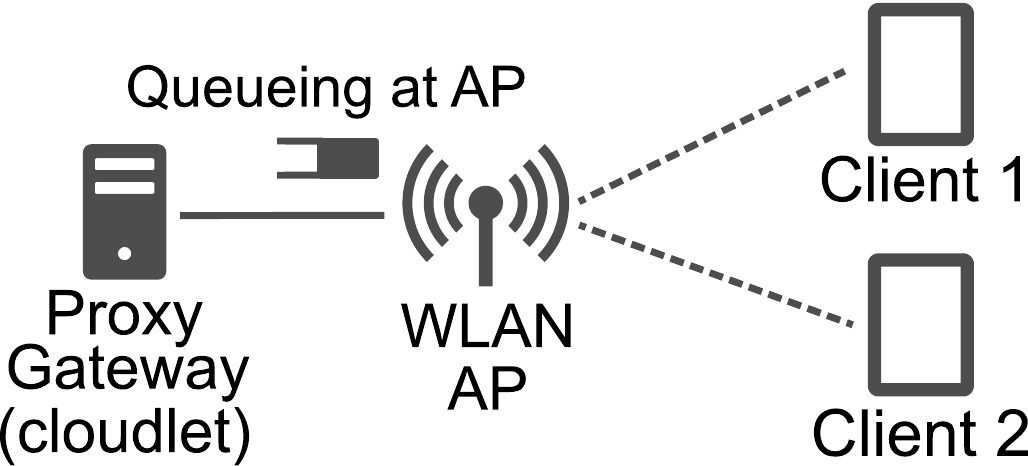}}
}
\subfigure[]{
\includegraphics[width=0.47\columnwidth]{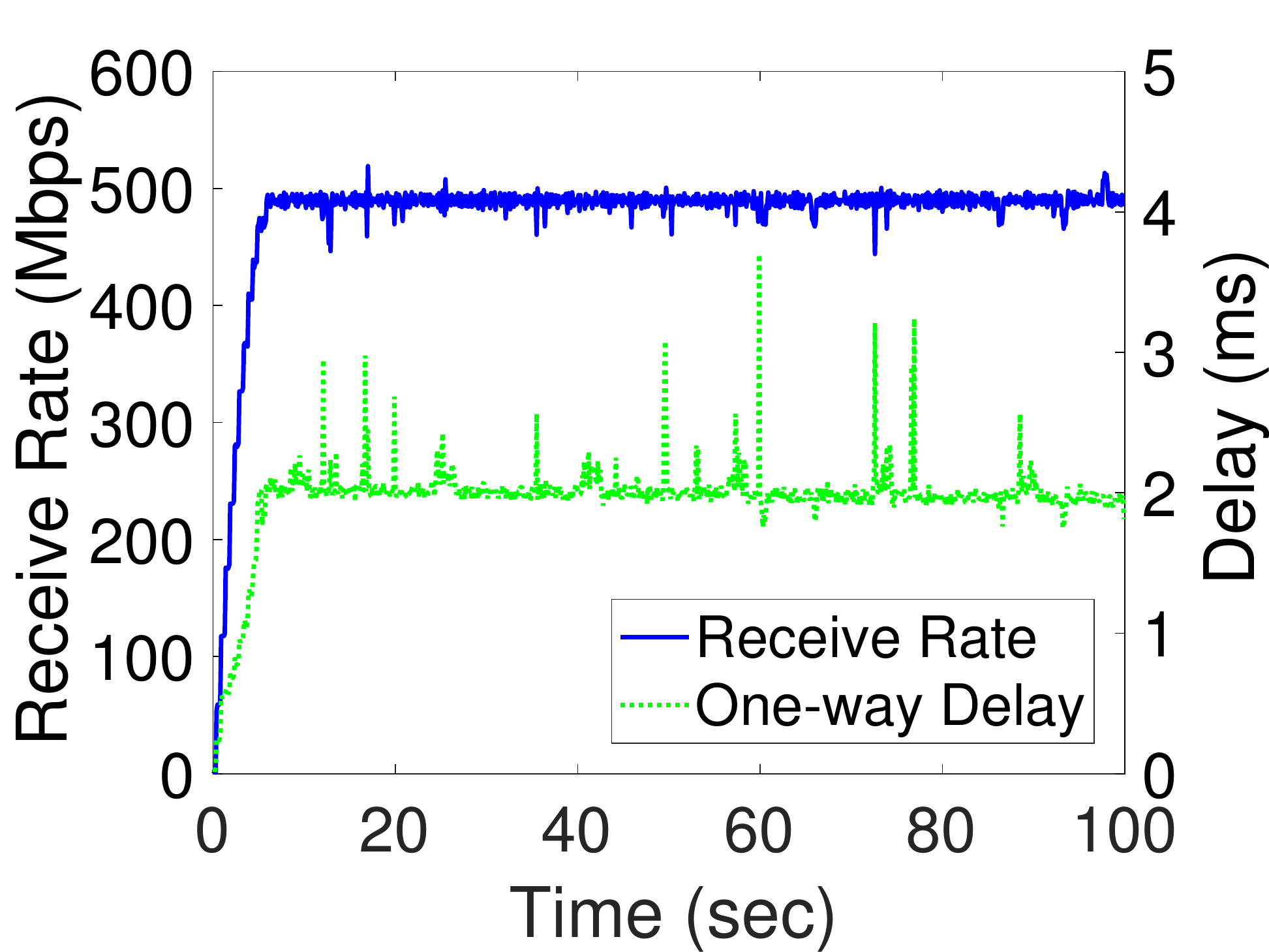}
}
\caption{(a) Cloudlet-based edge transport architecture with bottleneck in the WLAN hop (therefore queueing of downlink packets occurs at the AP as indicated on schematic) and (b) Illustrating low-latency high-rate operation in an 802.11ac WLAN (measurements are from a hardware testbed located in an office environment, see Appendix).}
\label{fig:edge}
\end{figure}

In this paper we consider next generation edge transport architectures of the type illustrated in Figure \ref{fig:edge}(a).   Traffic to and from client stations is routed via a proxy located close to the network edge (e.g. within a cloudlet).  This creates the freedom to implement new transport layer behaviour over the path between proxy and clients, which in particular includes the last wireless hop. One great advantage of this architecture is its ease of rollout since the new transport can be implemented as an app on the clients plus a proxy deployed in the cloud; no changes are required to existing AP's or servers.   

Our interest is in achieving high rate, low latency communication.  One of the most challenging requirements in 5G is the provision of connections with low end-to-end latency.  In most use cases the target is for $<$100ms latency, while for some applications it is $<$10ms \cite[Table 1]{ngmn}.  In part, this reflects the fact that low latency is already coming to the fore in network services, 
but the requirement for low latency also reflects the needs of next generation applications such as augmented reality and the tactile internet.

As can be seen from Figure \ref{fig:edge}(a), the transmission delay of a packet sent over the downlink is composed of two main components: (i) queueing delay at the AP and (ii)  MAC channel access time.  The MAC channel access time is determined by the WLAN channel access mechanism and is typically small, so the main challenge is to regulate the queueing delay.   We would like to select a downlink send rate which is as high as possible yet ensures that a persistent queue backlog does not develop at the AP.  

While measurements of one-way delay might be used to infer the onset of queueing and adjust the send rate, measuring one-way is known to be challenging\footnote{E.g. due to clock offset and skew between sender and receiver.} as is inference of queueing from one-way delay\footnote{For example, the transmission delay across a wireless hop can change significantly over time depending on the number of active stations, e.g. if a single station is active and then a second station starts transmitting the time between transmission opportunities for the original station may double, and it is difficult to distinguish changes in delay due to queueing and changes due to factors such as this.}.    Use of round-trip time to estimate the onset of queueing is also known to be inaccurate when there is queueing in the reverse path.   In this paper we avoid these difficulties by using measurements of the aggregation level of the frames transmitted by the AP.   Use of aggregation is ubiquitous in modern WLANs since it brings goodput near to line-rate by reducing the relative time spent in accessing the channel when transmitting several packets to the same destination.   As we will see, the number of packets aggregated in a frame is relatively easy to measure accurately and reliably at the receiver.   Intuitively, the level of aggregation is coupled to queueing. Namely, when only a few packets are queued then there are not enough to allow large aggregated frames to be assembled for transmission.  Conversely, when there is a persistent {backlog in the queue for a particular wireless client} then there is a plentiful supply of packets and large frames can be consistently assembled. We show that by regulating the downlink send rate so as to maintain an appropriate aggregation level it is possible to avoid queue build up at the AP and so to realise high rate, low latency communication in a robust and practical fashion.

Figure \ref{fig:edge}(b) shows typical results obtained by regulating the aggregation level.  These measurements are from a hardware testbed located in an office environment.  It can be seen that the one-way delay is low, at around 2ms, while the send rate is high, at around 500Mbps (this data is for an 802.11ac downlink using three spatial streams and MCS 9).  Increasing the send rate further leads to sustained queueing at the AP and an increase in delay, but the results in Figure \ref{fig:edge}(b) illustrate the practical feasibility of operation in the regime where the rate is maximised subject to the constraint that sustained queueing is avoided.

In summary, our main contributions are as follows.  Firstly, we establish that regulating send rate so as to maintain a target aggregation level can indeed be used to realise high rate, low latency communication over 802.11ac WLANs.  Secondly we address two important practical issues arising in production networks, namely that (i) many client devices are non-rooted mobile handsets/tablets and (ii) the bottleneck may lie in the backhaul rather than the WLAN, or indeed vary between the two over time.  We show that both these issues can be resolved by use of simple and robust machine learning techniques.  Thirdly, we present a prototype transport layer implementation of our low latency rate allocation approach and use this to evaluate performance under real radio channel conditions.

\section{Preliminaries}\label{sec:prelim}

\subsection{Aggregation in 802.11n, 802.11ac etc}

A feature shared by all WLAN standards since 2009 (when 802.11n was introduced) has been the use of aggregation to amortise PHY and MAC framing overheads across multiple packets.   This is essential for achieving high throughputs.   \edit{Since the PHY overheads are largely of fixed duration, increasing the data rate reduces the time spent transmitting the frame payload but leaves the PHY overhead unchanged.}  Hence, the efficiency, as measured by the ratio of the time spent transmitting user data to the time spent transmitting an 802.11 frame, decreases as the data rate increases unless the frame payload is also increased i.e. several data packets are aggregated and transmitted in a single frame. 


\subsection{Measuring Aggregation}\label{sec:expt}

\begin{figure}
\centering
\subfigure[10Mbps send rate]{
\includegraphics[width=0.46\columnwidth]{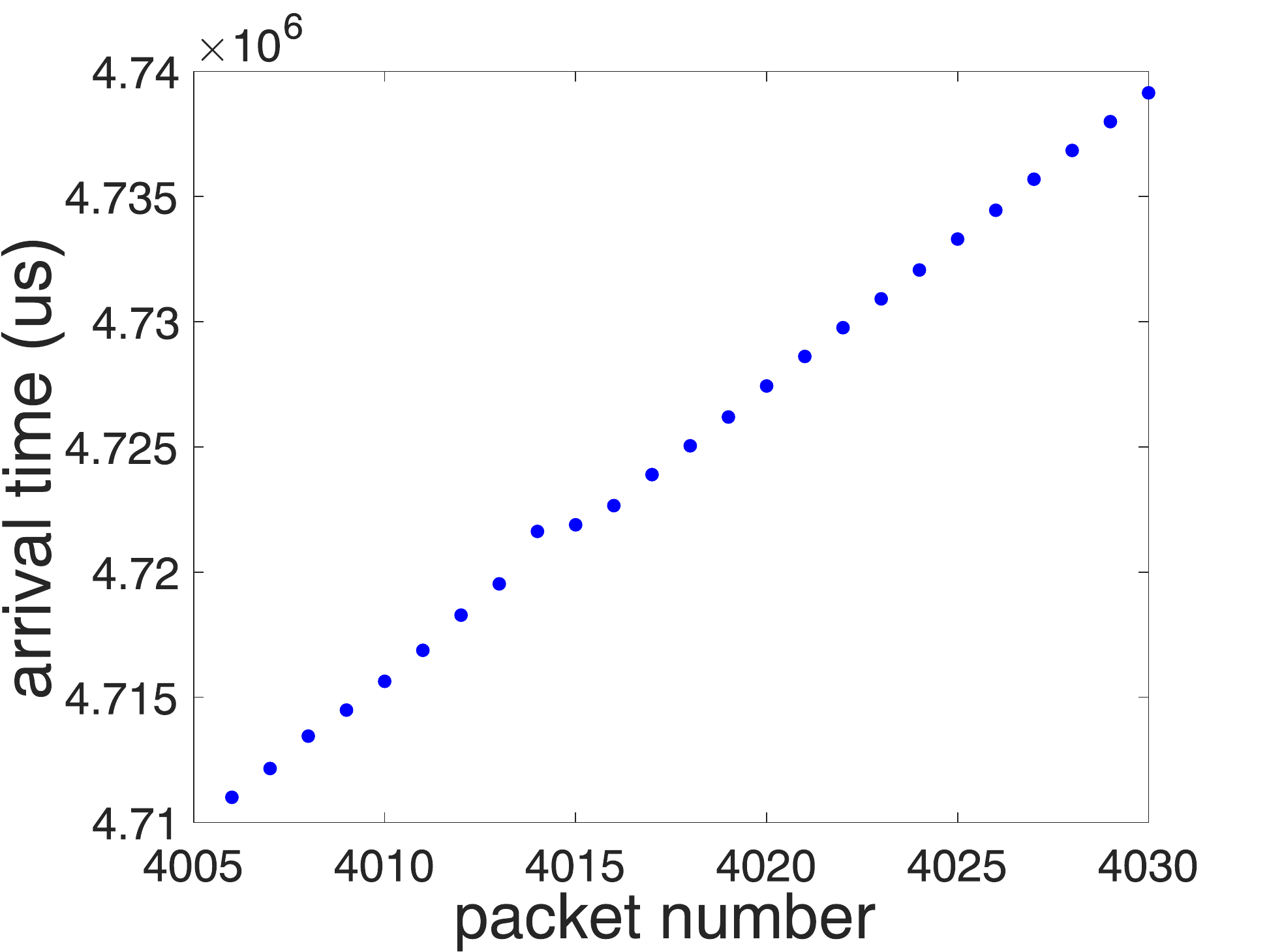}
}
\subfigure[200Mbps send rate]{
\includegraphics[width=0.46\columnwidth]{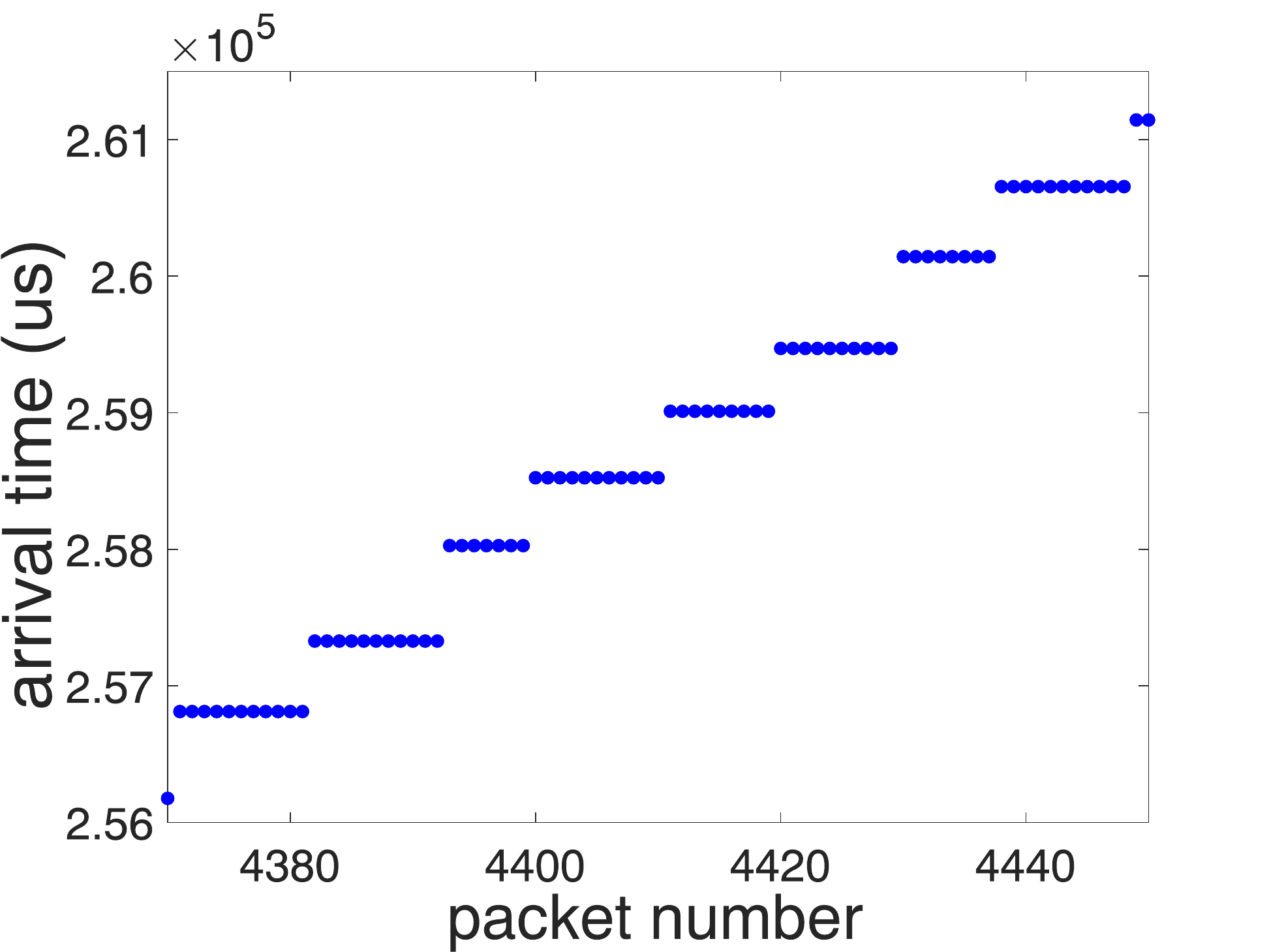}
}
\caption{MAC timestamp measurements for UDP packets transmitted over an 802.11ac downlink to two client stations.  Packets transmitted in the same frame have the same MAC timestamp, and it can be seen from (a) that while there tends to be only one packet per frame at a 10Mbps send rate this increases in (b) to around 10 packets per frame at 200Mbps.  The same downlink send rate is used for both client stations, data is shown for one client station.  Experimental data, setup in Appendix.}\label{fig:two}
\end{figure}

The level of aggregation can be readily measured at a receiver using packet MAC timestamps.   Namely, a timestamp is typically added by the NIC to each packet recording the time when it is received.  This timestamp is derived from the WLAN MAC and has microsecond granularity\footnote{Note that, as will be discussed in more detail later, a second timestamp is also added by the kernel but this is recorded at a later stage in the packet processing chain and so is significantly less accurate.}.  When a frame carrying multiple packets is received then those packets have the same MAC timestamp and so this can be used to infer which packets were sent in the same frame. 

For example, Figure \ref{fig:two} shows measured packet timestamps for two different downlink send rates.   The experimental setup used is described in the Appendix.  It can be seen from Figure \ref{fig:two}(a) that when the UDP arrival rate at the AP is relatively low each received packet has a distinct timestamp whereas at higher arrival rates, see Figure \ref{fig:two}(b), packets start to be received in bursts having the same timestamp.    This behaviour reflects the use by the AP of aggregation at higher arrival rates, as confirmed by inspection of the radio headers in the corresponding tcpdump data.  

\subsubsection{Link Layer Retransmission Book-keeping}
\begin{figure}
\centering
\includegraphics[width=0.46\columnwidth]{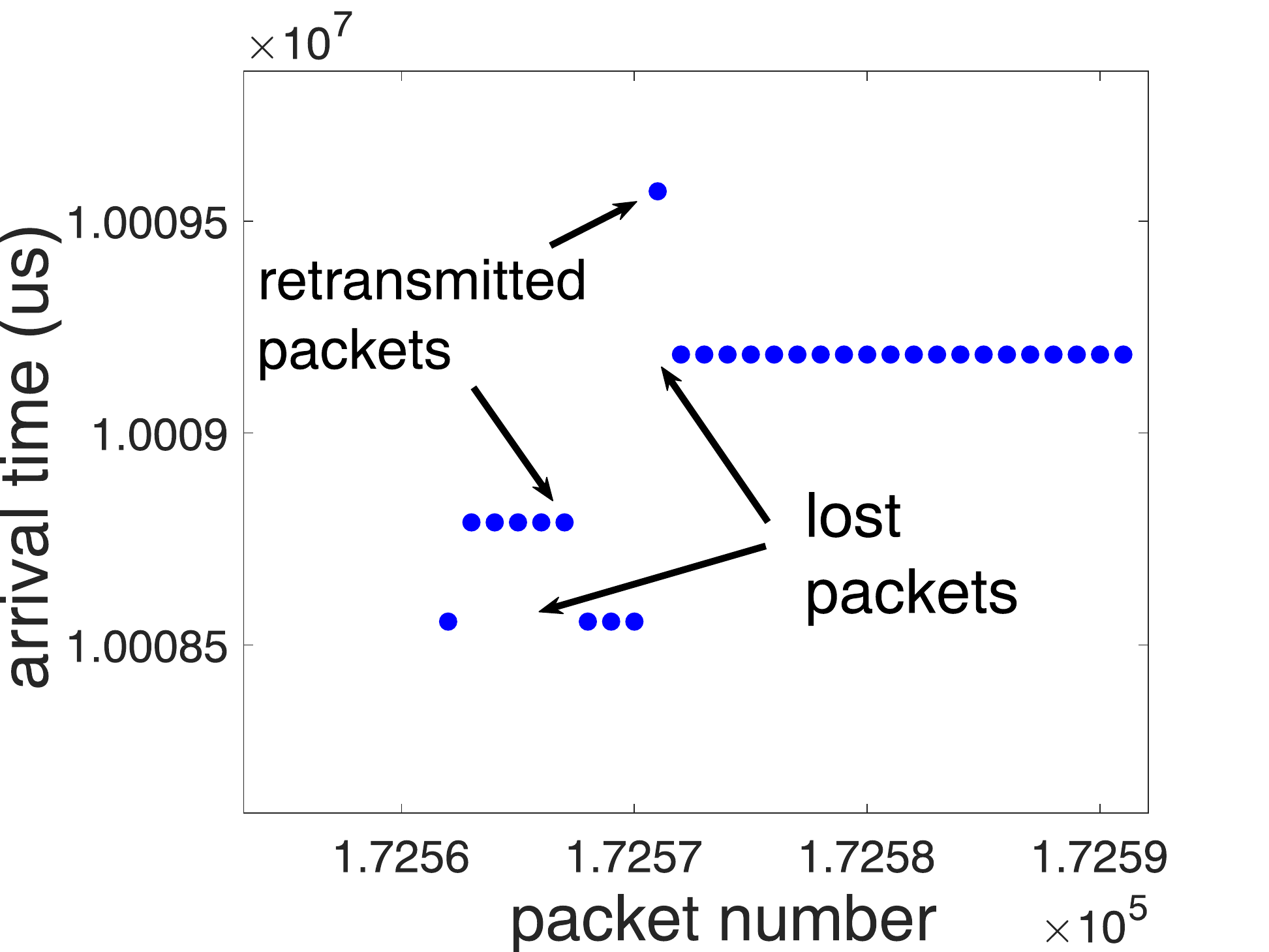}
\caption{Experimental measurements illustrating typical 802.11ac link layer packet loss and retransmission behaviour.  In the first frame a burst of five packets are lost and retransmitted in the second frame.  In the third frame the first packet is lost and retransmitted in the fourth frame. }\label{fig:loss}
\end{figure}

The 802.11ac link layer retransmits lost packets.  Our measurements indicate that these retransmissions usually occur in a dedicated frame in which case the aggregation level of that frame is often lower than for regular frames, e.g. see Figure \ref{fig:loss}.   Losses also mean that the number of received packets in a frame is lower than the number transmitted.  Fortunately, by inserting a unique sequence number into the payload of each packet we can infer both losses and retransmissions since they respectively appear as ``holes'' in the received stream of sequence numbers and as out of order delivery.  We can therefore adjust our book-keeping to compensate for these when estimating the aggregation level.


\section{Low Delay High-Rate Operation}\label{sec:low}

\begin{figure}
\centering
\subfigure[One station, NS3]{
\includegraphics[width=0.46\columnwidth]{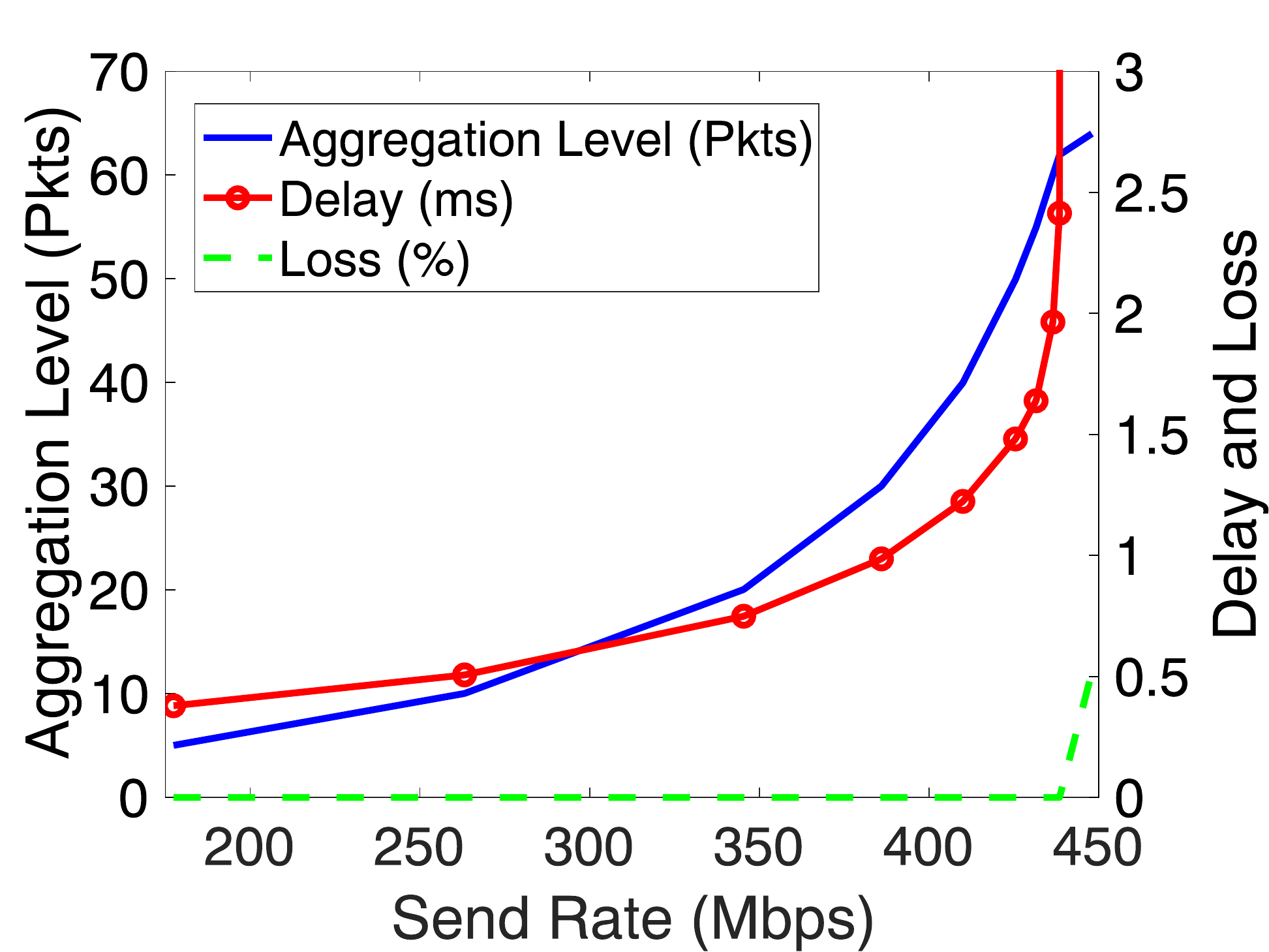}
}
\subfigure[10 stations, NS3]{
\includegraphics[width=0.46\columnwidth]{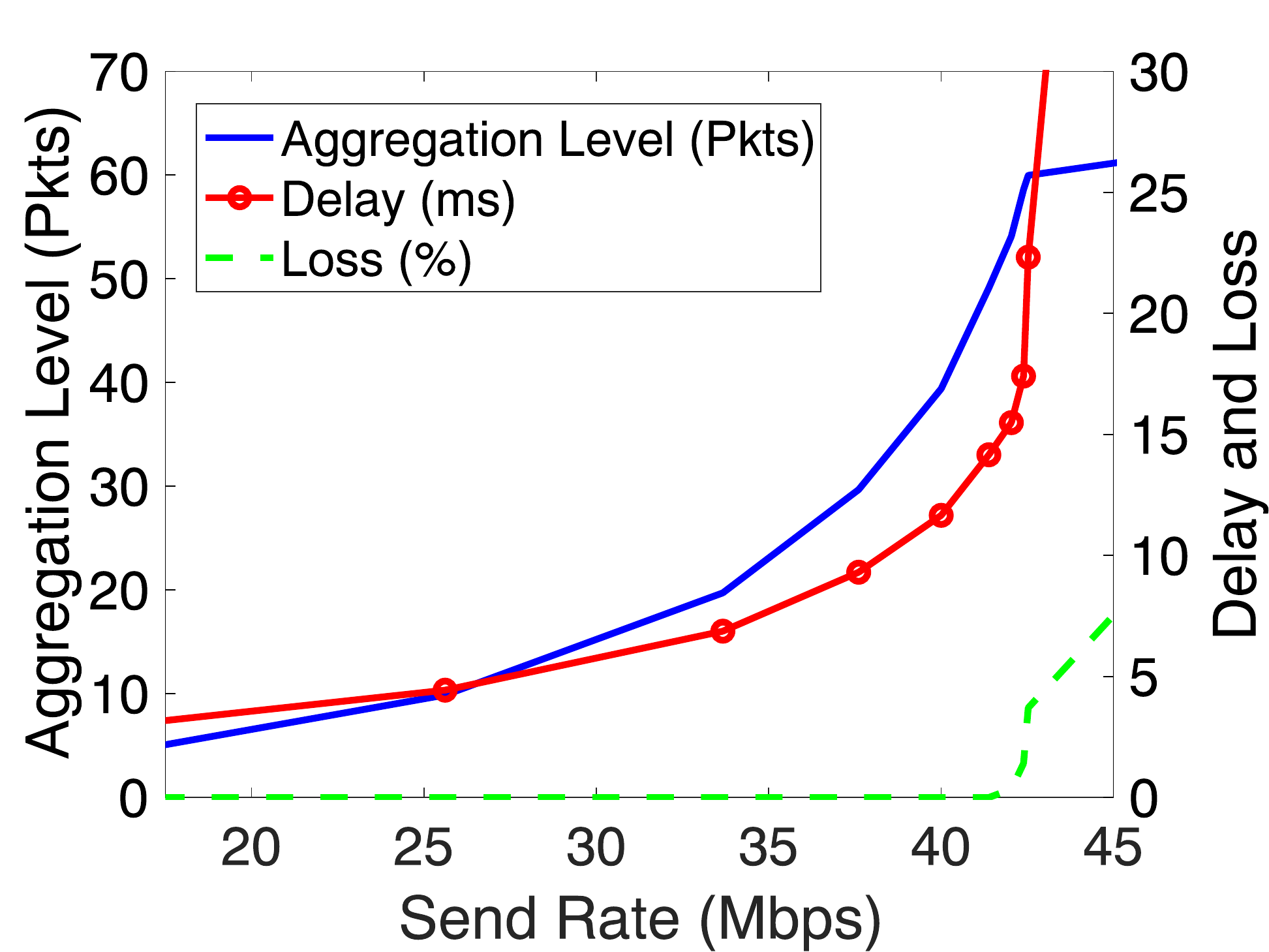}
}
\subfigure[One station \& uplink tx's, NS3]{
\includegraphics[width=0.46\columnwidth]{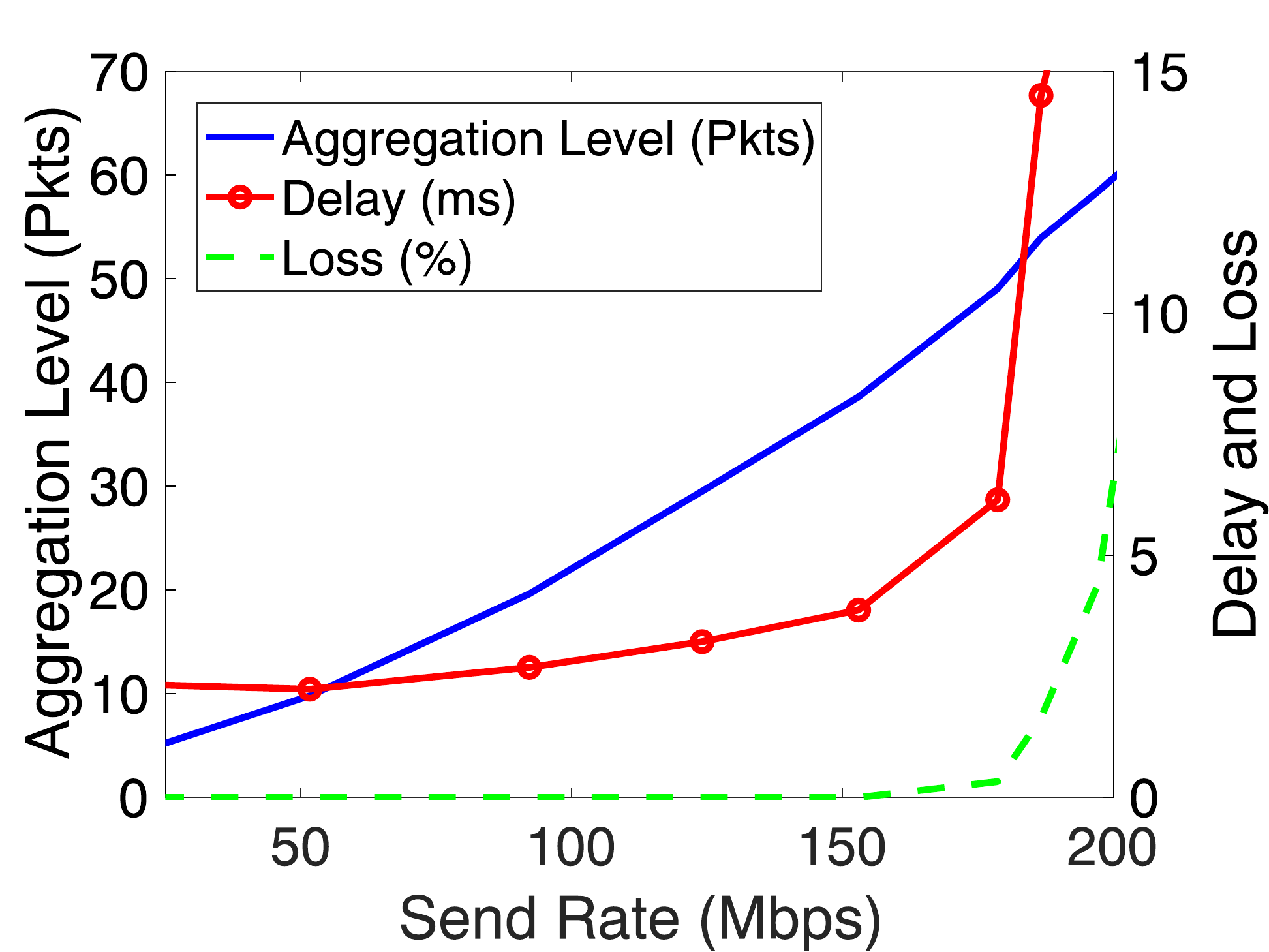}
}
\subfigure[One station, testbed]{
\includegraphics[width=0.46\columnwidth]{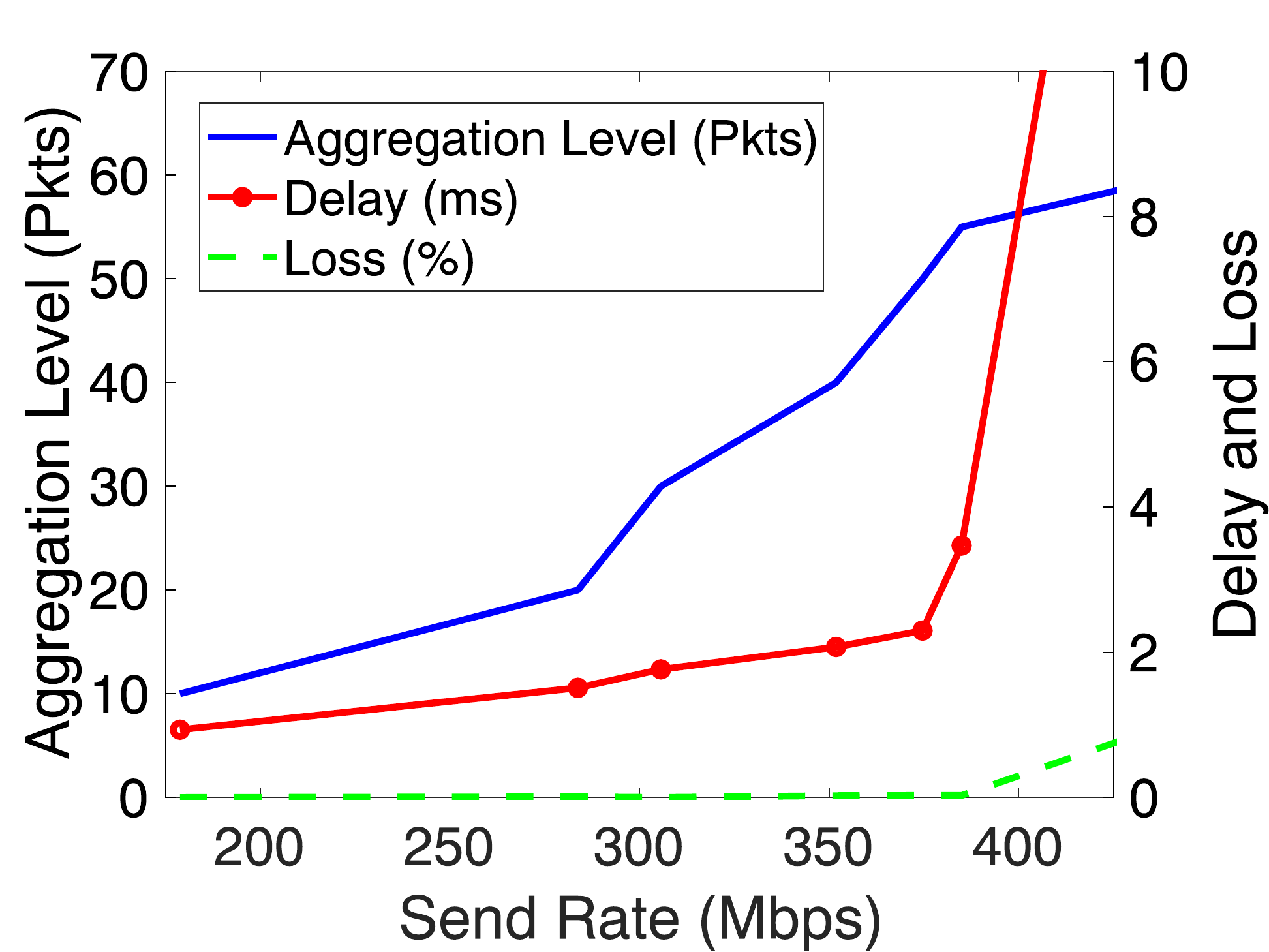}
}
\caption{Measurements of average aggregation level, one-way packet delay and packet loss vs the send rate for a range of network conditions.  (a) downlink flow to one client station, (b) 10 downlink flows to each of 10 client stations, data shown is for one of these flows, (c) setup as in (a) but with contention from an uplink flow, (d) setup as in (a) but measurements are from a hardware testbed located in an office environment. }\label{fig:intro}
\end{figure}

Figure \ref{fig:intro} shows measurements of the mean aggregation level, packet delay and loss vs the send rate to a client station for a range of network configurations.    A number of features are evident.  Firstly, as the send rate is increased the aggregation level increases monotonically until it reaches the maximum value $N_{max}$ supported by the MAC (for the data shown $N_{max}=64$ packets).   Secondly, the packet delay increases monotonically with send rate, initially increasing slowly but then increasing sharply as the send rate approaches the downlink capacity.  Observe that the sharp increase in delay coincides with aggregation level approaching its upper limit of 64 packets and with the onset of packet loss.  Note that all packet loss in this data is due to queue overflow since we verified that link layer retransmissions repair channel losses.   

\begin{figure}
\centering
\includegraphics[width=\columnwidth]{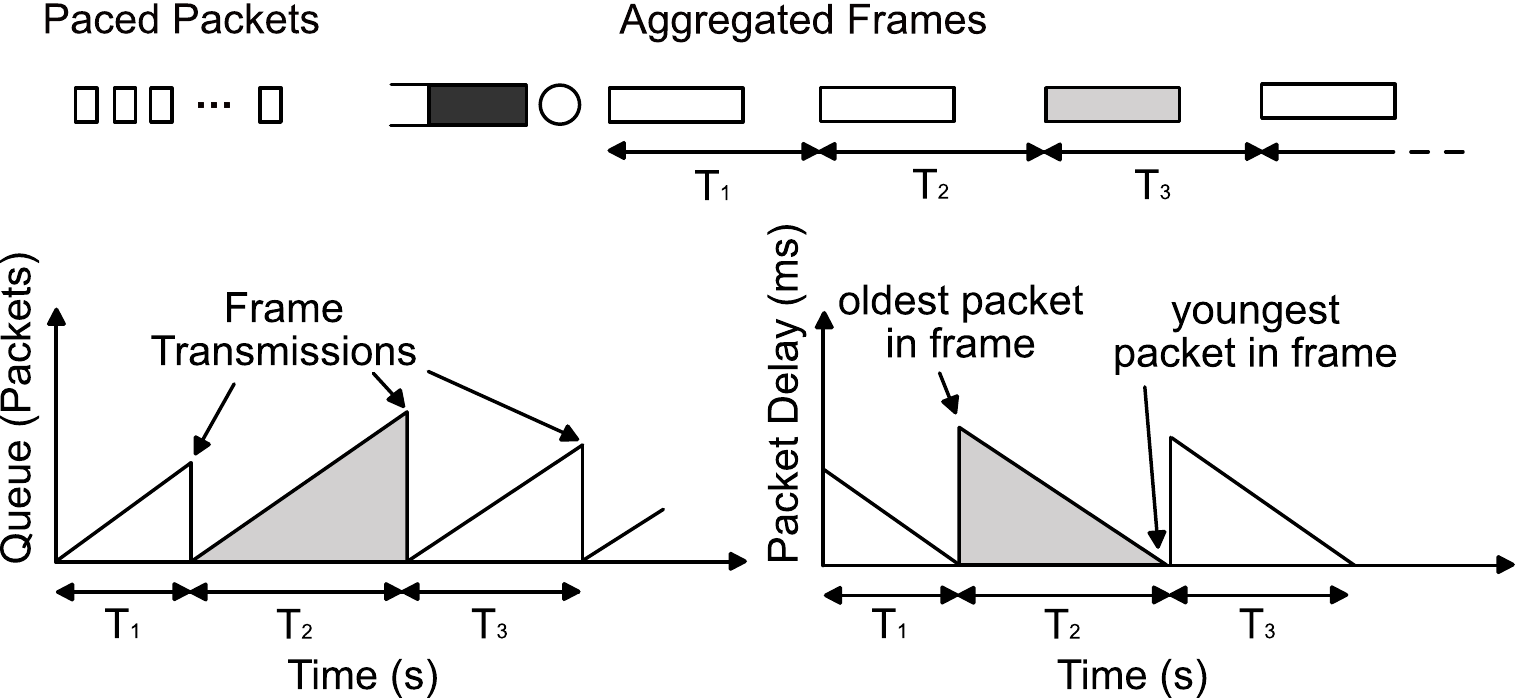} 
\caption{Illustrating connections between queueing, packet delay and frame aggregation at the AP.   Packets arriving at the AP are queued for transmission, the queue growing roughly linearly over time as packets arrive in a paced fashion.   When a transmission opportunity occurs an aggregated frame is constructed from the queued packets.  Provided the number of queued packets is less than the maximum frame aggregation level $N_{max}$ then the queue backlog is cleared by the transmission.  The delay of the oldest packet in a frame is upper bounded by the time between transmission opportunities.}\label{fig:delayschem}
\end{figure}

We can understand the behaviour in Figure \ref{fig:intro} in more detail by reference to the schematic in Figure \ref{fig:delayschem}.   Packets are transmitted by the sender in a paced fashion.  On arriving at the AP they are queued until a transmission opportunity occurs.  The queue occupancy increases roughly linearly since the arriving packets are paced (have roughly constant inter-arrival times).  Upon a transmission opportunity the queued packets are assembled into an aggregated frame and transmitted.   Provided the queue is less than $N_{max}$ the queue backlog is cleared by this transmission.  For example, consider the shaded frame in Figure \ref{fig:intro}.   This frame is transmitted at the end of time interval $T_2$ and the packets indicated by the shaded area on the queueing plot are aggregated into this frame.    The oldest packet in this frame could have arrived just after interval $T_1$ and so may have waited up to $T_2$ seconds before transmission.  Later arriving packets will, of course, experience less delay than this.   The intervals $T_1$, $T_2$ etc between frame transmissions are random variables due to the randomised channel access mechanism used by 802.11 transmitters.   Importantly, these intervals depend on the aggregation level, i.e. the duration $T_2$ depends on the time taken to send the frame aggregated from packets arriving in interval $T_1$ etc, and in turn the aggregation level depends on the interval duration since more packets arrive in a longer interval.   The delay and aggregation level are therefore coupled to one another and this is what we see in Figure \ref{fig:intro}.    Note that the intervals between transmissions may also vary due to contention with other transmitters (uplink transmissions by clients, transmissions by other WLANs sharing the same channel etc), link layer retransmissions, transmissions by the AP to other clients (recall modern APs use per station queueing so the coupling is only via these intervals) and so on but the basic setup remains unchanged and this is also reflected in Figure \ref{fig:intro}.    

The data in Figure \ref{fig:intro} suggests that if we could operate the system at an aggregation level of, for example, around 32 packets then we can obtain a high transmit rate while maintaining low delay.   It is this observation that underlies the approach we propose here.   Note that the AP transmit efficiency increases with the aggregation level since the overheads on a frame transmission are effectively fixed and so sending more packets in a frame increases efficiency.  Hence, operating at less than the maximum possible aggregation level $N_{max}$ incurs a throughput cost and there is therefore a trade-off between delay and rate.  However, Figure \ref{fig:intro} indicates that this trade-off is quite favourable, namely that low delay comes at the cost of only a relatively small reduction in rate compared to the maximum possible.

%

\subsection{Controlling Delay}


We proceed by introducing a simple feedback loop that adjusts the sender transmit rate (corresponding to the AP arrival rate, assuming no losses between sender and AP) to maintain a specified target aggregation level.   Namely, time is partitioned into slots of duration $\Delta$ seconds and we let $\T_{i,k}$ denote the set of frames transmitted to station $i$ in slot $k$.  Station $i$ measures the number of packets $N_{i,f}$ aggregated in frame $f$ and reports the average $\mean{N}{_i}(k):=\frac{1}{|\T_{i,k}|}\sum_{f\in\T_{i,k}}N_{i,f}$ back to the sender.    The sender then uses proportional-integral (PI) feedback\footnote{While design of more sophisticated control strategies is of interest, this is an undertaking in its own right and we leave this to future work.} to increase its transmit rate $x_{i}$ if the observed aggregation level $\mean{N}{_i}(k)$ is less than the target value $N_\epsilon$ and decrease it if $\mean{N}{_i}(k)>N_\epsilon$.   This can be implemented using the pseudo-code shown in Algorithm \ref{algo1}.   Note that this feedback loop involves three design parameters, update interval $\Delta$, feedback gain $K$ and target aggregation level $N_{\epsilon}$.  We consider the choice of these parameters in more detail shortly but typical values are $\Delta=500$ms or $1000$ms, $K=K_0/n$ with $K_0=1$ (where $n$ is the number of client stations in the WLAN) and $N_{\epsilon}=32$ packets.

\begin{algorithm}[H]
\begin{algorithmic}
\STATE $k=1$
\WHILE{1}
\STATE $\mean{N}{_i}\leftarrow\frac{1}{|\T_{i,k}|}\sum_{f\in\T_{i,k}}N_{i,f}$
\STATE $x_{i} \leftarrow x_{i} - K(\mean{N}{_{i}}-N_\epsilon)$  
\STATE $k\leftarrow k+1$
\ENDWHILE
\end{algorithmic}
\caption{Feedback loop adjusting transmit rate $x_i$ to regulate aggregation level $\mean{N}{_i}$.}\label{algo1}
\end{algorithm}

\noindent We implemented this feedback loop in our experimental testbed, see Appendix for details, and Figure \ref{fig:edge}(b) shows typical results obtained by regulating the aggregation level.   

\subsection{Multiple Stations: Equal Airtime Fairness}

When there are multiple client stations we can modify Algorithm \ref{algo1} as follows to allocate roughly equal airtime to each station.  Recall that the airtime used to transmit the payload of station $i$ is $T_i=\mean{N}{_i}L/\mean{R}{_i}$, where $L$ is the number of bits in a packet, $\mean{N}{_i}$ is the number of packets in a frame and $\mean{R}{_i}$ is the MCS rate used to transmit the frame in bits/s.   So selecting $\mean{N}{_i}=\mean{R}{_i}/\mean{R}{_{i^*}}$ makes the airtime equal with $T_i=T_{i^*}$ for all stations.   Letting 
$\vv{x}$ denote the vector of downlink send rates to ensure equal airtimes we therefore increase the rate $x_{i^*}$ of the station $i^*$ with highest MCS rate $\mean{R}{_i}(k)$ when its observed aggregation level $\mean{N}{_i}(k)$ is less than the target value $N_\epsilon$ and decreases  $x_{i^*}$ when $\mean{N}{_i}(k)>N_\epsilon$, i.e. at slot $k$
\begin{align}
x_{i^*}(k+1) = x_{i^*}(k) - K(\mean{N}{_i}(k)-N_\epsilon)\label{eq:loop0}
\end{align}
The rates of the other stations are then assigned proportionally,
\begin{align}
x_i(k+1) = x_{i^*}(k+1)\frac{\mean{R}{_i}}{\mean{R}{_{i^*}}},\ i=1,\dots,n\label{eq:loop0a}
\end{align}
%

Note that the update (\ref{eq:loop0})-(\ref{eq:loop0a}) only uses readily available observations.  Namely, the frame aggregation level $N_{i,f}$ and the MCS rate $R_{i,f}$, both of which can be observed in userspace by packet sniffing on client $i$.  

\subsection{Selecting Design Parameters $\Delta$ and $K_0$}

\subsubsection{Convergence Rate}
\begin{figure}
\centering
\subfigure[Impact of $K_0$ (with $\Delta=1000ms$)]{
\includegraphics[width=0.46\columnwidth]{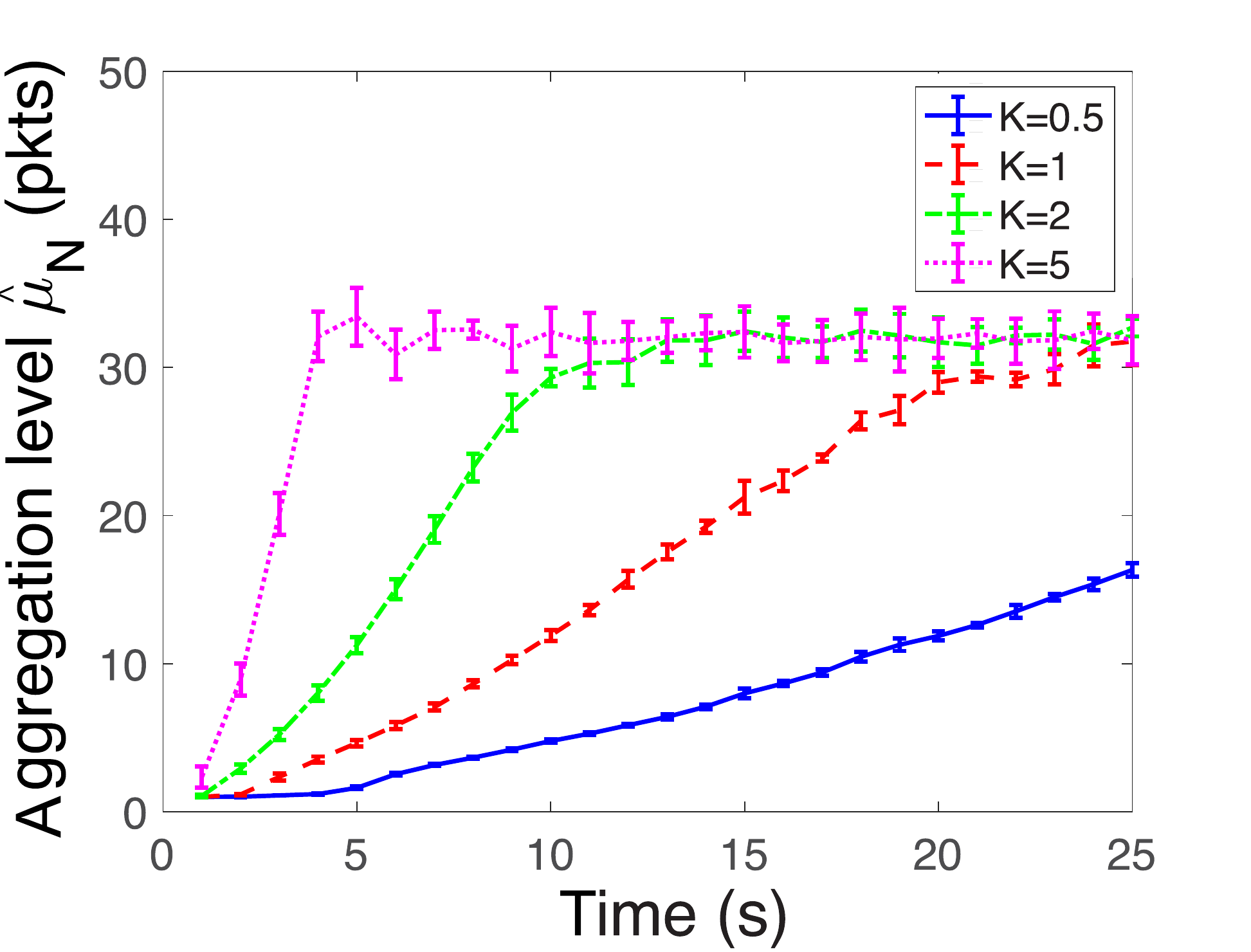}
}
\subfigure[Impact of $\Delta$ (with $K_0=1$)]{
\includegraphics[width=0.46\columnwidth]{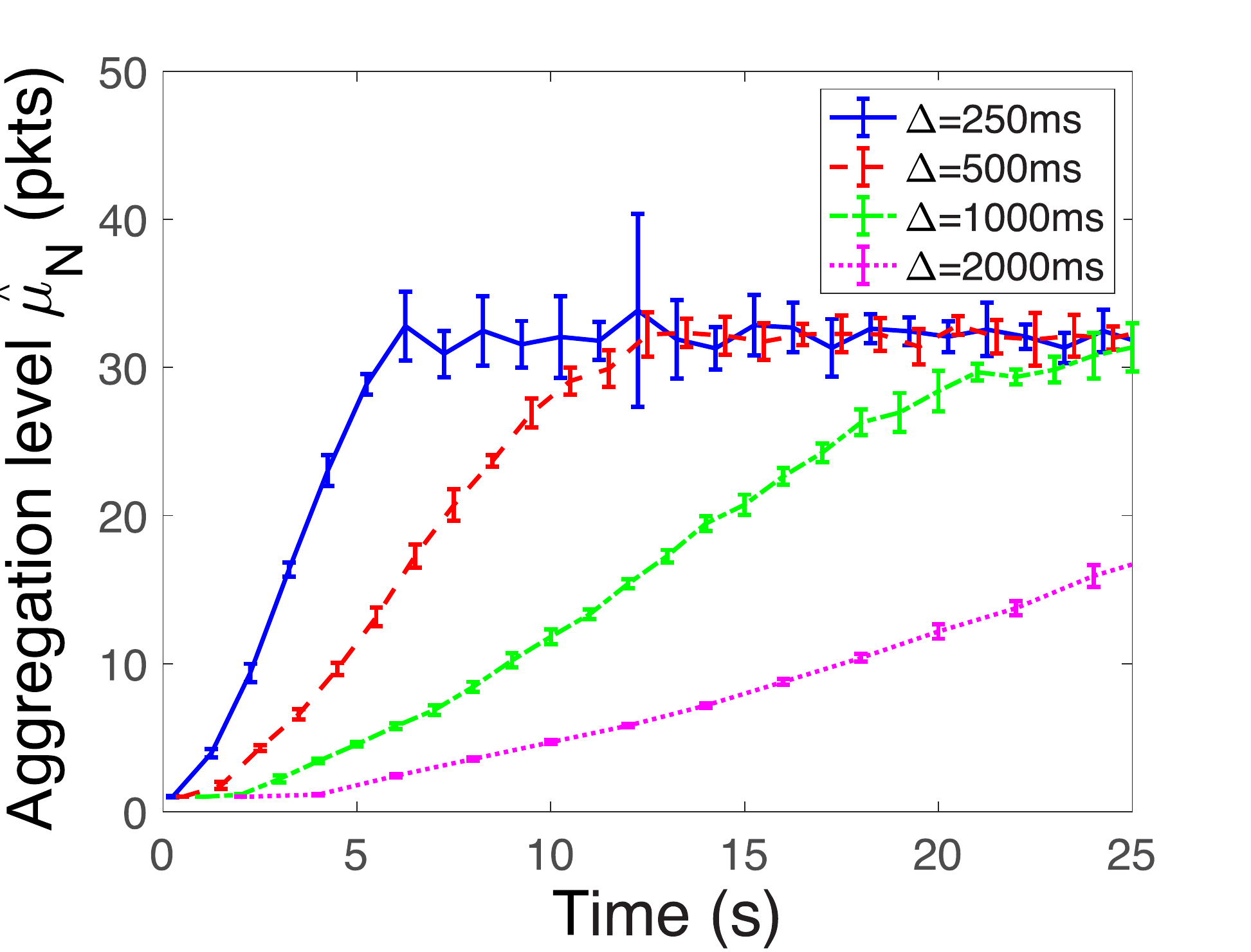}
}
\caption{Convergence rate vs feedback gain $K_0$ and update interval $\Delta$.  Mean and standard deviation from 10 runs at each parameter value.  One client, 802.11ac, $N_\epsilon=32$. Experimental data, setup in Appendix, $N_{max}=64$. }\label{fig:convergence1}
\end{figure}

We expect that the speed at which the aggregation level and send rate converge to their target values when a station first starts transmitting is affected by the choice of feedback gain $K_0$ and update interval $\Delta$.  Figure \ref{fig:convergence1} plots measurements showing the transient following startup of a station vs the choice of $K_0$ and $\Delta$.  

It can be seen from Figure \ref{fig:convergence1}(a) that as gain $K_0$ is increased (while holding $\Delta$ fixed) the time to converge to the target aggregation level $N_\epsilon=32$ decreases.    However, as the gain is increased the feedback loop eventually becomes unstable.  Indeed, not shown in the plot is the data for $K_0=10$ which shows large, sustained oscillations that would obscure the other data on the plot.   Similarly, it can be seen from Figure \ref{fig:convergence1}(b) that as the update interval $\Delta$ is decreased the convergence time decreases.  

\subsubsection{Disturbance Rejection}
Observe in Figure \ref{fig:convergence1}(a) that while the convergence time decreases as $K_0$ is increased the corresponding error bars indicated on the plots increase.   As well as the convergence time we are also interested in how well the controller regulates the aggregation level about the target value $N_\epsilon$.  Intuitively, when the gain $K_0$ is too low then the controller is slow to respond to changes in the channel and the aggregation level will thereby show large fluctuations.   When $K_0$ is increased we expect feedback loop is also able to respond more quickly to genuine changes in channel behaviour.   

\begin{figure}
\centering
\subfigure[Impact of $K_0$ ($\Delta=1000ms$).]{
\includegraphics[width=0.46\columnwidth]{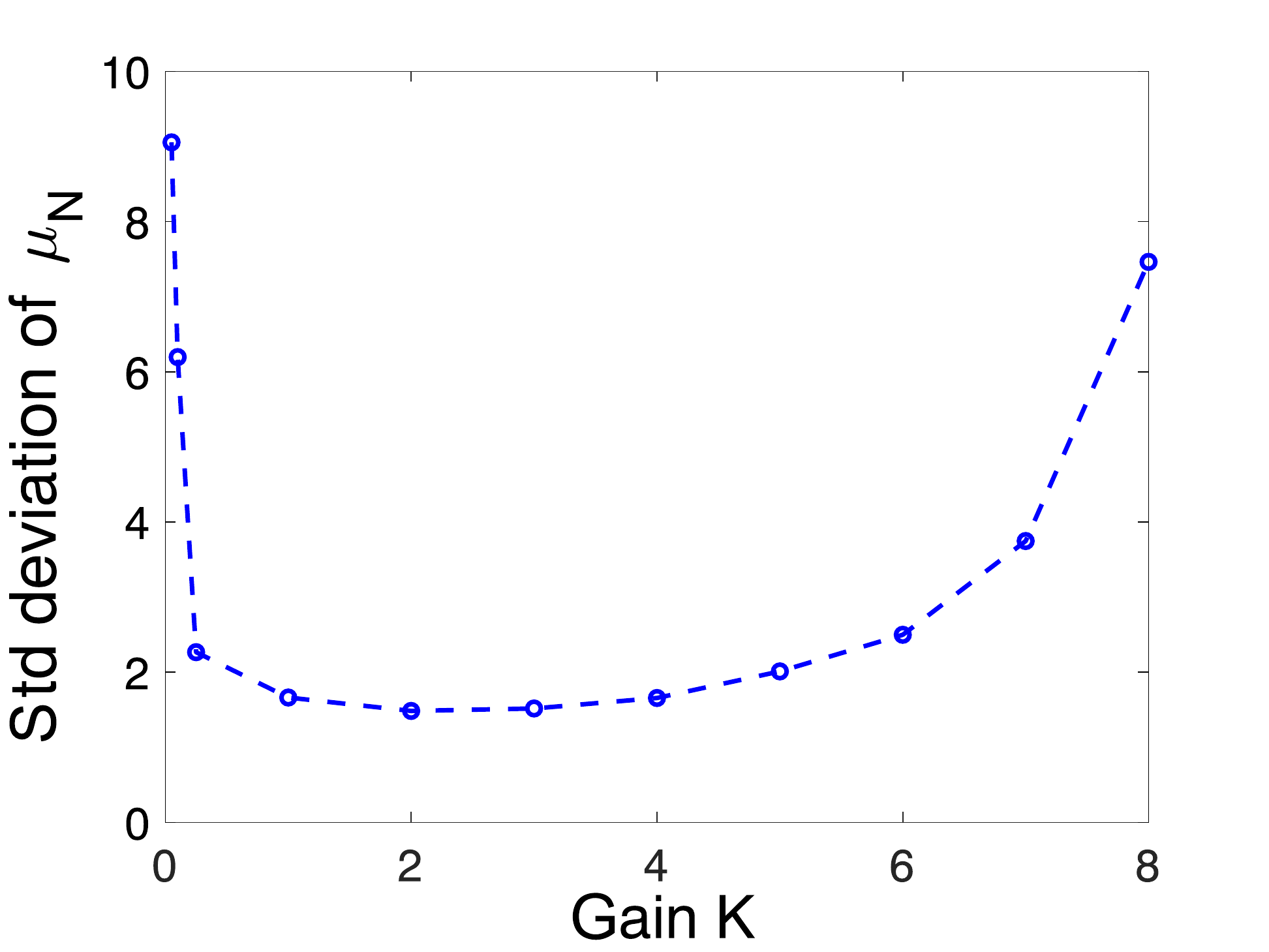}
}
\subfigure[Impact of $\Delta$ ($K_0=1$).]{
\includegraphics[width=0.46\columnwidth]{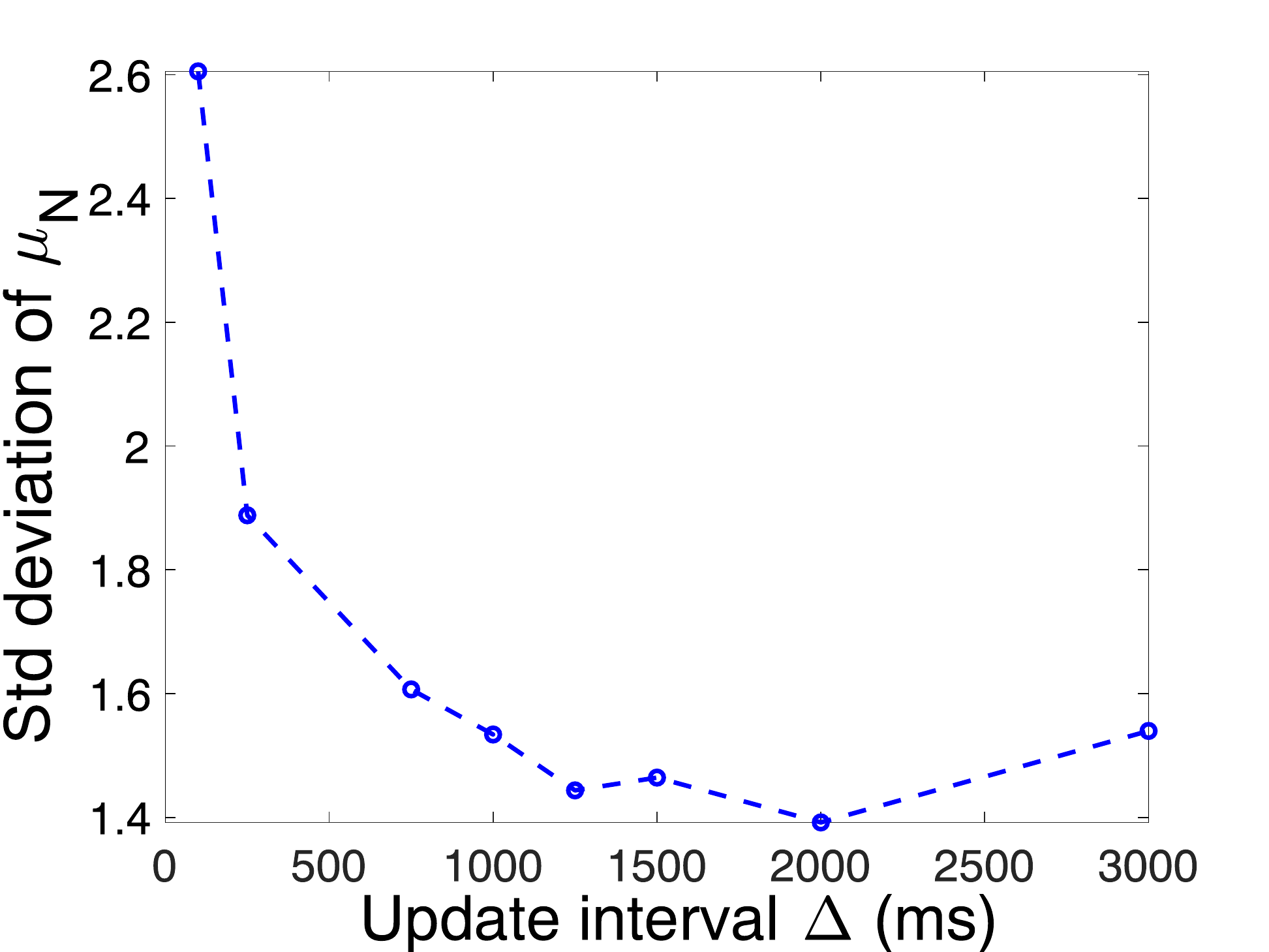}
}
\caption{Noise rejection (as measured by standard deviation of  $\mean{N}{}$) vs feedback gain $K_0$ and update interval $\Delta$.   One client, 802.11ac, $N_\epsilon=32$. Experimental data, setup in Appendix, $N_{max}=64$. }\label{fig:convergence2}
\end{figure}

This behaviour can be seen in Figure \ref{fig:convergence2}(a) which plots measurements of the standard deviation of the aggregation level $\mean{N}{}$ (where the empirical mean is calculated over the update interval $\Delta$ of the feedback loop) as the control gain $K_0$ is varied. When the update interval $\Delta$ is made smaller we expect that the observations $\mean{N}{_i}(k)$ and $\mean{R}{_i}(k)$ will tend to become more noisy (since they are based on fewer frames) which may also tend to cause the aggregation level to fluctuate more.   However, the feedback loop is also able to respond more quickly to genuine changes in channel behaviour.   Conversely, as $\Delta$ increases the estimation noise falls but the feedback loop becomes more sluggish.  

Figure  \ref{fig:convergence2}(b) plots measurements of the standard deviation of the aggregation level $\mean{N}{}$ as $\Delta$ is varied.   It can be seen that due to the interplay between these two effects the standard deviation of $\mean{N}{}$ increases when $\Delta$ selected too small or too large, with a sweet spot for $\Delta$ around 1250-2000ms.
 
 \begin{figure}
\centering
\includegraphics[width=0.95\columnwidth]{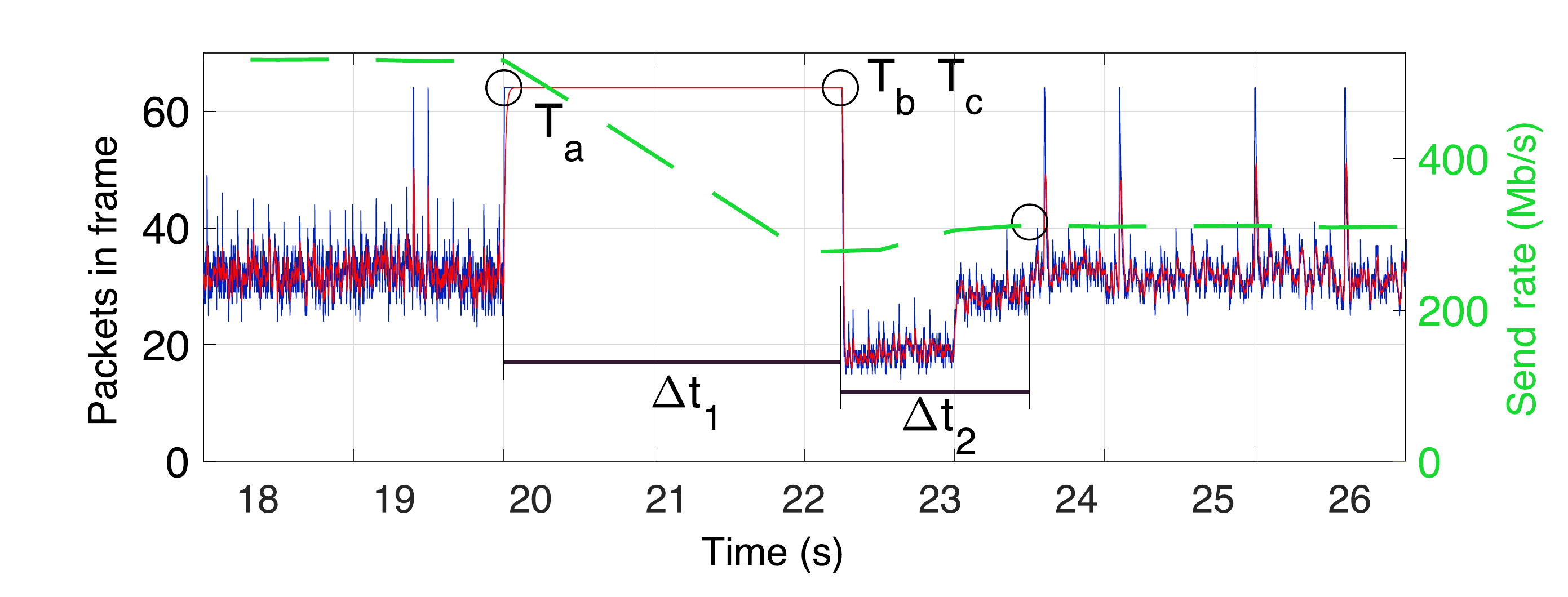}
\caption{Illustrating adaption of send rate by feedback algorithm in response to a  change in channel conditions (from use of 2 spatial streams down to 1 spatial stream).  NS3 simulation, single client station, MCS 9, $K_0=1$, $\Delta=500$ms, $N_\epsilon=32$, $N_{max}=64$. }\label{fig:ns3third}
\end{figure}

\subsubsection{Responding to Channel Changes}
The feeback algorithm used by the sender regulates the send rate to maintain a target aggregation level.  It therefore adapts automatically to changing channel conditions.   To investigate this we change to using NS3 simulations since this allows us to change the channel in a controlled manner (we also have experimental measurements showing adaptation to changing channel conditions, not shown here, but in this case we do not know ground truth).  

Fig.~\ref{fig:ns3third} illustrates typical behaviour of the feedback algorithm.   Initially the AP uses 2 spatial streams and then at $t = T_a = 20s$ it switches to 1 spatial stream.  For $\Delta t_1 = 2.24s$ all AMPDUs hit the maximum aggregation level of 64 packets and we start observing losses.  During this time it can be seen that the algorithm, which updates the send rate twice per second ($\Delta=500ms$), is slowing down the sending rate. It takes four rounds to reach a rate compatible with the channel, but it takes a little bit more to stabilise the aggregation level at the AP in $t = T_b$. After another three rounds ({for approximately $\Delta t_2 = 1.26s$}) it can be seen that the sending rate settles at its new value in $t = T_c$.  

\subsection{Fairness With High Rate \& Low Delay}

We now explore performance with multiple client stations.  We begin by considering a symmetric network configuration where client stations are all located at the same distance from the AP.  We then move on to consider asymmetric situations where the channel between AP and client is different for each client.   Again we use NS3 simulations in this section since this facilitates studying the performance with larger numbers of clients.

\subsubsection{Clients Same Distance From AP}
\begin{figure}
\centering
\includegraphics[width=0.8\columnwidth]{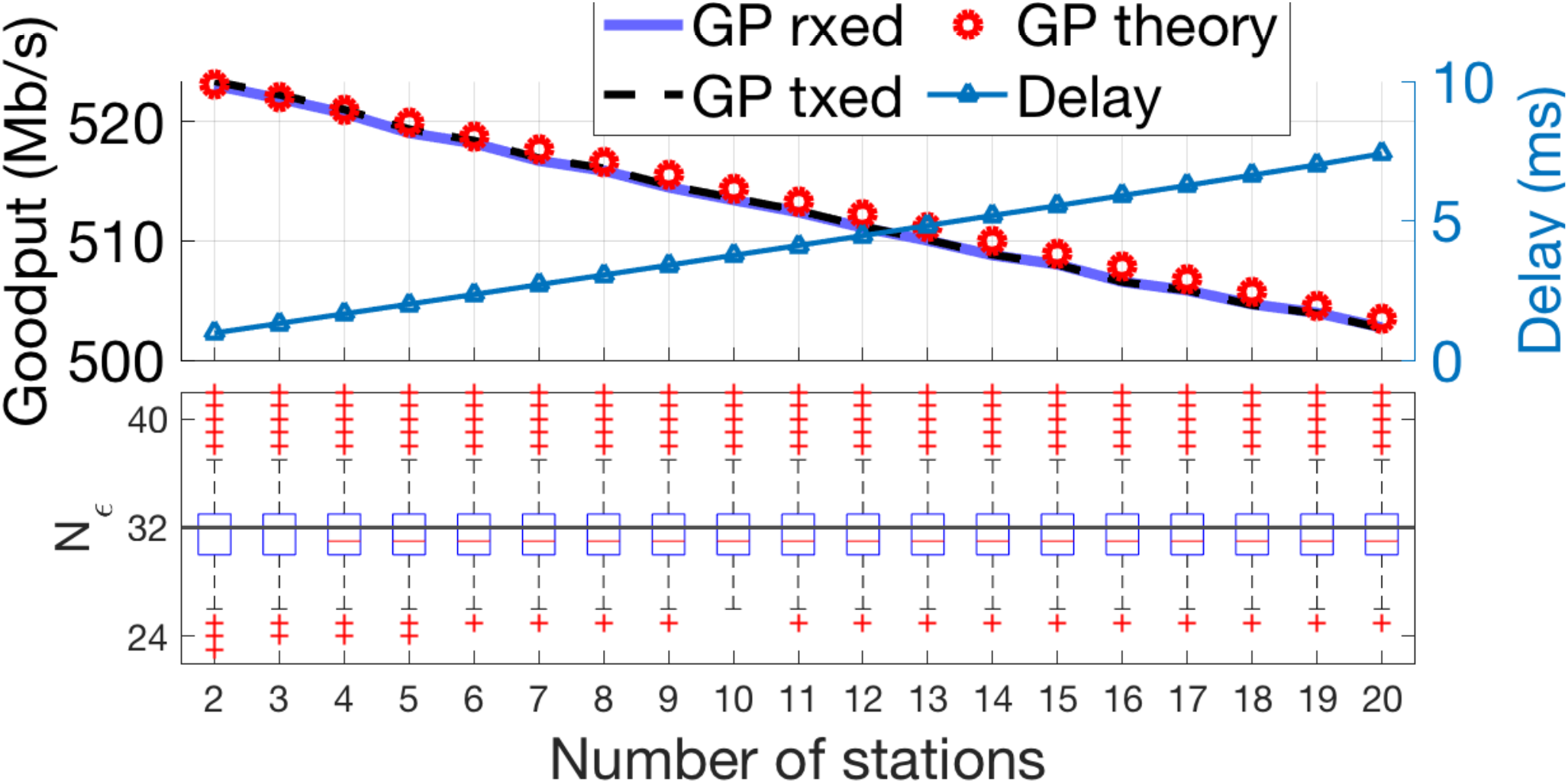}
\caption{Sum-goodput and delay vs number of receivers (top) and corresponding distribution of aggregation level about the target value of $N_{\epsilon}=32$ (bottom).  In the top plot the \textit{GP theory} line (GP, goodput) is a theoretical upper limit computed by assuming an AMPDU with $N_\epsilon=32$ packets, 10 feedback packets per second per receiver, 10 beacons per second, and no collisions. NS3 simulations, $K_0=1$, $\Delta=500$ms, $N_{max}=64$.}
\label{fig:ns3first}
\end{figure}

We begin by considering a network configuration where client stations are all located two meters from the AP.   Fig.~\ref{fig:ns3first} (top) shows measurements of the aggregated application layer goodput and average delay vs the number of receivers.
It can be seen that the aggregated goodput measured at the receivers  is close to the {theoretical limit supported by the channel (MCS) configuration}, being only a few Kb/s below this for 20 receivers.   This goodput is evenly shared by the receivers (the measured Jain's Fairness Index is always $1$).  The average delay increases almost linearly at the rate of $~350\mu s$ per additional station.   The lower plot in Fig.~\ref{fig:ns3first} shows the measured distribution of frame aggregation level, with the edges of the boxes indicating the 25th and 75th percentiles.   It can be seen that the feedback algorithm tightly concentrates the aggregation level around the target value of $N_{\epsilon}=32$.    As expected, since the delay is regulated to a low level we did not observe any losses.   

\begin{figure}
\centering
\includegraphics[width=0.95\columnwidth]{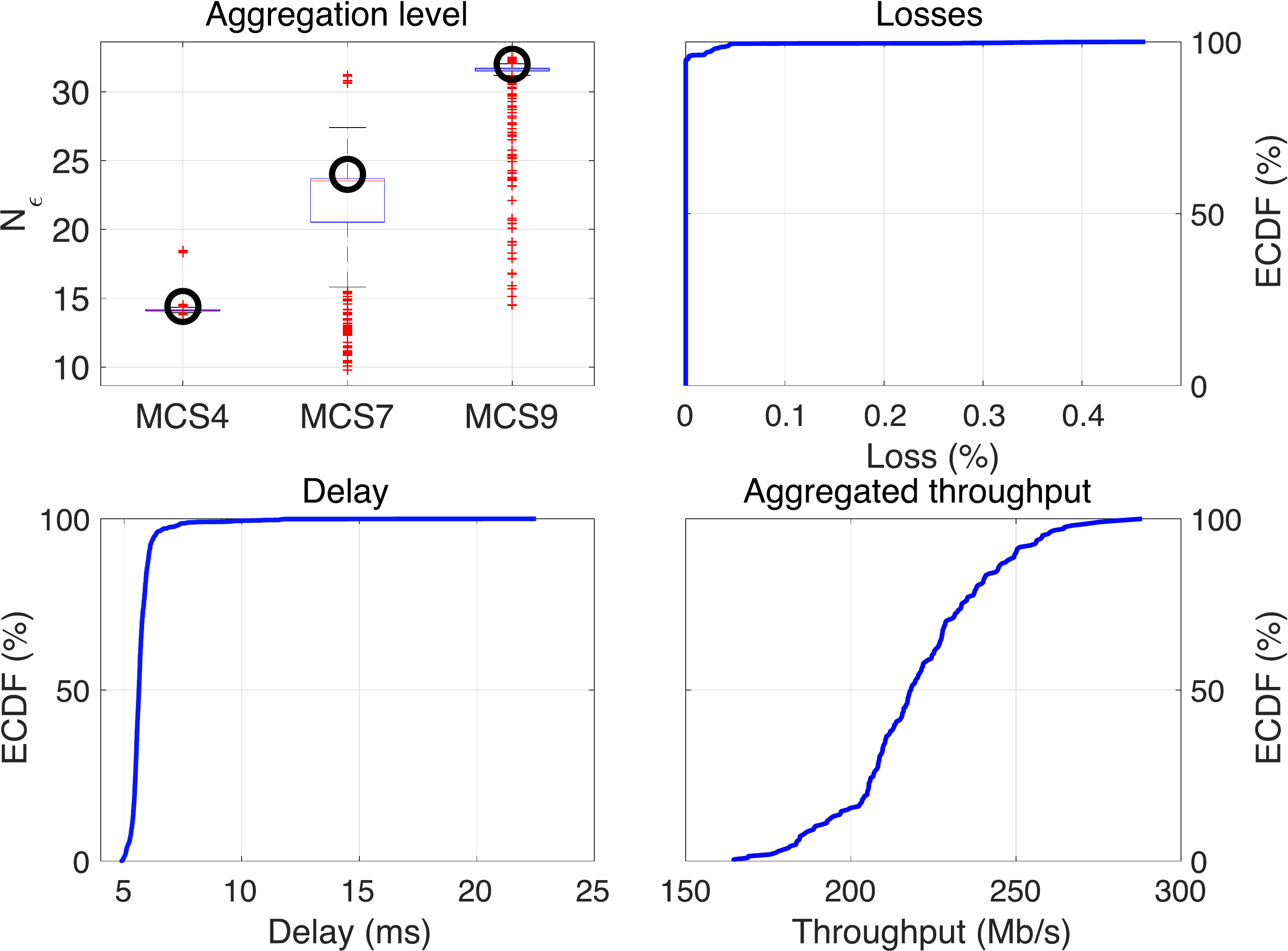}
\caption{Performance with 8 receivers placed randomly in a square of side 40m: MCS is chosen by MinstrelHT algorithm, NSS = 1.   NS3 simulations, $K_0=1$, $\Delta=500$ms, $N_{max}=64$.}
\label{fig:ns3second}
\end{figure}

\subsubsection{Randomly Located Clients} 
We now consider a scenario where the client stations are randomly located in a square of side 40m and the AP is located in its centre.     We configured MinstrelHT algorithm as the rate controller, this adjusts the MCS used by each client station based on its channel conditions (better for stations closer to the AP, worse for those further away). {To ease visualisation  we use NSS=1 which helps to reduce the} MCS fluctuations generated by Minstrel.   We ran experiments with eight receivers until we collected 200 points where the rate controller converged to a stable choice for all receivers, i.e., with more than $85\%$ of frames received with the same MCS.  We group receivers by MCS and report statistics on $N_\epsilon$ for each group as boxplots in the top-left plot in {Figure \ref{fig:ns3second}}.    The thick circles indicate the choice of rate allocation that assigns equal airtime to all receivers, and it can be seen that the measured rate allocation {is} maintained close to those values by the feedback algorithm.

In clockwise order {Figure \ref{fig:ns3second}} then shows the ECDF of losses, aggregated goodput and delay.   Observe that losses occur in this configuration because of far away nodes not being able to correctly decode all packets.
The aggregate goodput can drop as low as $150$Mb/s when MinstrelHT selects MCS4 for all receivers, but converges to $300$Mb/s with MCS 9 (the theoretical maximum goodput with MCS 9, NSS 1 and $N_\epsilon=32$ is $307$Mb/s).
Delay is consistently less than 6ms. 


\subsubsection{Coexistence With Legacy WLANS}
We next analyse the performance of stations which regulate the aggregation level when they co-exist with legacy stations.

\begin{figure}
 \centering
 \subfigure[]{
 \includegraphics[width=0.4776\columnwidth]{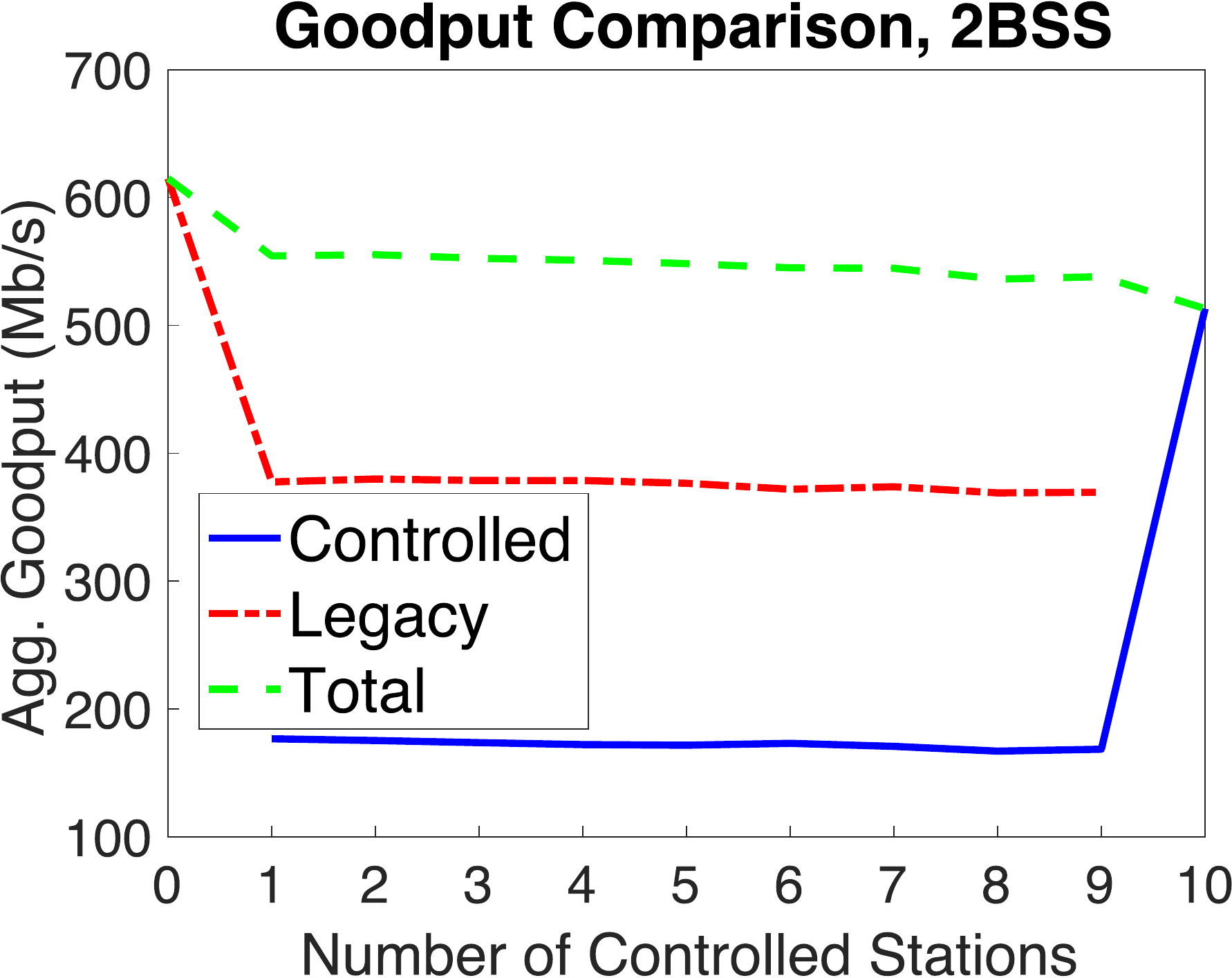} 
 }
  \subfigure[]{
 \includegraphics[width=0.46\columnwidth]{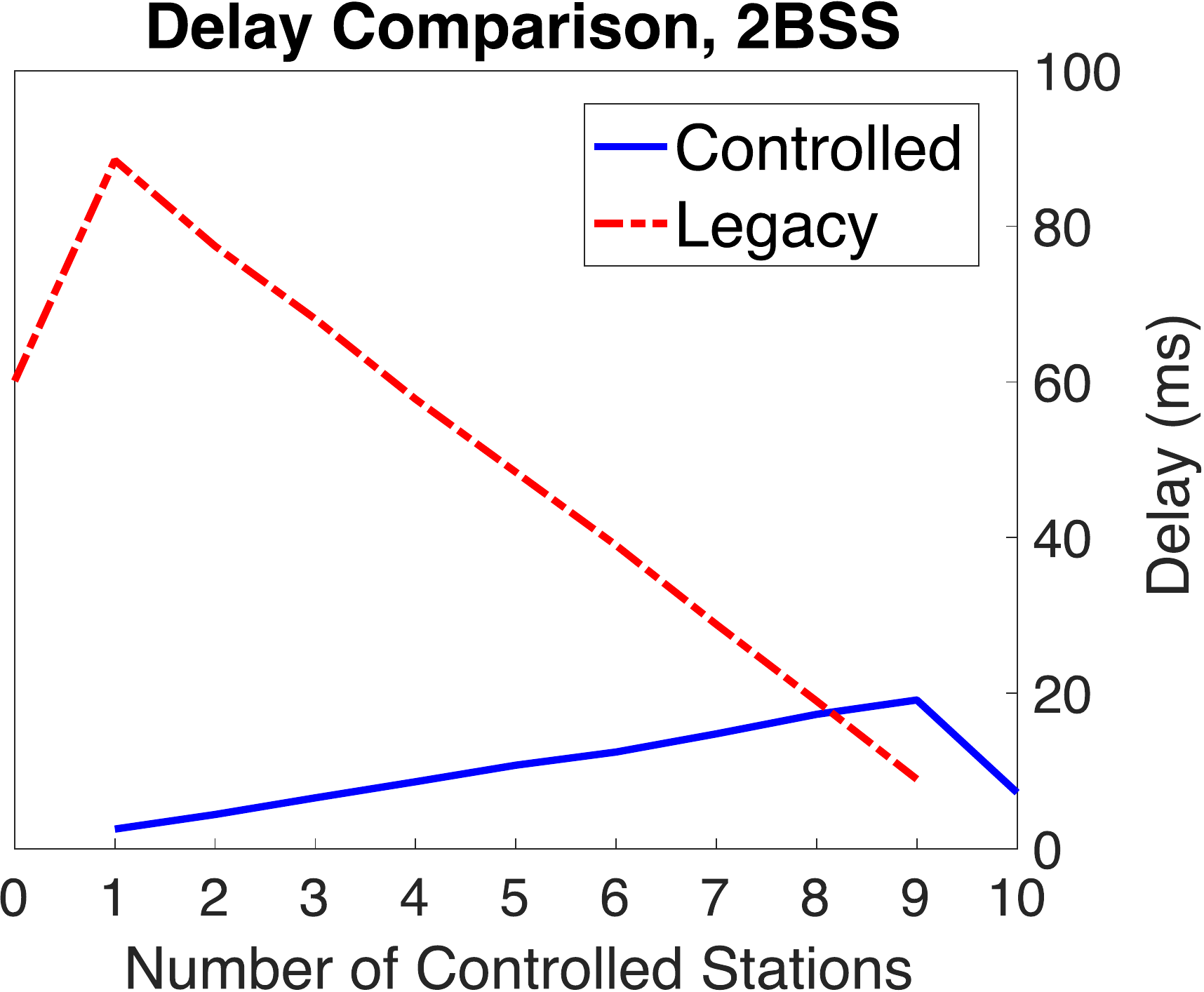} 
 }
  \subfigure[]{
 \includegraphics[width=0.4776\columnwidth]{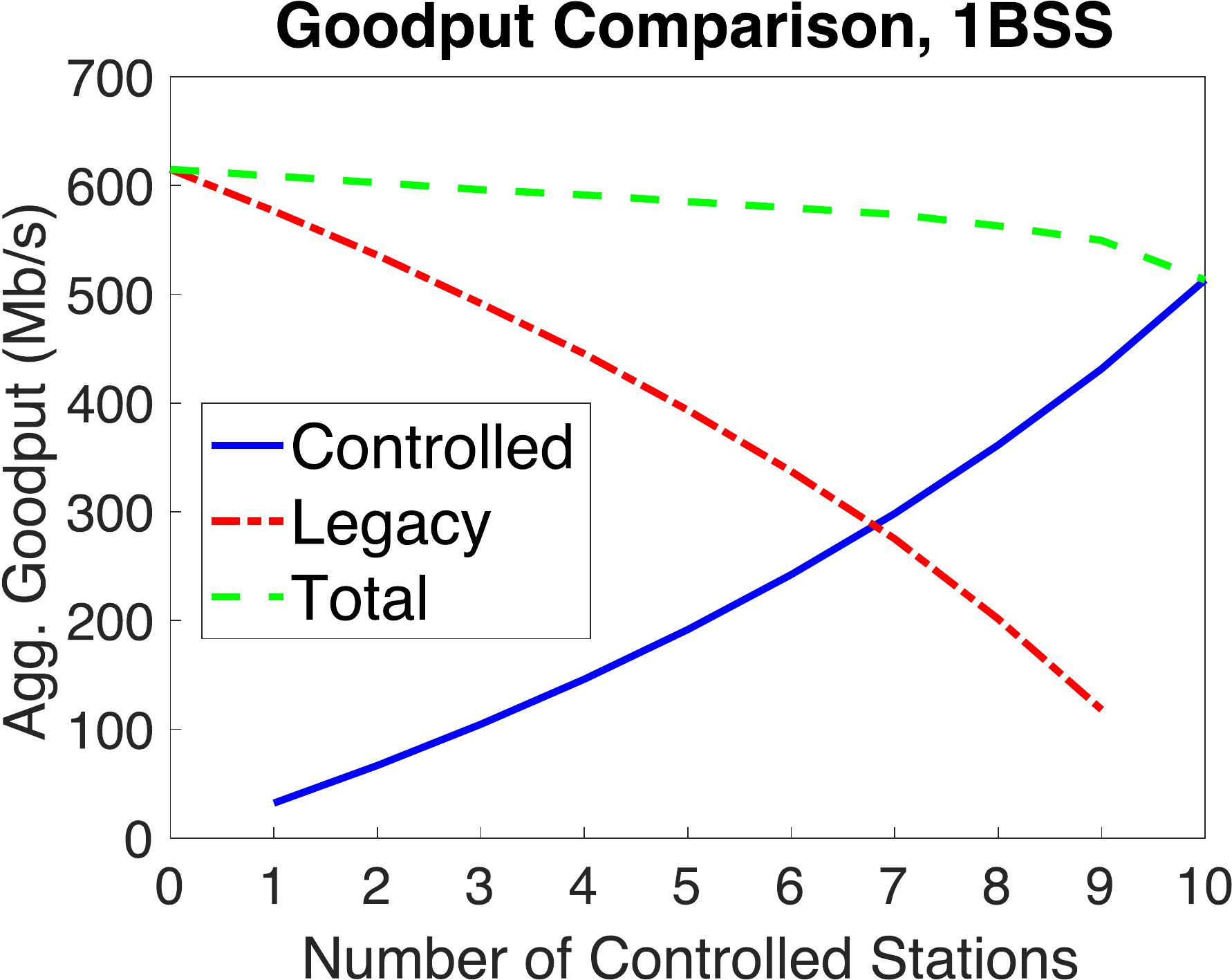} 
 }
  \subfigure[]{
 \includegraphics[width=0.46\columnwidth]{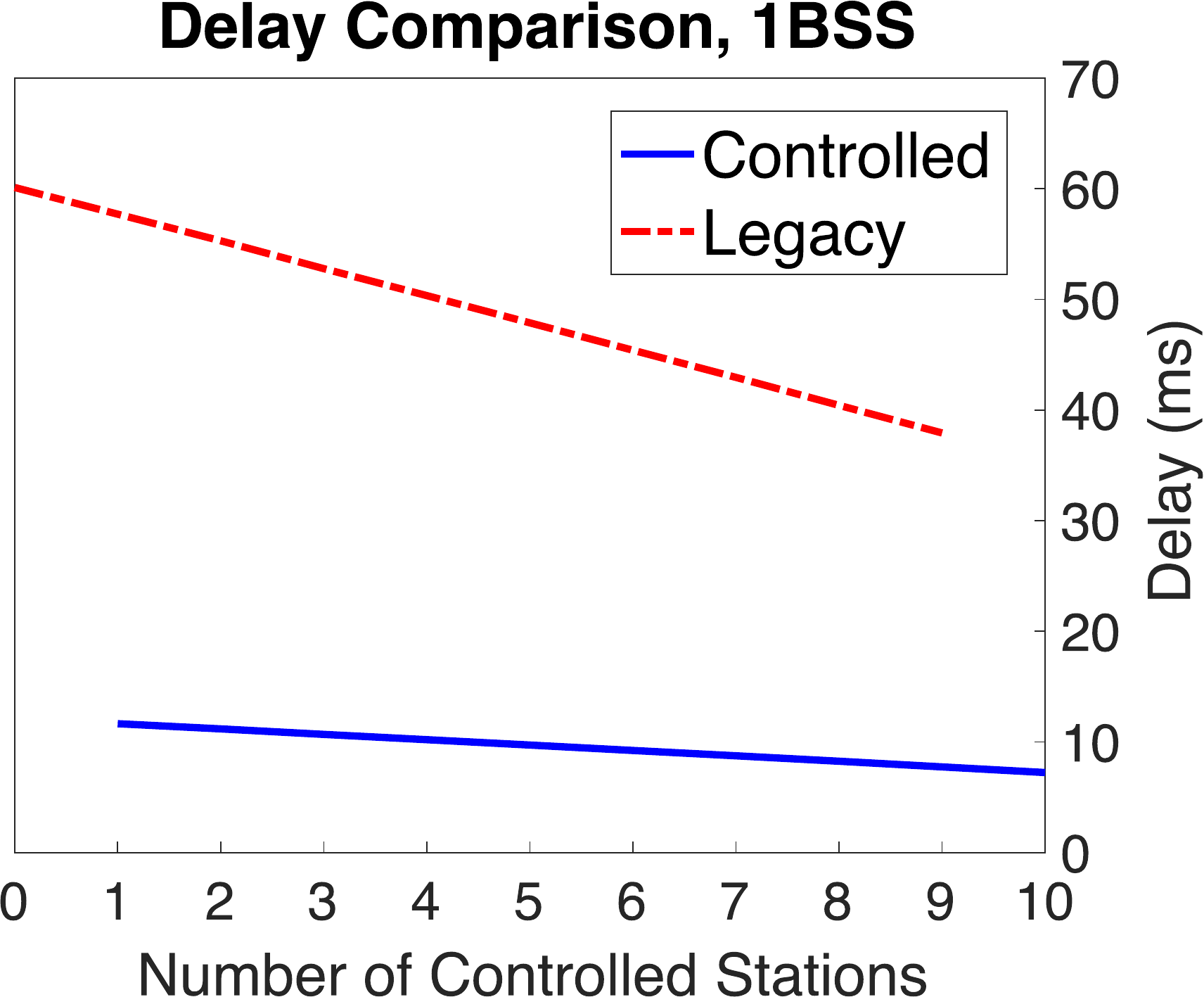} 
 }
 \caption{Top: comparison of Goodput (left) and Delay (right) when Controlled and Legacy stations form two BSSs with separate APs. Bottom: same comparison when all stations are joined to the same AP. NS3 simulations, $K_0=1$, $\Delta=500$ms, $N_{max}=64$. }\label{fig:figcomp}
\end{figure}

Figures~\ref{fig:figcomp}(a)-\ref{fig:figcomp}(b) show the measured delay and performance for a setup with two WLANs sharing the same channel.  The first WLAN contains stations that regulate the aggregation level on the downlink while in the second WLAN the AP has a persistent backlog and so always has packets to send to each station.   We hold the total number of client stations constant at $10$ but vary the fraction which regulate the aggregation level, as indicated on the x-axis of the plots.   

When the number of new controlled stations is zero, i.e. there are 10 saturated legacy stations sharing the same AP, then the aggregated {goodput} shown in Figures~\ref{fig:figcomp}(a) is close to the theoretical prediction for when the aggregation level is 64 packets (approximately $615$Mb/s).   When the number of controlled stations is 10, i.e. there are 10 controlled stations sharing the same AP, then it can be seen that the aggregated {goodput} is around 515Mbps, close to the theory value when the aggregation level is 32 packets (which is the target value for the controlled stations).  

When there is a mix of controlled and legacy stations it can be seen that the legacy stations gain a higher fraction of the total {goodput} than the stations which control aggregation level, as expected since the legacy stations use the maximum possible aggregation level $N_{max}=64$.  {However, this gain in {goodput} comes at the price of higher delays for the legacy stations. If we compare the delay achieved for the same number of stations in the two groups, it can be seen that the delay of the legacy stations is approximately four times that of stations that regulate the throughput.}  The {goodput} fraction is (almost) invariant with the number of stations since the 802.11 MAC assigns an equal number of transmission opportunities to each AP regardless of the number of clients in each WLAN.

This data confirms that WLANs where stations regulate their aggregation level can coexist successfully with legacy WLANS.   Namely, legacy WLANs are not penalised and the new controlled stations are still able to achieve low delay at reasonably high rates.

Figures~\ref{fig:figcomp}(c)-\ref{fig:figcomp}(d) show corresponding measurements when the legacy stations share the same WLAN as the new controlled stations.
\
%
 Now the fraction of {goodput} allocated to each class of station changes as the number of controlled stations is varied.  This is because both legacy and controlled stations share the same AP and this uses round-robin scheduling.  
Once again, as expected legacy stations achieve higher {goodput} than stations which regulate the aggregation level.  The high aggregation level used by the legacy stations means that their transmissions take more airtime. 
{This induces delay for the new controlled stations since they must wait for the legacy station transmissions to complete due to the round-robin scheduling used by the AP: still the delay for the controlled stations is always less than $12ms$ even in the worst case with a single controlled station against 9 legacy stations and it falls to $7ms$ when all stations are controlled.}



 \section{Non-rooted Mobile Handsets}\label{sec:noroot}
 
The results in the previous section make use of MAC timestamps to measure the aggregation level of frames received at WLAN clients.  However, access to MAC timestamps is via prism/radiotap libpcap headers and typically requires root privilege.   While this is fine for devices running operating systems such as Linux or OS X, it is problematic for mobile handsets and tablets since root privilege is generally not enabled for users on Android and iOS.   Mobile handsets/tablets are, of course, the primary means of accessing the internet for many users and so for our low latency approach to be of practical use it is important that it is compatible with these.
 
Potentially we can sidestep this constraint by use of a separate network sniffer with root access.  But this is unappealing for at least two major reasons.  Firstly, it entails installation of additional infrastructure, with associated cost and maintenance and unfavourable economics as cell sizes shrink.    Secondly, sniffing in monitor mode is itself becoming increasingly complex due to use of MIMO (the directionality makes it difficult to achieve monitor coverage) and very high rate PHYs (making sniffing liable to error/corruption).    Similarly, deploying measurement mechanisms on the AP may not be an option: manufacturers, in fact, restrict access to their devices and the update cycle can be much slower than in the case of a proxy software running on an edge- or cloud- server.

With the above in mind, we note that the kernel adds timestamps to received packets and these are accessible on mobile handsets via the {\tt SO\_TIMESTAMP} socket option without root privilege.  However, the kernel timestamp for a packet is recorded when the received packet is moved to the receive ring buffer and so is subject to significant ``noise'' associated with driver scheduling e.g. interrupt mitigation may lead to a switch to polled operation when under load.   

\begin{figure}
 \centering
 \subfigure[100Mbps]{
 \includegraphics[width=0.46\columnwidth]{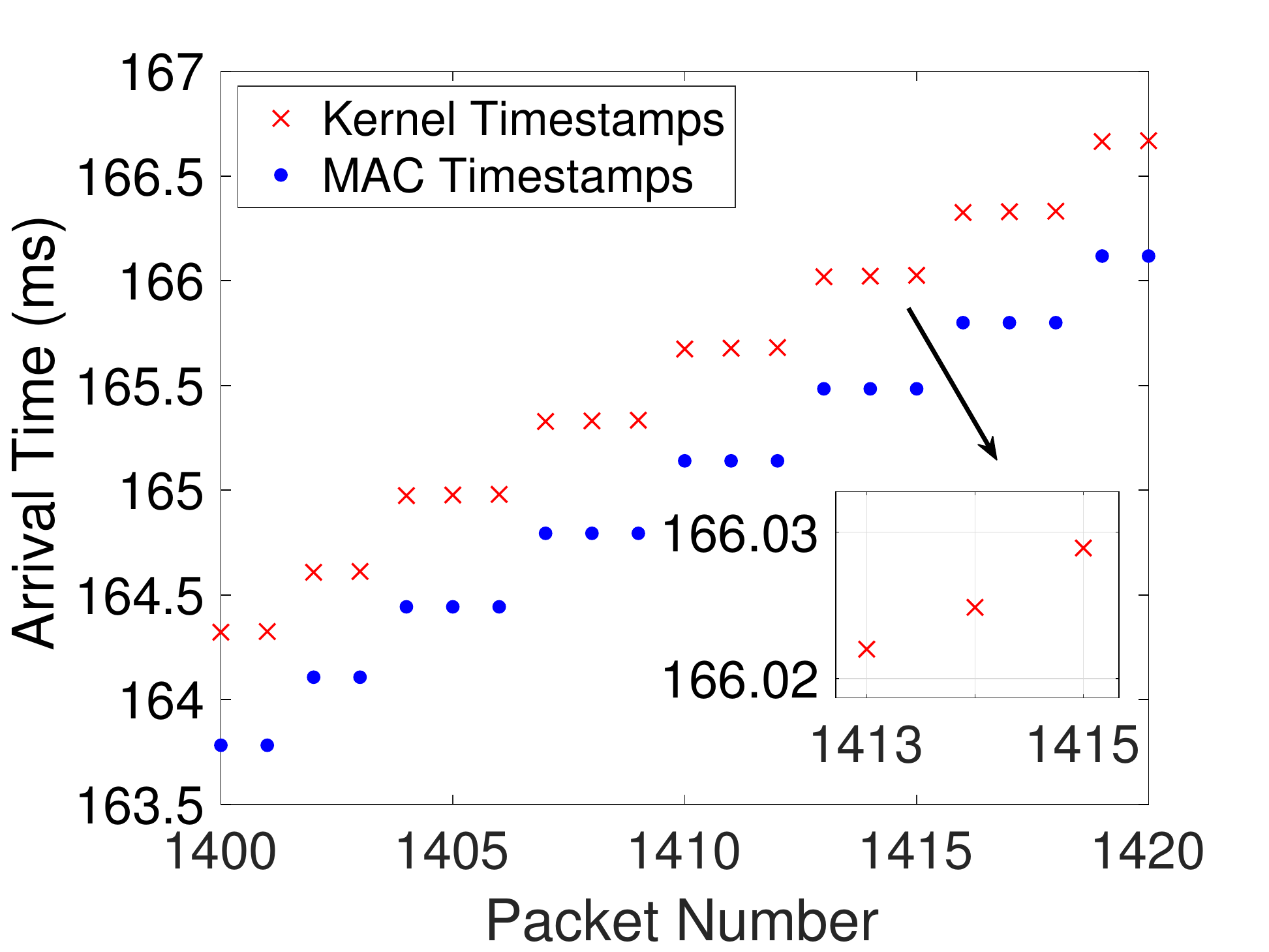}
 }
\subfigure[300Mbps]{
 \includegraphics[width=0.46\columnwidth]{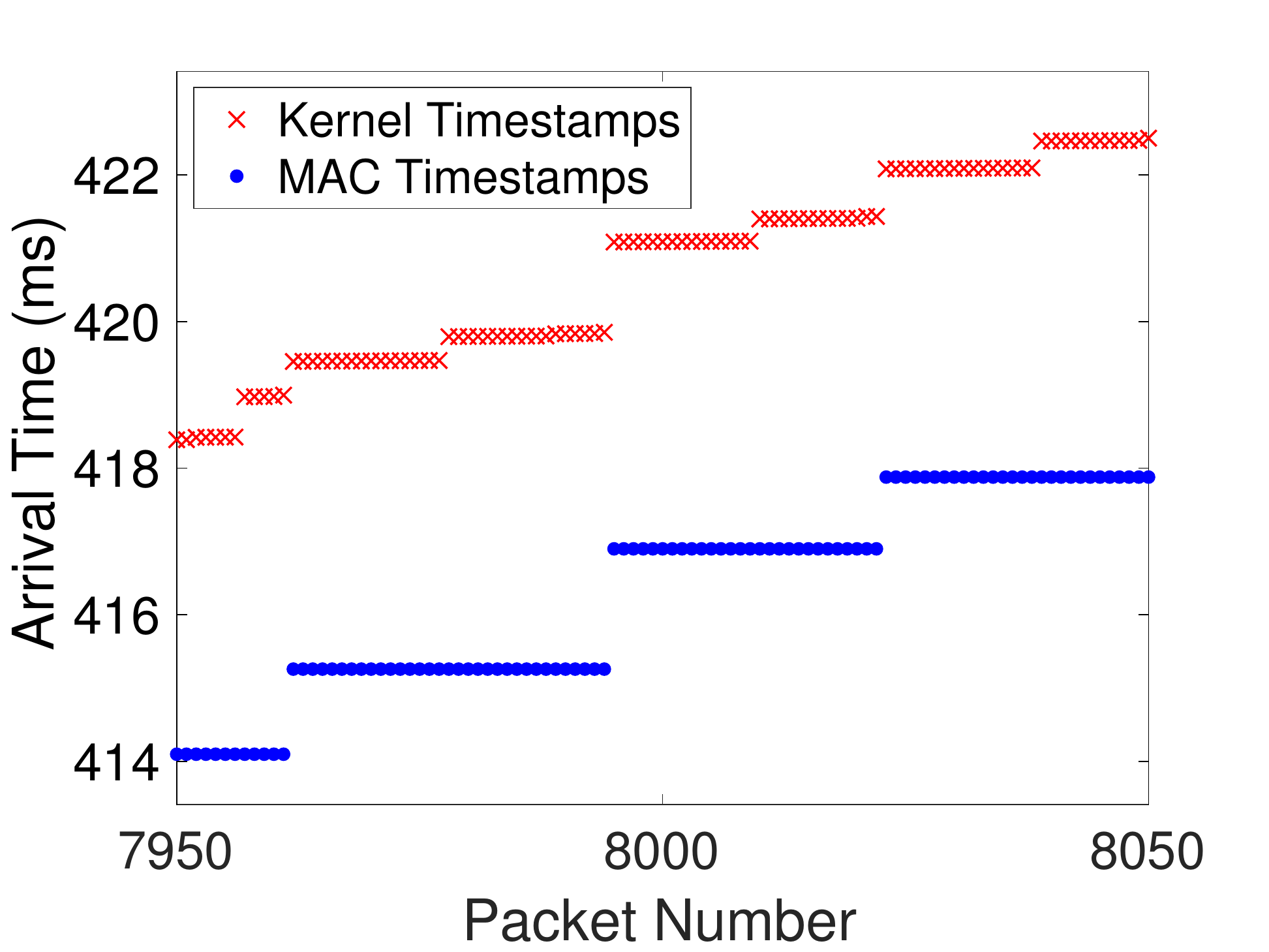}
 }
 \caption{Kernel and MAC timestamp measurements for UDP packets transmitted to a mobile handset over 802.11ac.  In (a) the arrival rate of packets at the AP is 100Mbps and in (b) 300Mbps. Experimental data, setup in Appendix, AMDSU aggregation ($N_{max}=128$). }\label{fig:fig2}
\end{figure}

Typical  kernel timestamp ``noise'' is shown, for example, in Figure \ref{fig:fig2}.   When the arrival rate at the AP is relatively low, it can be seen from Figure \ref{fig:fig2}(a) that two or three packets share each MAC timestamp and so are aggregated into the same frame.  While the kernel timestamps differ for packets transmitted in the same frame (see plot inset), there is nevertheless a clear jump in the kernel timestamps between frames and this can be used to cluster packets sharing the same frame.   Figure \ref{fig:fig2}(b) shows corresponding measurements taken at a higher network load.  It can be seen that now many more packets are aggregated into each MAC frame, as might be expected.  However, there is now no longer a clear pattern between the jumps in kernel timestamp values and the frame boundaries:  sometimes there are large jumps within a frame.  Although not shown in this plot, it can also happen that no clear jump in kernel timestamps is present at boundary between frames.   We believe this is due to the action of NAPI interrupt mitigation, which causes the kernel to switch from interrupt to driver polling at higher network loads. 

In this section we explore whether, despite their noisy nature, kernel packet timestamps can still be successfully used to estimate the aggregation level of frames received on a mobile handset.

\subsection{Estimating Aggregation Level: Logistic Regression}
To estimate the aggregation level using noisy kernel timestamps we adopt a supervised learning approach.   For training purposes we have ground truth available via the sniffer, although we require the resulting estimator to be robust enough to be used across a wide range of network conditions without retraining.   

As our main input features we use the inter-packet arrival times of last $m$ received packets, derived from their kernel timestamps, plus their standard deviation.   Parameter $m$ is a design parameter that we will select shortly.  The input feature vector $X^{(i)}$ associated with the $i$'th packet is therefore:
\begin{align}
X^{(i)} = [\begin{matrix}
  t_{i-m+1} & t_{i-m+2} & \dotsc & t_i & \sigma_i
 \end{matrix}]^T
\end{align}
where $t_i$ is the arrival time difference (in microseconds) between $i$th and ($i - 1$)th packets received and $\sigma_i$ is the standard deviation of the last $m$ inter-packet arrival times.  We define target variable $Y^{(i)}$ as taking value $1$ when the $i$th packet is the first packet in an aggregated frame and $0$ otherwise.   

While the OS polling noise is challenging to model and the relation between MAC and kernel timestamps is complex, surprisingly it turns out that we can quite effectively estimate $Y^{(i)}$ using the following simple logistic model:
\begin{align}
P(Y^{(i)} = 1 | X^{(i)}=X) = \frac{1}{(1 + e^{-\theta_0 - \vv{\theta}^T X})}\label{eq:logit1}
\end{align}
where $\vv{\theta} = (\theta_1\dots, \theta_{m + 1})^T\in\mathbb{R}^{m + 1}$ plus $\theta_0$ are the $m + 2$ (unknown) model parameters.

To train this model we use timestamp data collected for a range of send rates from 100Mbps to 600Mbps where at each rate 250,000 packets are collected. We use the F1 metric, which combines accuracy and precision, to measure performance at predicting label $Y^{(i)}$ for each packet.   We use the Scikit-learn library \cite{scikit-learn} to train the model using this data, applying 20-fold cross-validation to estimate error bars.   We found the standard deviation to be consistently less than 0.01 and since it is hard to see such small error bars on our plots these are omitted.    

\begin{figure}
\centering
   \subfigure[F1 Score vs. $m$]{
  \includegraphics[width=0.47\columnwidth]{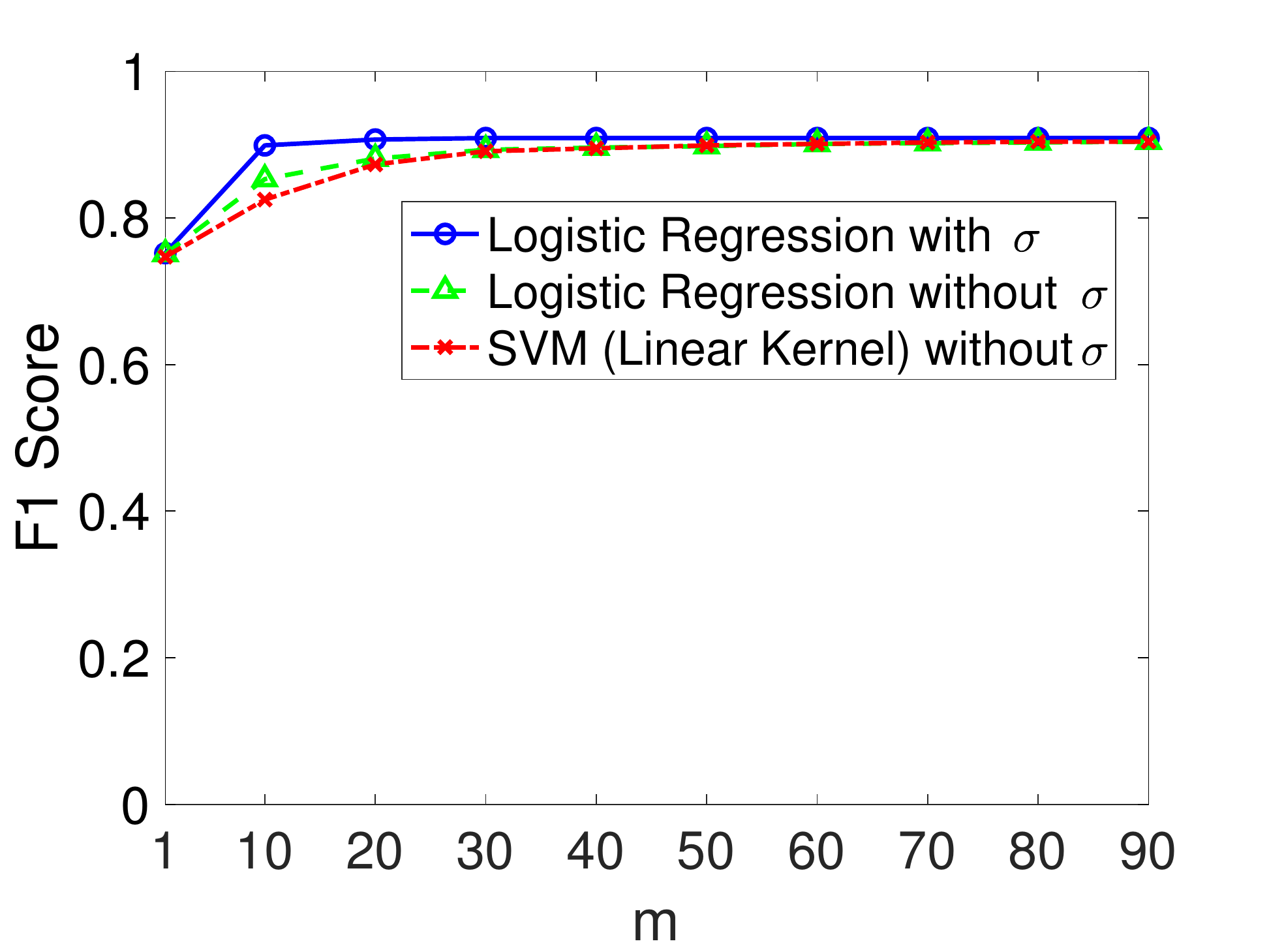}
  }
   \subfigure[F1 Score vs. Send Rate]{
 \includegraphics[width=0.47\columnwidth]{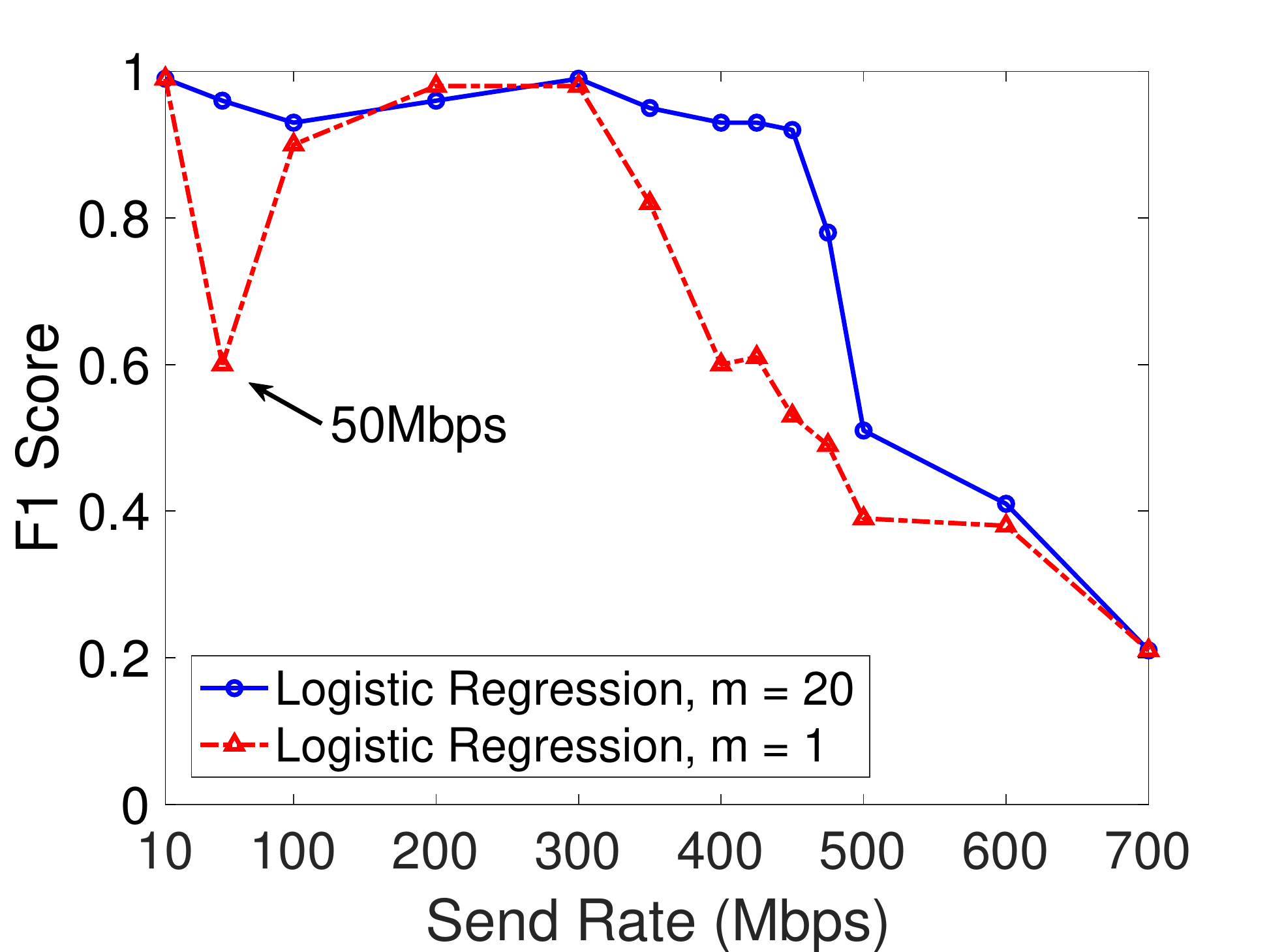}
 }
 \subfigure[RMSE of Predicted Aggregation Level vs. Send Rate]{
 \includegraphics[width=0.47\columnwidth]{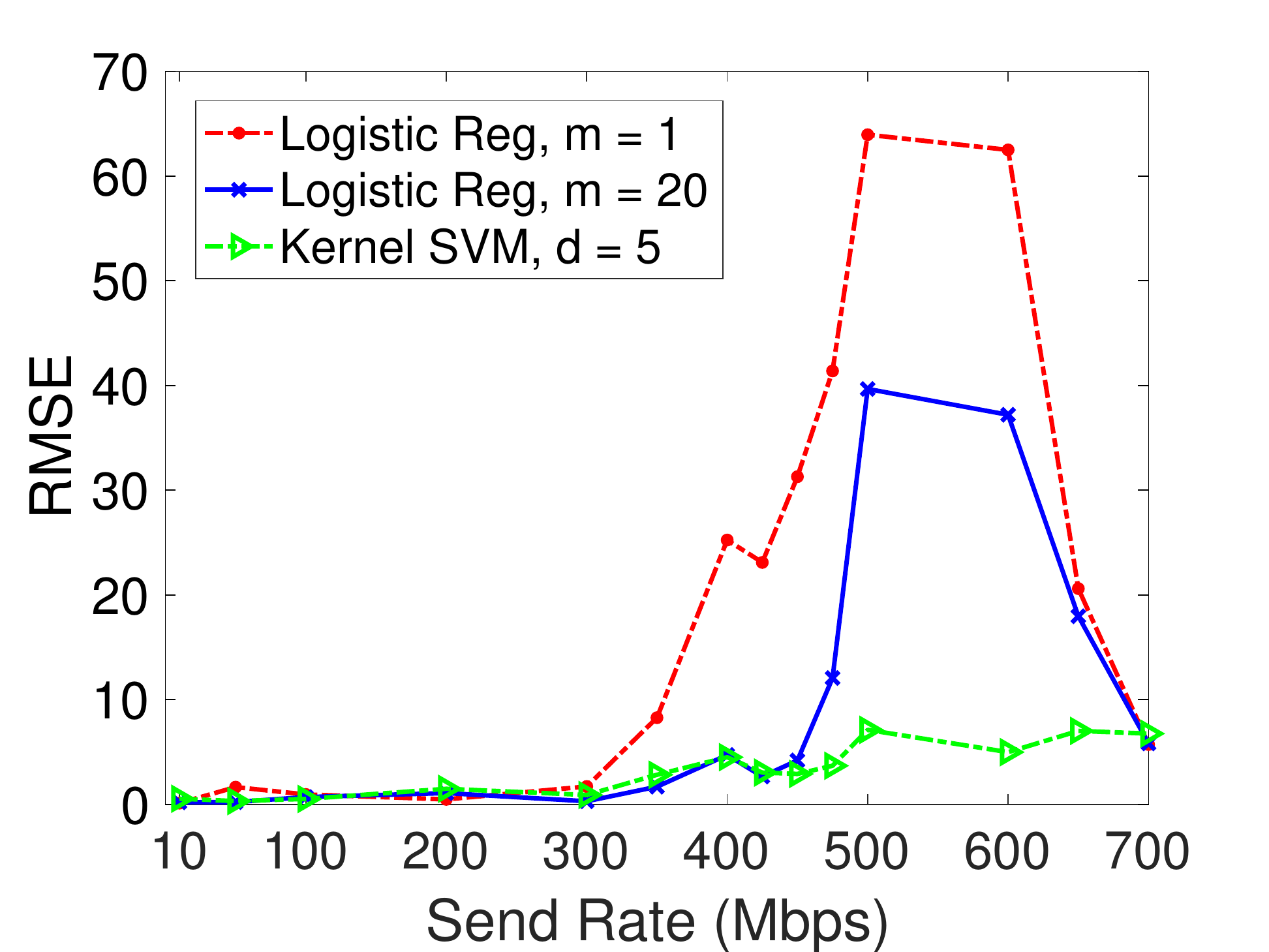}
 }
 \subfigure[]{
 \includegraphics[width=0.47\columnwidth]{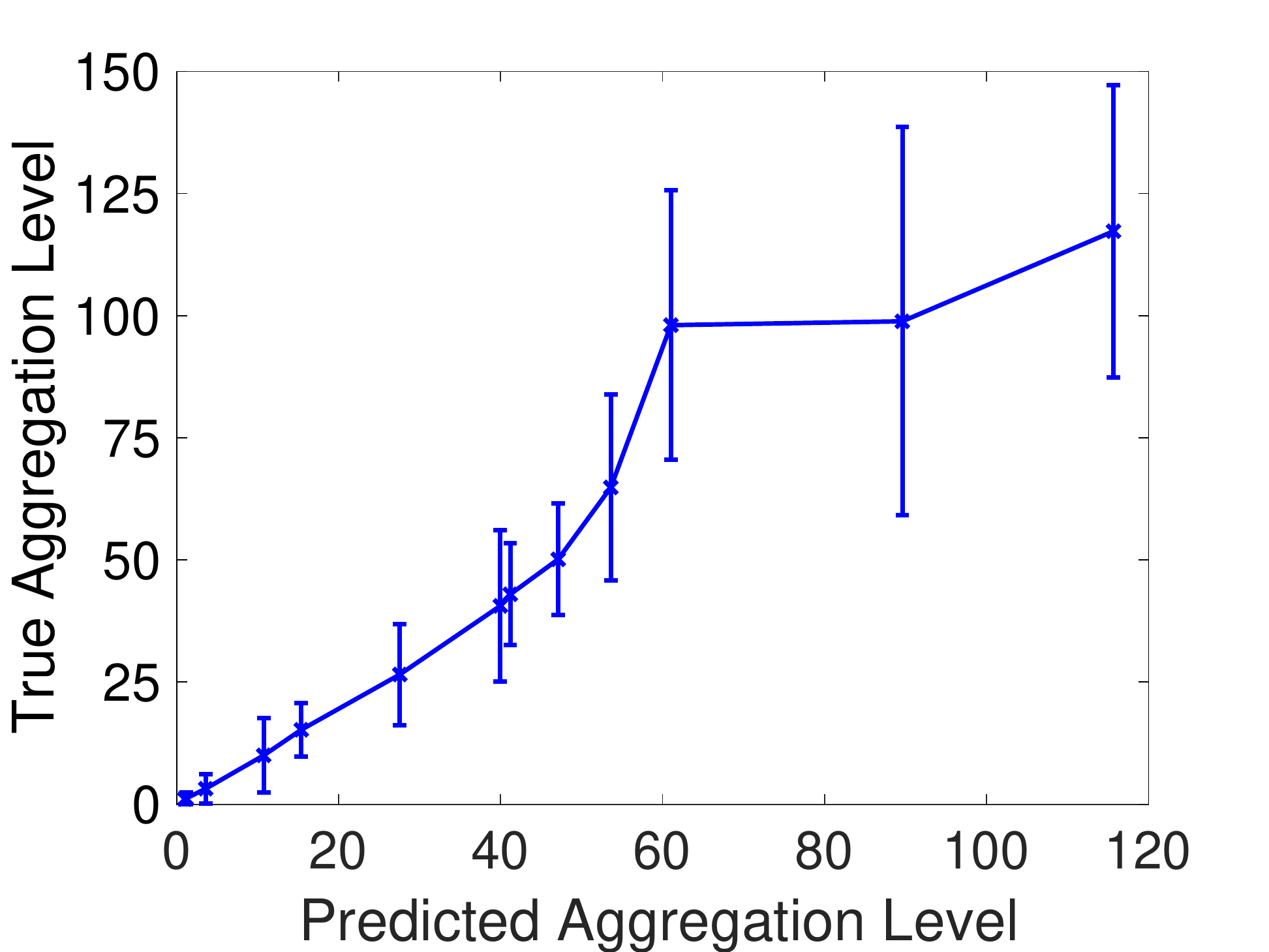}
 }
  \caption{Performance of logistic regression estimator vs (a) number $m$ of features and (b), (c) send rate; (d) actual vs predicted aggregation level.  Experimental data, Samsung Galaxy handset, setup in Appendix, AMDSU aggregation (so $N_{max}=128$).}
  \label{fig:fig4}
\end{figure}

Figure \ref{fig:fig4}(a) plots the measured F1 score vs the number $m$ of input features.  Data is shown both for logistic regression and SVM cost functions and also when the standard deviation $\sigma_i$ is and is not included in the feature vector.   From this data we can make the following observations.   For $m$ greater than about $40$ features the performance of all of the estimators is much the same, but for $m$ less than $40$ addition of $\sigma_i$ boosts performance by around $5\%$.    Note that use of a small value of $m$ is desirable since we need to wait for $m$ packets in order to generate an estimate for $Y^{(i)}$ and so the latency of the estimator increases with $m$.    The performance with the logistic regression and SVM cost functions is similar, as might be expected, but we adopt the logistic regression choice of parameters as the predictions have slightly better performance.   Unless otherwise stated, hereafter we use $m=20$ plus feature $\sigma_i$.   Note also that the same set of parameter values $(\vv{\theta}, \theta_0)$ is used for all send rates i.e. a single estimator is used across the full range of operation.  

Figure \ref{fig:fig4}(b) plots the performance of this estimator vs the downlink send rate.   It can be seen that the prediction accuracy is high for rates up to about 500Mbps, but then starts to drop sharply.   This is the accuracy of predicting the label $Y^{(i)}$ of each packet, but of course our real interest is in predicting the aggregation level.  The aggregation level can be directly derived from the $Y^{(i)}$ labels (its just the number of packets between those labelled with $Y^{(i)}=1$, capped at $N_{max}$).   Figure \ref{fig:fig4}(c) shows the measured aggregation level prediction accuracy vs the downlink send rate.  It can be seen that, as might be expected, it shows quite similar behaviour to Figure \ref{fig:fig4}(b).  A notable exception is at send rates above 700Mbps where the accuracy of the aggregation level improves.  This is because at such high rates the aggregation level has hit the upper limit $N_{max}$ and the estimator simply predicts $N_{max}$ as the aggregation level.   We will consider the causes for the drop in accuracy at rates between 500-700Mbps in more detail shortly, but note briefly that it is directly related to the load-related ``noise'' on kernel timestamps (recall Figure \ref{fig:fig2}(b)).

As a baseline for comparison Figures \ref{fig:fig4}(b)-(c) also shows the performance of the logistic regression estimator when $m=1$ (and without $\sigma_i$).   The latter corresponds to an estimator that labels packets by simply thresholding on the inter-packet arrival time i.e. when the time between the current and previous packet exceeds a threshold we label the current packet as the first in a new MAC frame.   The threshold level used is optimised to maximise prediction accuracy on the training data.   From Figure \ref{fig:fig2}(a) we can expect that under lightly loaded conditions this approach is quite effective, but Figure \ref{fig:fig2}(b) also tells us that it is likely to degrade as the load increases and indeed it can be seen from Figures \ref{fig:fig4}(b)-(c) that the performance of this baseline estimator degrades for rates above about 300Mbps (compared to rates above about 500Mbps with $m=20$).   It can also be seen that the accuracy drops sharply at a rate of 50Mbps.  This is because at this send rate the inter-packet send time is close to the simple threshold used in the baseline estimator.   Hence the logistic regression estimator with $m=20$ offers significant performance gains over this baseline estimator.

\subsection{Improving Accuracy At High Network Loads: SVM}

\begin{figure}
 \centering
 \subfigure[300Mbps]{
 \includegraphics[width=0.46\columnwidth]{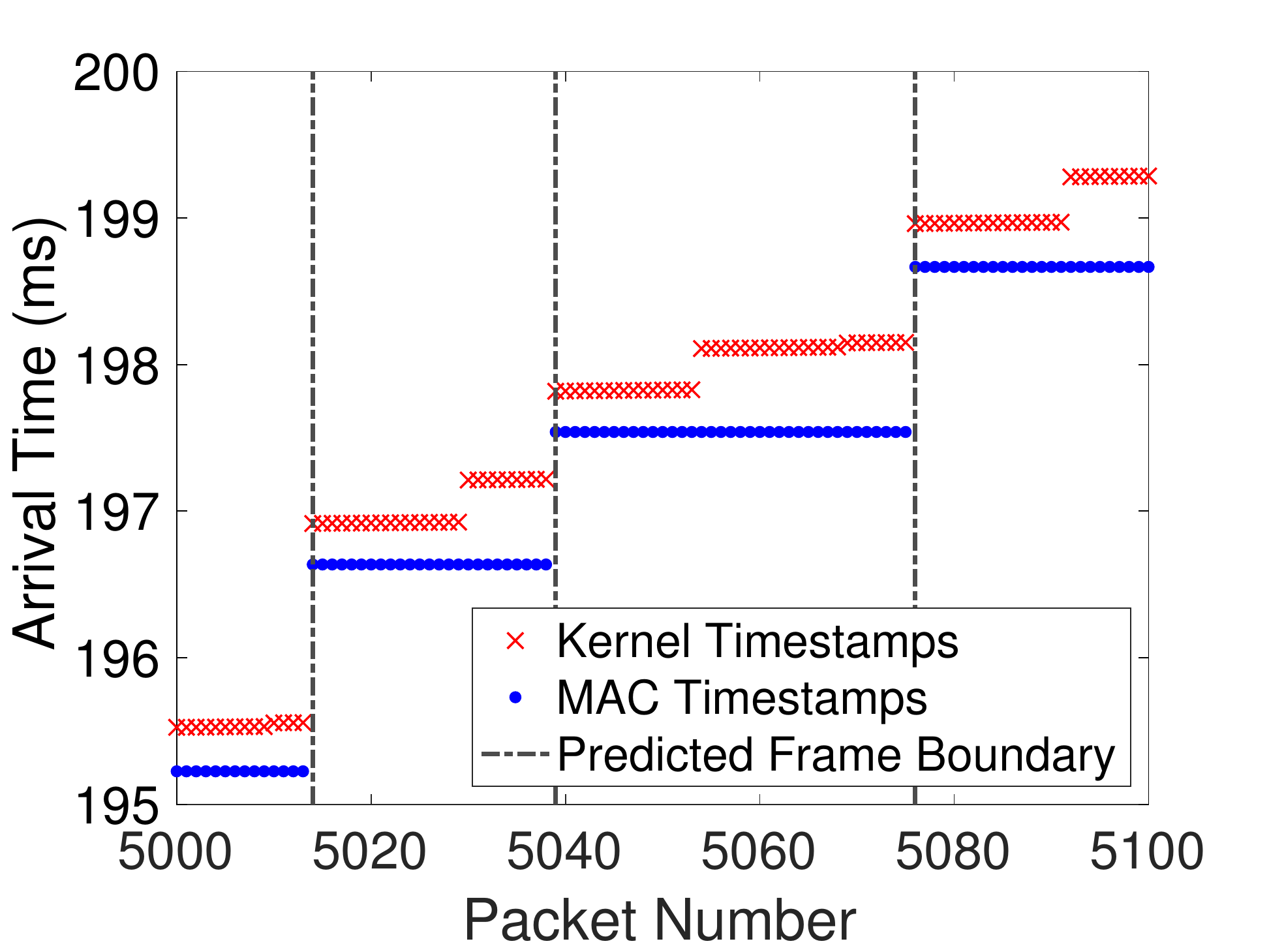}
 }
  \subfigure[600Mbps]{
 \includegraphics[width=0.46\columnwidth]{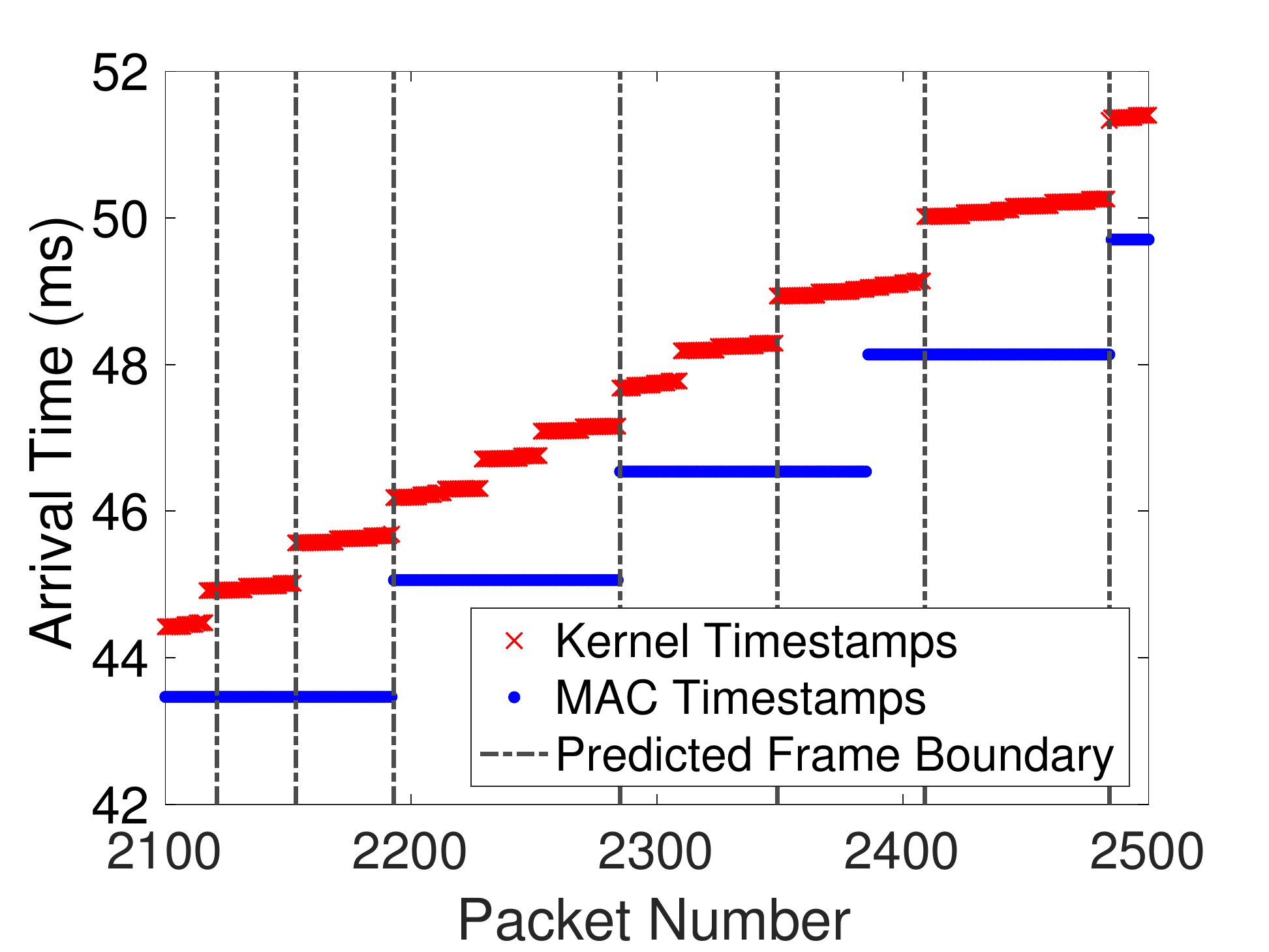}
 }

  \subfigure[300Mbps]{
 \includegraphics[width=0.46\columnwidth]{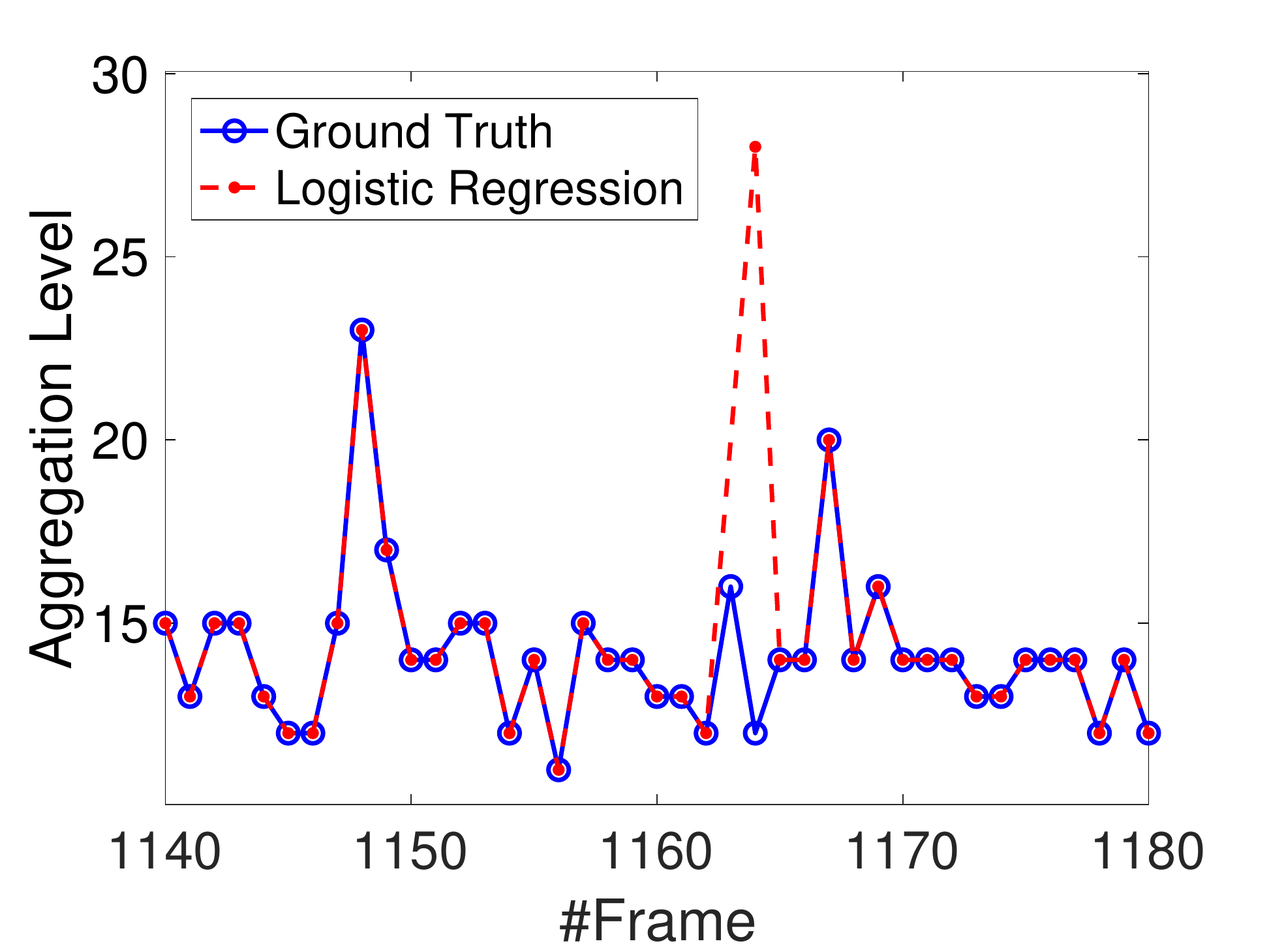}
 }
\subfigure[600Mbps]{
 \includegraphics[width=0.46\columnwidth]{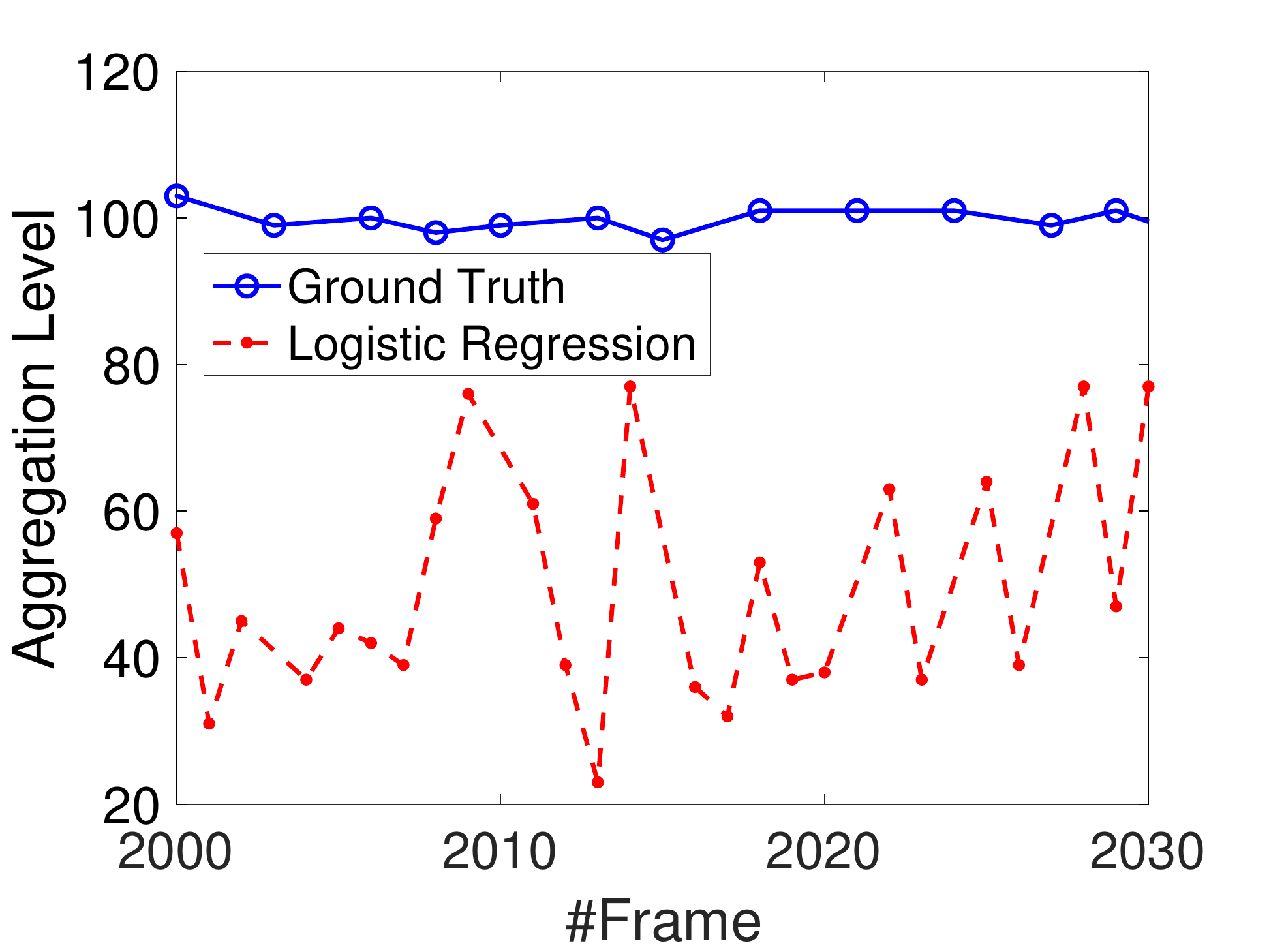}
 }
 \caption{Frame boundaries predicted by the logistic regression estimator for (a) medium and (b) high network loads, and corresponding kernel and MAC timestamp measurements.   Also time histories of estimated aggregation level for (a) medium and (b) high network loads.  Experimental data, Samsung Galaxy handset, setup in Appendix, AMDSU aggregation ($N_{max}=128$).}\label{fig:fig6}
\end{figure}

As already noted, the accuracy of the logistic regression estimator falls for send rates in the range 500-700Mbps, see Figures \ref{fig:fig4}(b)-(c).   The reason for this can be seen from Figure \ref{fig:fig6}.  Figure \ref{fig:fig6}(a) shows the frame boundaries predicted by the estimator when the arrival rate at the AP is 300Mbps.  The true frame boundaries can be inferred from the MAC timestamps, which are also shown in this plot. Observe that even although there are jumps between the kernel timestamps of packets sharing the same frame the estimator is still able to accurately predict the frame boundaries.  Figure \ref{fig:fig6}(b) shows the corresponding data when the arrival rate is increased to 600Mbps.  It can be seen that there are now many jumps in the kernel timestamps of packets sharing the same MAC frame and sometime no jump in timestamps between packets transmitted in different frames (e.g. see the frame towards the right-hand edge of the plot).  As a result the estimator makes many mistakes when trying to predict the frame boundaries.  

%

Figures \ref{fig:fig6}(c)-(d) show time histories of the estimated aggregation level.   Observe in Figure \ref{fig:fig6}(d) that there is a fairly consistent offset between the predicted and actual aggregation level.  While this figure is for a single send rate of 600Mbps, Figure \ref{fig:fig4}(d) plots the relationship between predicted and actual aggregation level for a range of send rates.  Since the error is fairly consistent the potential exists to improve the estimator for send rates in the 500-600Mbps range.   We explored various approaches for this and found the most effective is to combine the logistic regression estimator with a radial-basis function kernel SVM with the following input features,
\begin{align}
X^{(i)} = [\begin{matrix}
  \mu_{\widehat{N}_{\delta_{i - d + 1}}} & \dotsc & \mu_{\widehat{N}_{\delta_i}} & \sigma_{_{\delta_i}} & A_{_{\delta_i}}
 \end{matrix}]^T
\end{align}
where we partition time into 100ms slots and $\mu_{\widehat{N}_{\delta_i}}$ is the empirical average of the aggregation level predicted by the logistic regression estimator over the $i$ slot, $\sigma_{_{\delta_i}}$ the empirical variance and  $A_{_{\delta_i}}$ the number of frames.   The averaging over slots reduces the noise and significantly improves performance.   The output of the SVM is the predicted aggregation level.   We trained this SVM using the measured aggregation level averaged over each slot (again to reduce noise during training) and using cross-validation selected $d = 5$.   Figure \ref{fig:fig4}(c) plots the RMSE of the predictions vs send rate when the logistic regression estimator is augmented with this SVM.  It can be seen that the performance is considerably improved for send rates in the 500-600Mbps range, with the the RMSE now no more than 8 packets compared with a value of around 40 packets when the logistic regression estimator is used on its own.    

Note that since it operates over 100ms slots the SVM estimator is less responsive that the logistic regression estimator, but since the controller only updates the send rate every $\Delta$ seconds with $\Delta$ typically 0.5 or 1s then the 100ms delay introduced by the SVM estimator is minor.   However, since the SVM estimator is relatively computationally expensive and the logistic regression estimator is sufficiently accurate for the low delay operating regime of interest here (where the rates are less than 500Mbps), so in the rest of the paper we confine use to the logistic regression estimator unless otherwise stated.

\subsection{Effect of CPU Load On Estimator Performance}
\begin{figure}
 \centering
 \subfigure[400Mbps - Low Load ($\approx30\%$)]{
 \includegraphics[width=0.46\columnwidth]{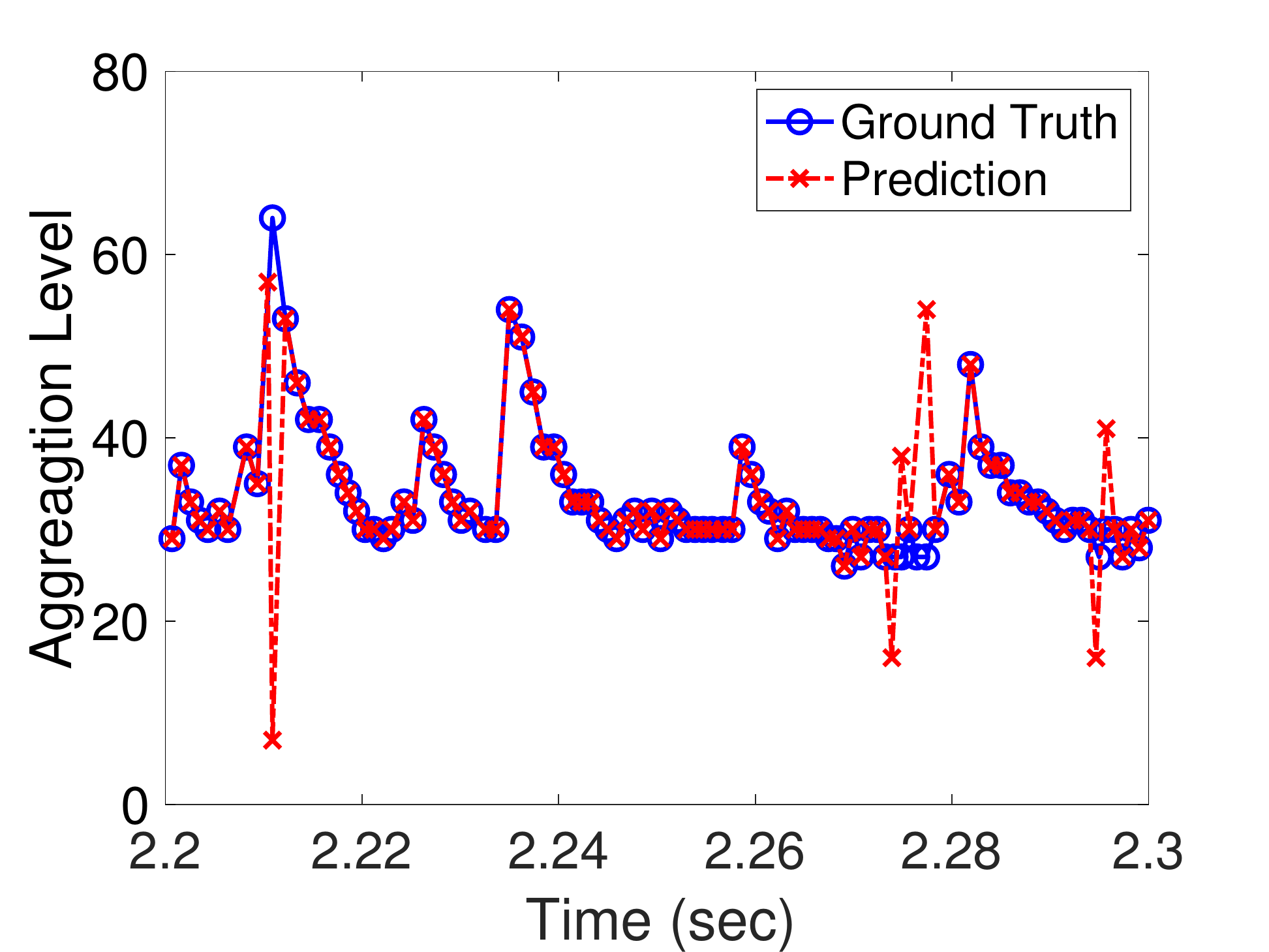}
 }
\subfigure[400Mbps - High Load ($\approx55\%$)]{
 \includegraphics[width=0.46\columnwidth]{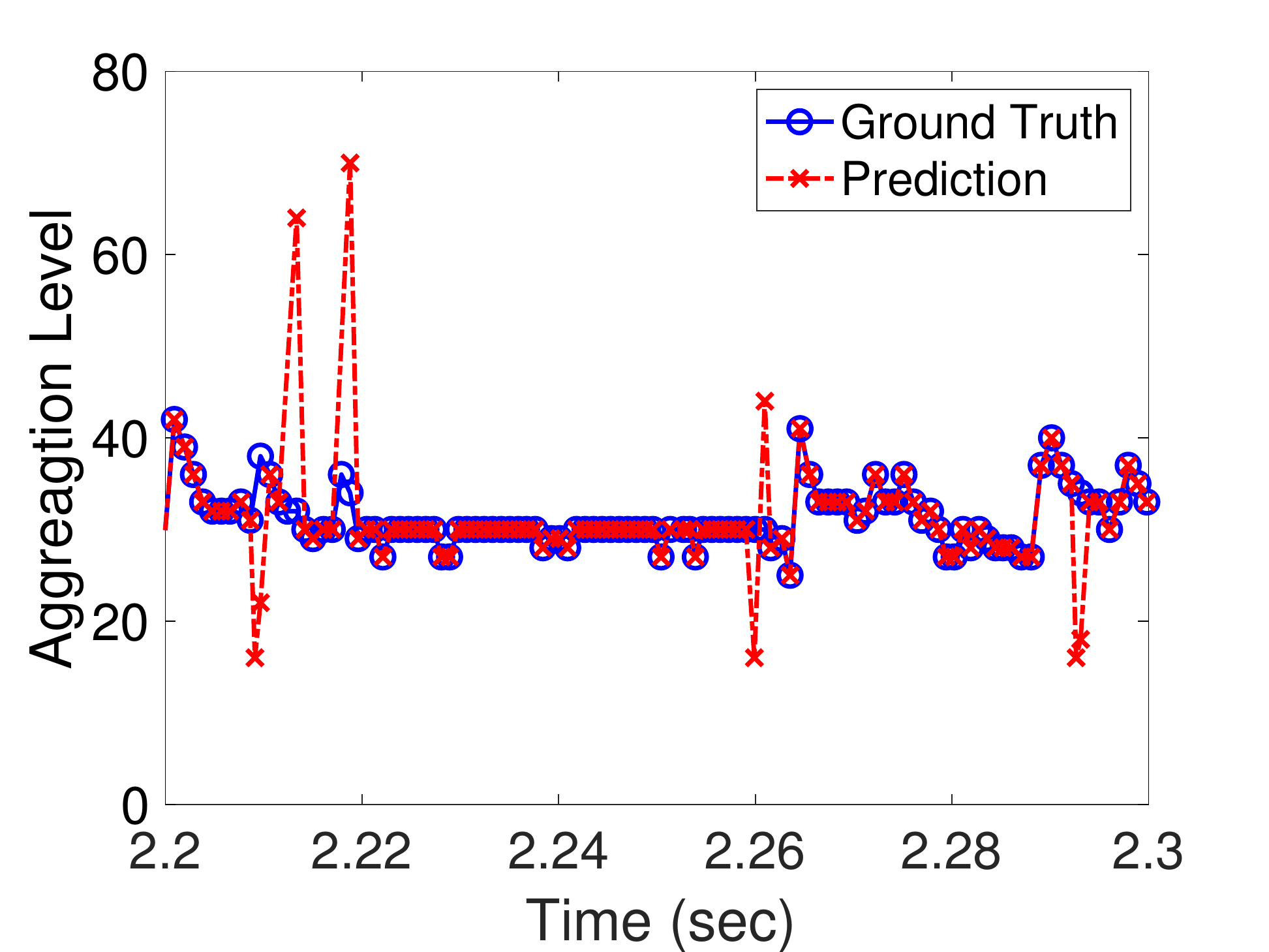}
 }
 \caption{Illustrating the accuracy of prediction for different CPU loads. Experimental data, Samsung Galaxy handset, setup in Appendix, AMDSU aggregation ($N_{max}=128$).}\label{fig:fig8}
\end{figure}

To understand whether the noise on kernel timestamps is affected by system CPU load as well as network load we collected packet timestamp measurements while varying the CPU load by playing a 4K video in full screen mode.   We found that CPU load makes little difference to the accuracy of the aggregation level estimator.  For example, Figure \ref{fig:fig8} shows two typical time histories of measured and estimated aggregation level.  Figure \ref{fig:fig8}(a) is when the CPU load is around $30\%$ and Figure \ref{fig:fig8}(a) when the CPU load is around $55\%$.   Even with a fairly high network load of 400Mbps it can be seen that the aggregation levels predicted by the estimator agree well with the actual aggregation level regardless of the CPU load.

\subsection{Robustness of Estimator}
While the foregoing measurements are for a Samsung Galaxy tablet we obtained similar results (using the same trained estimator, without changing the parameter values) when using a Google Pixel 2 handset.   We also obtained similar results when there are multiple WLAN clients, as might be expected since per client queueing is used by 802.11ac APs i.e packets to different clients queued separately.

\subsection{Performance Comparison With TCP Cubic \& BBR}
We extended the client-side code in our prototype implementation of the rate allocation approach in Section \ref{sec:low} to make use of kernel timestamps and the logistic regression estimator of aggregation level.  The code is written in C and could be directly cross-compiled for use on the Samsung Galaxy tablet.   

\begin{figure}
 \centering
 \subfigure[Receive Rate]{
 \includegraphics[width=0.46\columnwidth]{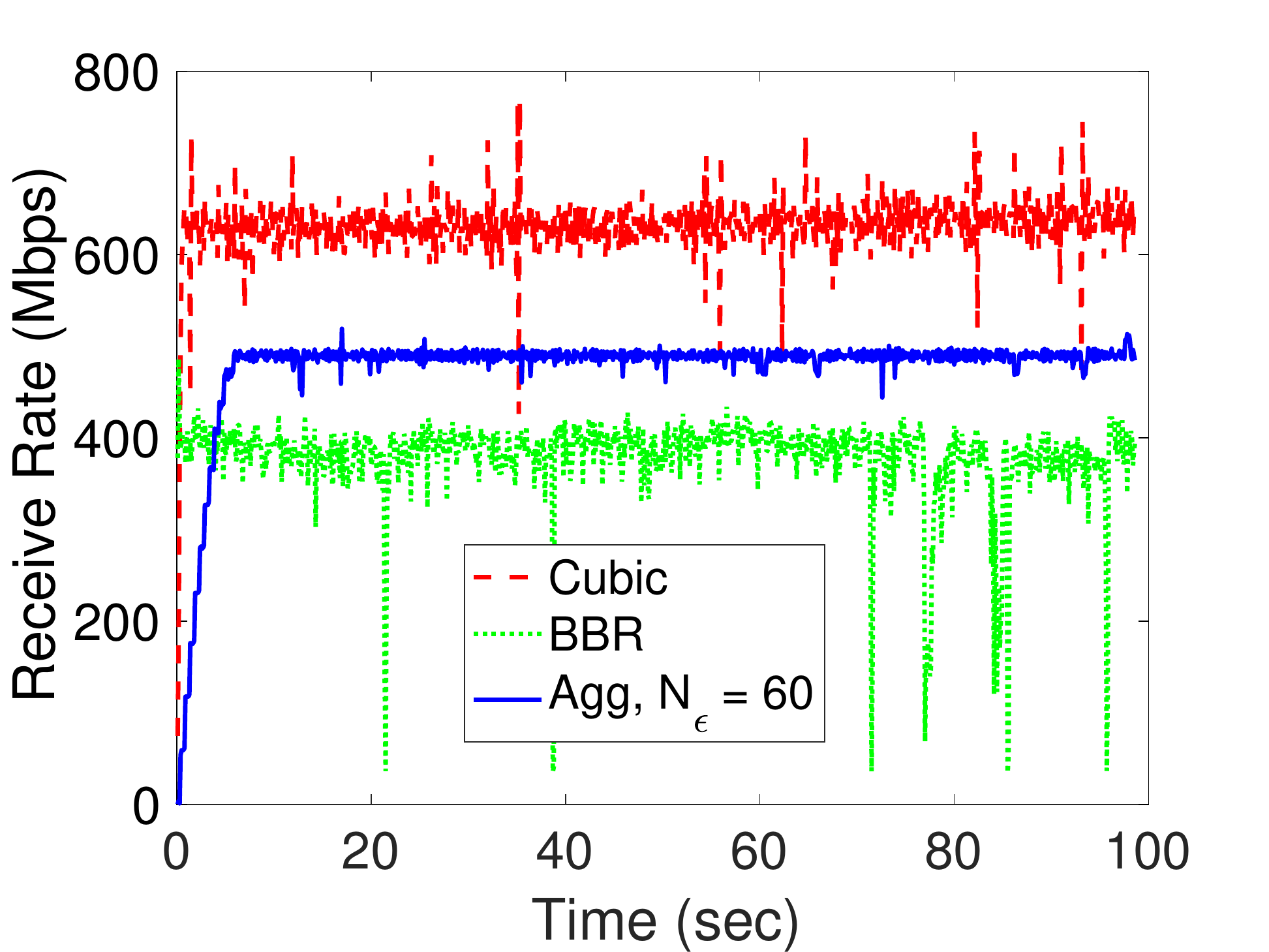}
 }
\subfigure[One-way Delay]{
 \includegraphics[width=0.46\columnwidth]{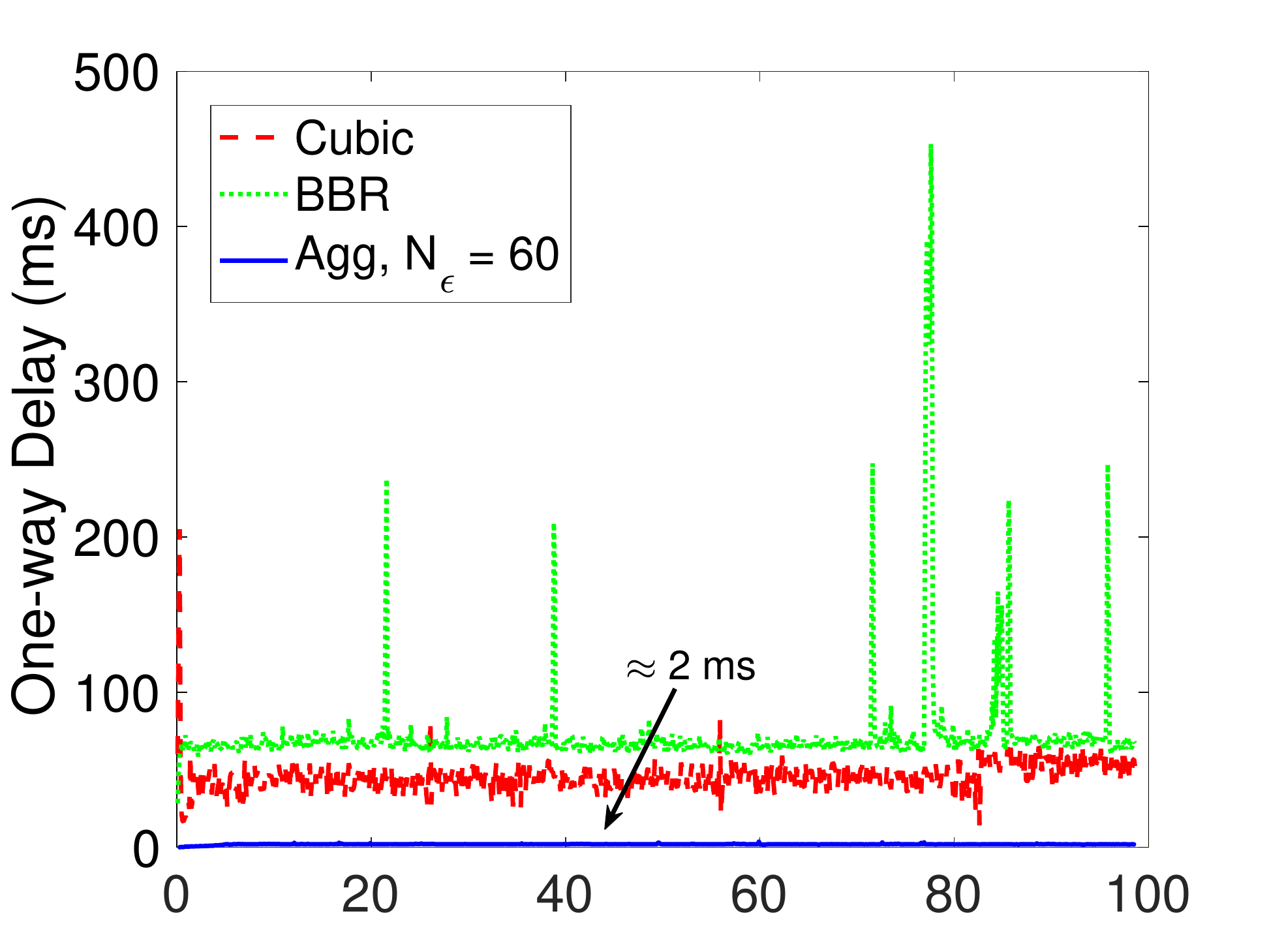}
 }
 \subfigure[\#Packet loss]{
 \includegraphics[width=0.46\columnwidth]{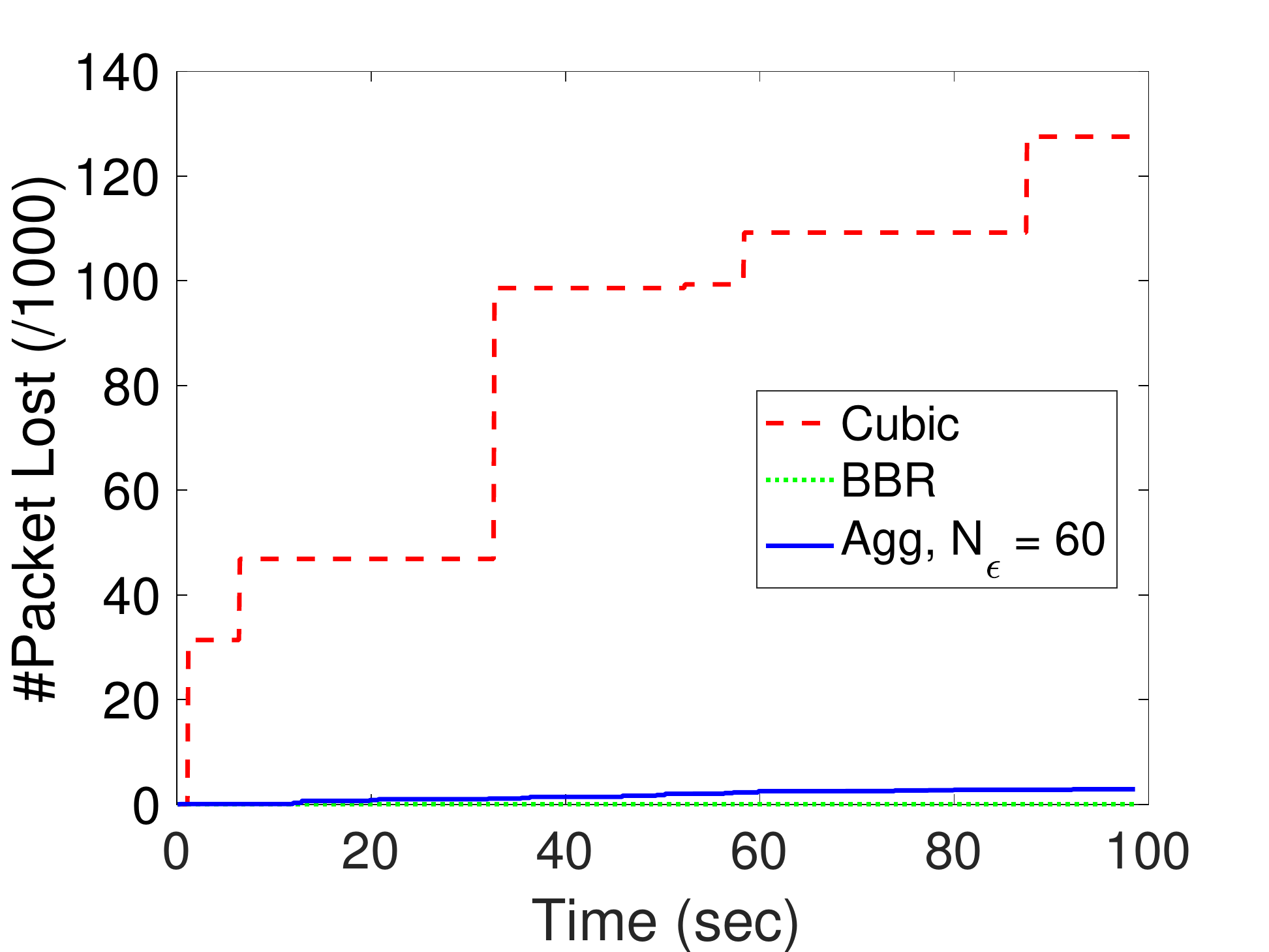}
 }
 \caption{Compare the performance of aggregation-based rate control algorithm with TCP Cubic and BBR.   The one-way delay in (b) is averaged over 100ms intervals.  $K_0 = 1$, $\Delta = 1000$ms, $N_\epsilon=60$. Experimental data, Samsung Galaxy handset, setup in Appendix, AMDSU aggregation ($N_{max}=128$).}\label{fig:fig23}
\end{figure}

As might be expected, we obtain similar results to those shown in Section \ref{sec:low} when using MAC timestamps and so do not reproduce these here.  Instead we take the opportunity to compare the performance of our proposed aggregation-based rate control algorithm with TCP Cubic~\cite{Cubic}, the default congestion control algorithm used by Linux and Android.  In addition, we compare performance against TCP BBR \cite{BBR} since this is a state-of-the-art congestion control algorithm currently being developed by Google and which also targets high rate and low latency.

Since TCP Cubic is implemented on Android we use the Samsung Galaxy as client. However, TCP BBR is not currently available for Android and so we use a Linux box (Debian Stretch, 4.9.0-7-amd64 kernel) as the BBR client. 

Figure \ref{fig:fig23} shows typical receive rate and one-way delay time histories measured for the three algorithms.  It can be seen from Figure \ref{fig:fig23}(a) that Cubic selects the highest rate (around 600Mbps) but from Figure \ref{fig:fig23}(b) that this comes at the cost of high one-way delay (around 50ms).  This is as expected since Cubic uses loss-based congestion control and so increases the send rate until queue overflow (and so a large queue backlog and high queueing delay at the AP) occurs.   As confirmation, Figure \ref{fig:fig23}(c) plots the number of packet losses vs time and it can be seen that these increase over time when using Cubic, each step increase corresponding to a queue overflow event followed by backoff of the TCP congestion window.

BBR selects the lowest rate (around 400Mbps) of the three algorithms, but surprisingly also has the highest end-to-end one-way delay (around 75ms).  High delay when using BBR has also previously been noted by e.g. \cite{Parisa} where the authors propose that high delay is due to end-host latency within the BBR kernel implementation at both sender and receiver.   However, since our focus is not on BBR we do not pursue this further here but note that the BBR Development team at Google is currently developing a new version of BBR v2.

 Our low delay aggregation-based approach selects a rate (around 480 Mbps), between that of Cubic and BBR, consistent with the analysis in earlier sections.   Importantly, the end-to-end one-way delay is around 2ms i.e. more than 20 times lower than that with Cubic and BBR.   It can also be seen from Figure \ref{fig:fig23}(c) that it induces very few losses (a handful out of the around 4M packets sent over the 100s interval shown).
 \section{Detecting Bottleneck Location}
 
 The foregoing analysis applies to edge networks where the wireless hop is the bottleneck.   For robust deployment, however, we need to be able to detect when this is violated i.e. when the bottleneck is the backhaul link.   In this section we show how measurement of the aggregation level can be used for this purpose also.   Note that this is of interest in its own right  for network monitoring and management, separately from its use with our aggregation-based low delay rate control approach.
 
The basic idea is as follows.  When the AP is the bottleneck then a queue backlog will develop there and so an elevated level of aggregation will be used in transmitted frames.   Conversely, when the bottleneck is the backhaul link then we will observe delay and loss but a low level of packet aggregation.   Hence we can use aggregation level and loss/delay as input features to a bottleneck location classifier.   Note that this passive probing approach creates no extra load on the network (unlike active probing methods) and can also cope with bridged Layer-2 devices (such as a bridged AP) which are invisible to ICMP probes.


\subsection{Experimental Setup}\label{sec:classsetup}

\begin{figure}
 \centering
  \includegraphics[width=0.8\linewidth]{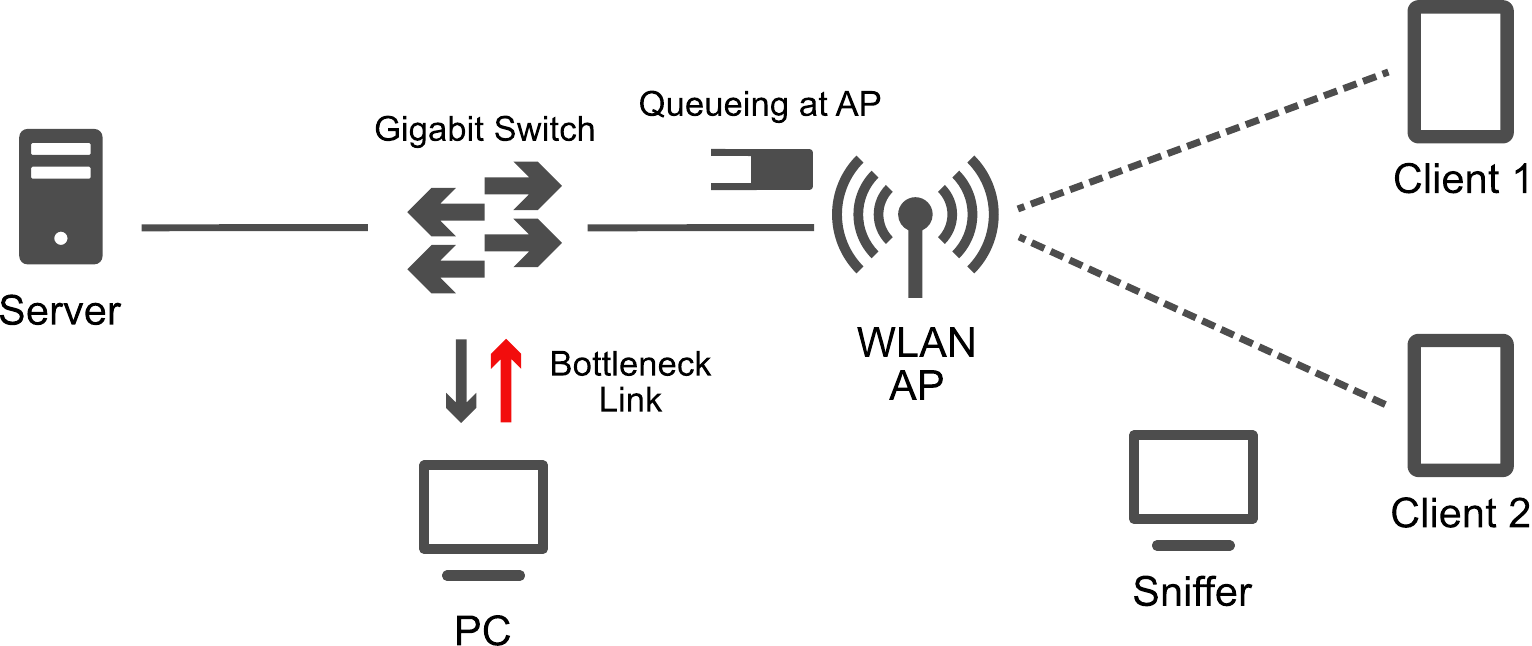}
  \caption{System model used in our testbed to do measurement when the backhaul is bandwidth bottleneck, 802.11 ac. }
  \label{fig:fig14}
\end{figure}

To adjust the bottleneck location we modify the experimental setup described in the Appendix so that packets between the sender and the AP are now routed via a middlebox which allows us to adjust the bandwidth of the backhaul link, see Figure \ref{fig:fig14}.   We adjust the bandwidth using two different techniques: (i) by forcing the ethernet link to operate at either 100Mbps or 1000Mbps (using {\tt ethtool}) and (ii) by generating cross-traffic on the backhaul link. When adjusting the link rate we also adjust the link queue size corresponding e.g. when changing to 100Mbps we set {\tt txqueuelen} to 100 packets.


As client stations within the WLAN we use a Samsung Galaxy Tab S3 (Client 1) and a Google Pixel 2 (Client 2).


\subsection{Bottleneck Classification: Ethernet Rate Limiting}

\subsubsection{Feature Selection}
We begin by considering when the bandwidth is limited by the ethernet link rate.  We use a supervised learning approach to try to build a bottleneck classifier.   To proceed we collect training data for a range of send and link rates (the link ethernet rate is varied between 100Mbps and 1000Mbps using {\tt ethtool}).  Using timestamps for each 802.11ac frame we extract the aggregation level and the number of packets lost.  For the latter we insert a packet id into the body of each packet and count ``holes'' in the sequence of received id's as losses -- this is after accounting for link layer retransmissions, so the losses are due to queue overflow.  We use these values for the last $n$ frames as input features.  That is, the input feature vector associated with the $i$ frame is:
\begin{align}
X^{(i)} &= 
[\begin{matrix} N_{i-n+1} & N_{i-n} & \dotsc & N_i & L_i^p \end{matrix} ]^T
\end{align}
where $N_j$ is aggregation of the $j$'th frame and $L_i^p$ is the fraction of observed packets lost out of the last $p$ packets received. We define target variable $Y^{(i)}$  as taking value 1 when the bottleneck is in the backhaul link and 0 otherwise.  

\begin{figure}
 \centering
 \subfigure[F1 Score vs. $n$]{
 \includegraphics[width=0.46\columnwidth]{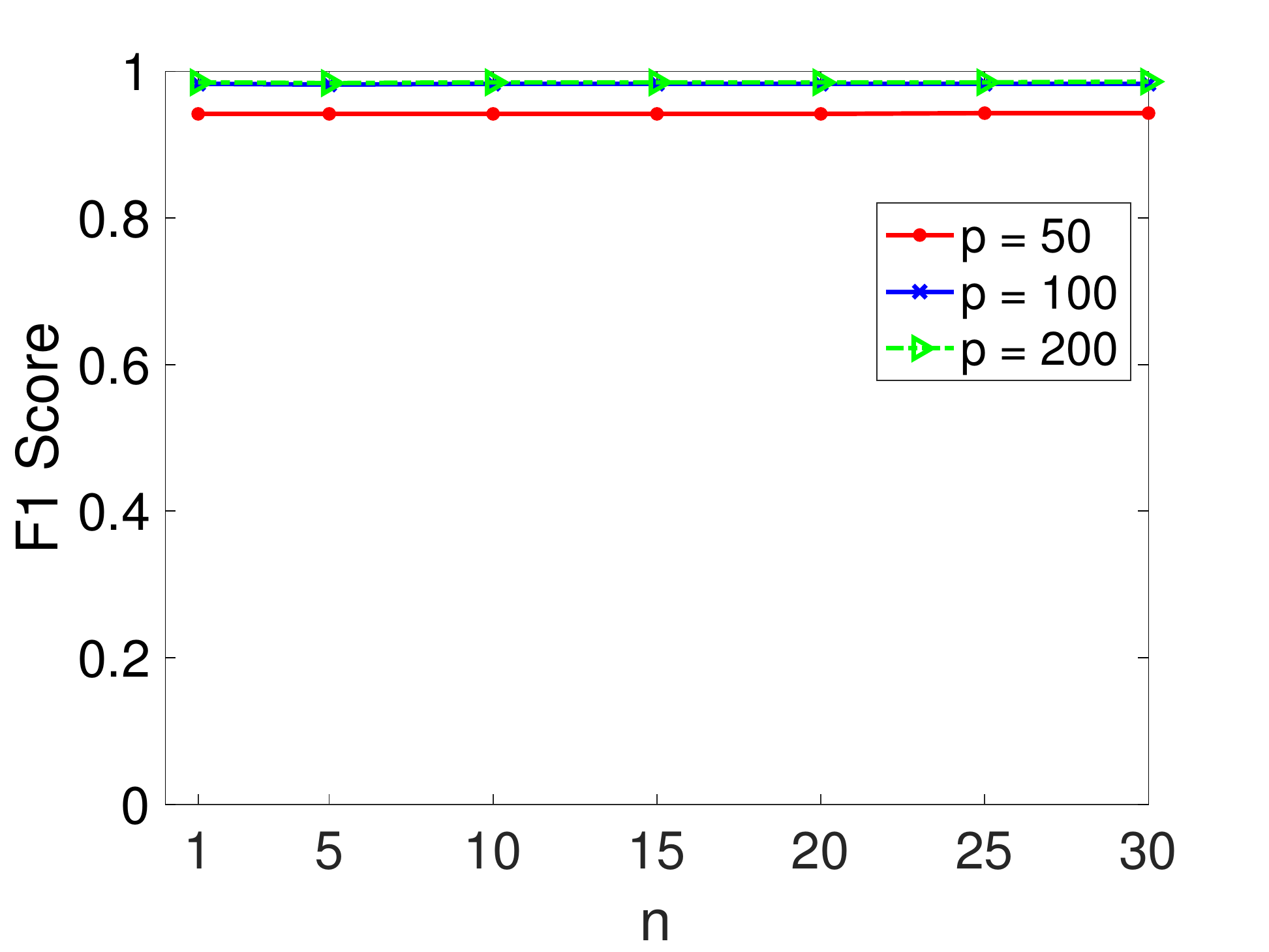}
}
\subfigure[F1 Score vs. $p$]{
 \includegraphics[width=0.46\columnwidth]{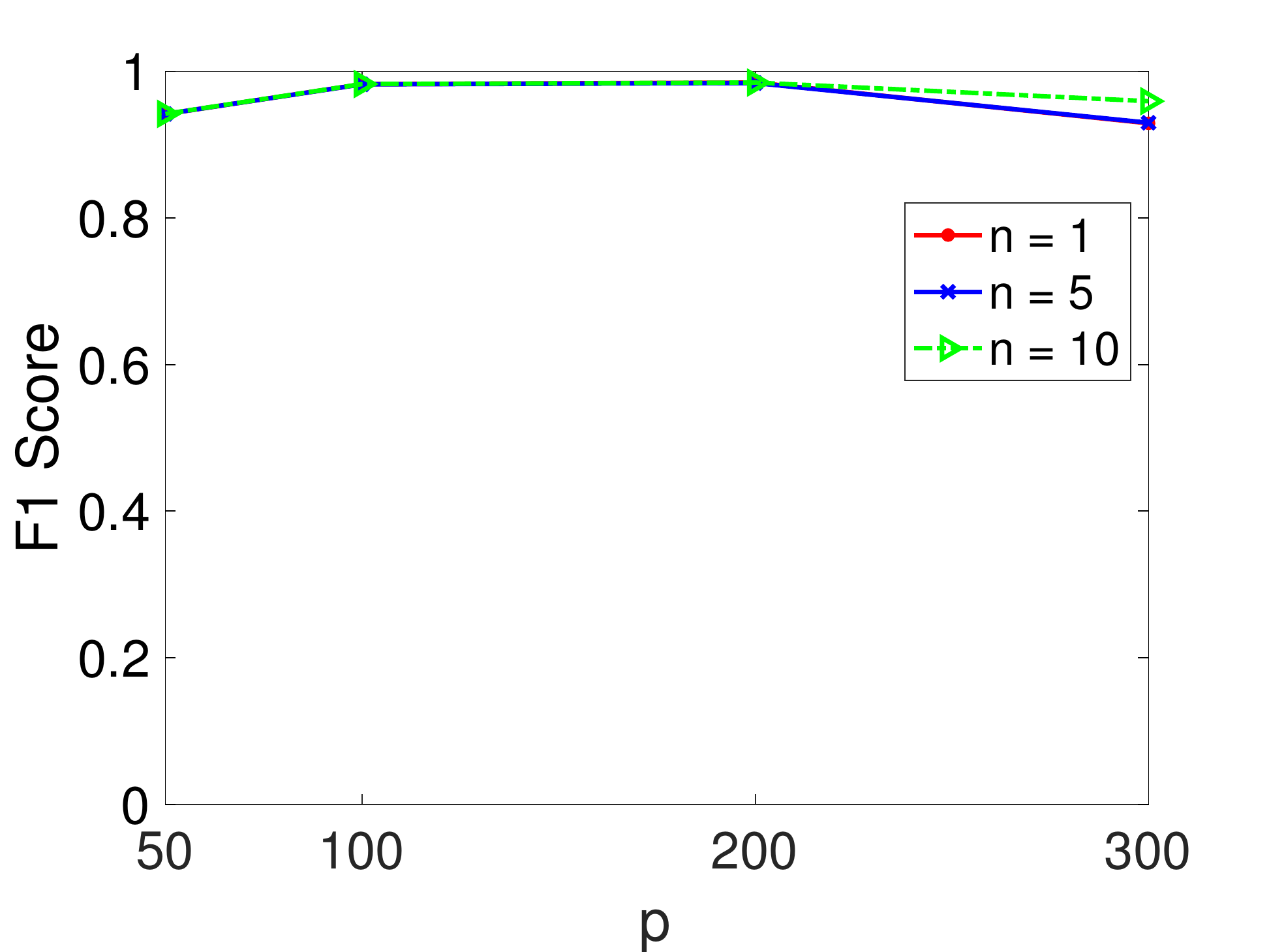}
 }
 \caption{Performance of the logistic regression classifier vs the number $n$ of input features and packets $p$ used. MAC timestamps, experimental data, Samsung Galaxy handset, setup in Section \ref{sec:classsetup}, AMDSU aggregation ($N_{max}=128$).  Ethernet rate limiting (100/1000Mbps). }\label{fig:fig15}
\end{figure}

Once again we try to use logistic regression to perform the classification.   Performance is measured as the F1 score, with 20-fold cross-validation used.   Figure \ref{fig:fig15} plots the measured performance vs the number $n$ of input features used and the number of packets $p$.   Note that for each value of $n$ and $p$ we hold the classifier parameters fixed i.e we use the same classifier for the full range of send rates and network configurations.  Data is shown for when MAC timestamps are used to measure the aggregation level $N_j$ but the performance is similar when kernel timestamps are used.   The F1 score is above $90\%$ for all values of $n$ and $p$ but is slightly higher for $p$ in the range 100-200 packets.  Note that the case of $n=1$ corresponds to simply thresholding on the aggregation level and loss rate of the current frame.    We use $n=5$ and $p=100$ in the following, but it can be seen that the results are not sensitive to these choices.

 \subsubsection{Classifier Performance}
\begin{figure}
 \centering
 \subfigure[F1 Score vs. Send Rate]{
 \includegraphics[width=0.47\columnwidth]{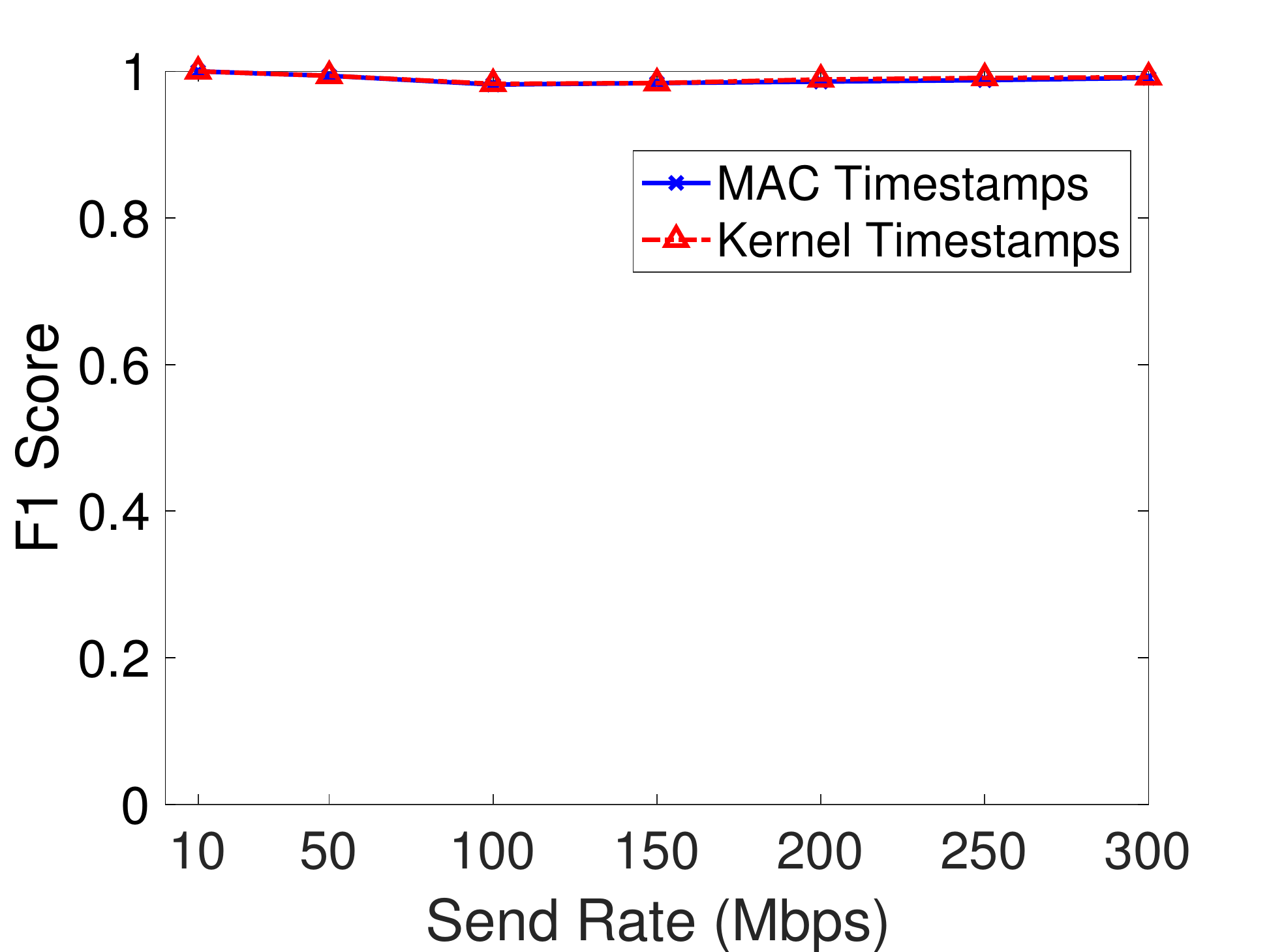}
}
\subfigure[]{
 \includegraphics[width=0.47\columnwidth]{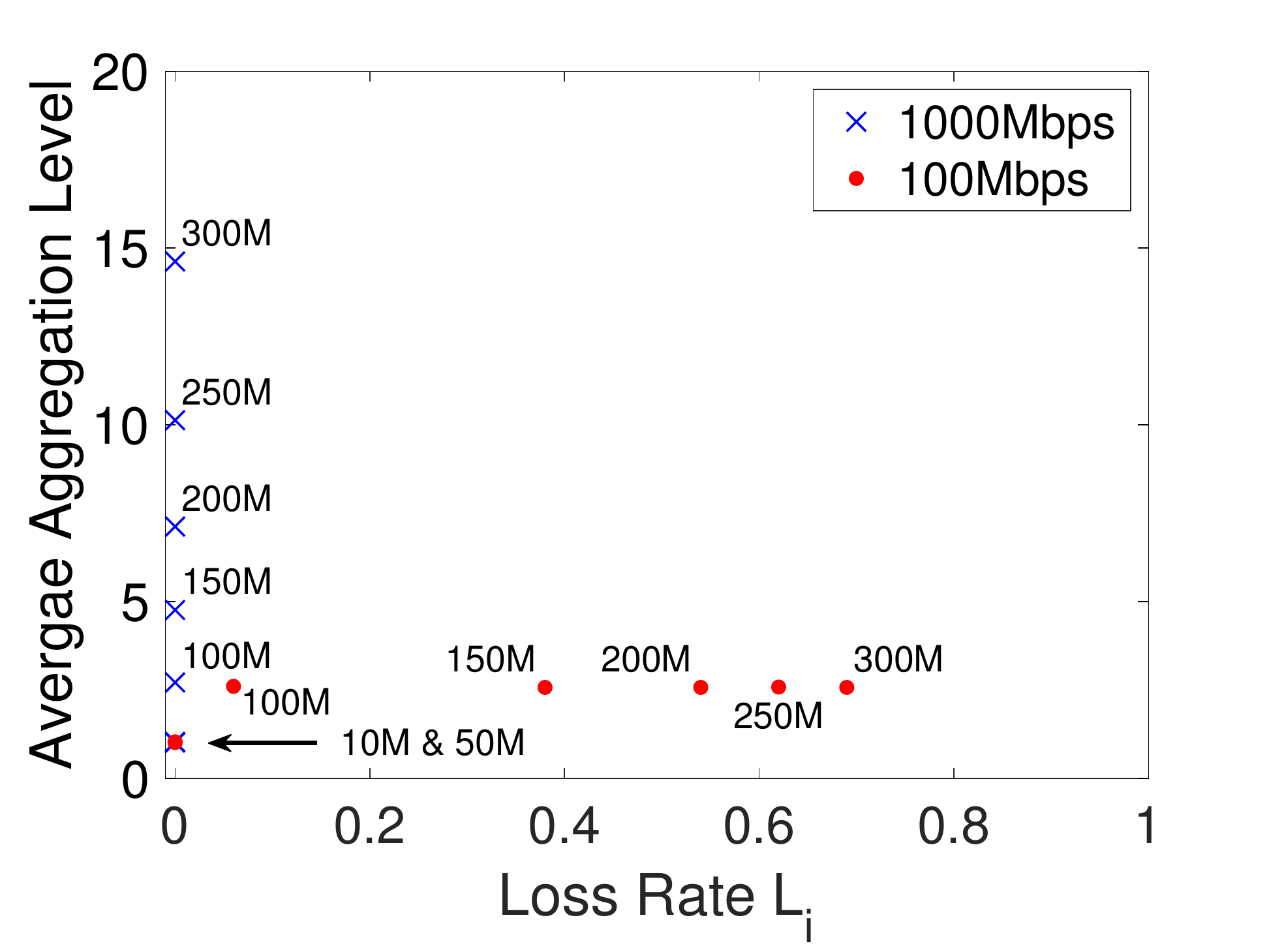}
}
 \caption{Performance of the logistic regression classifier, $n = 5$, $p = 100$.   Experimental data, Samsung Galaxy handset, setup in Section \ref{sec:classsetup}, AMDSU aggregation ($N_{max}=128$).  Ethernet rate limiting (100/1000Mbps).}\label{fig:fig15b}
\end{figure}

Figure \ref{fig:fig15b}(a) plots the measured classification accuracy vs the send rate when $n=5$ and $p=100$.  Each point includes data collected for 100Mbps and 1000Mbps link rates.  When the link rate is 100Mbps the backhaul link acts as the bottleneck for send rates of 100Mbps and above, when the link rate is 1000Mbps the WLAN acts as the bottleneck for send rates of around 500Mbps and above.  It can be seen the classification accuracy is very high, close to $100\%$ (we do not show data for send rates above 300Mbps in Figure \ref{fig:fig15b}(a) but for higher send rates the accuracy is also close to $100\%$).   

We can gain some insight into this good performance from Figure \ref{fig:fig15b}(b), which plots the aggregation level vs the loss rate $L_i$ for a range of send rates.  The send rate is indicated beside each point and the points marked by an $\times$ are when the backhaul is 1000Mbps while those marked by $\bullet$ are when the backhaul is 100Mbps.   When the backhaul is 1000Mbps it can be seen that the loss rate $L_i$ stays close to zero while the aggregation level increases with send rate.  When the backhaul is 100Mbps it can be seen that the loss rate increases for send rates above 100Mbps while the aggregation level remains low at all rates.  Hence, we can easily separate the points at the top left and bottom right corners of this plot, corresponding to higher send rates.  At send rates around 100Mbps the points for 100Mbps and 1000Mbps backhaul are quite close but the classifier is still accurate.    

Similar performance is observed when the transmitting to two WLAN clients as might be expected (as already noted, there is per station queueing at the AP).

\subsection{Bottleneck Classification: Cross-Traffic}
We now consider situations where there is cross-traffic sharing the backhaul link to the AP.   When this cross-traffic is sufficiently high then the bottleneck for the WLAN traffic shifts from the WLAN to the backhaul, and vice versa when the level of cross-traffic falls.	

\begin{figure}
 \centering
 \subfigure[$p = 100$]{
 \includegraphics[width=0.46\columnwidth]{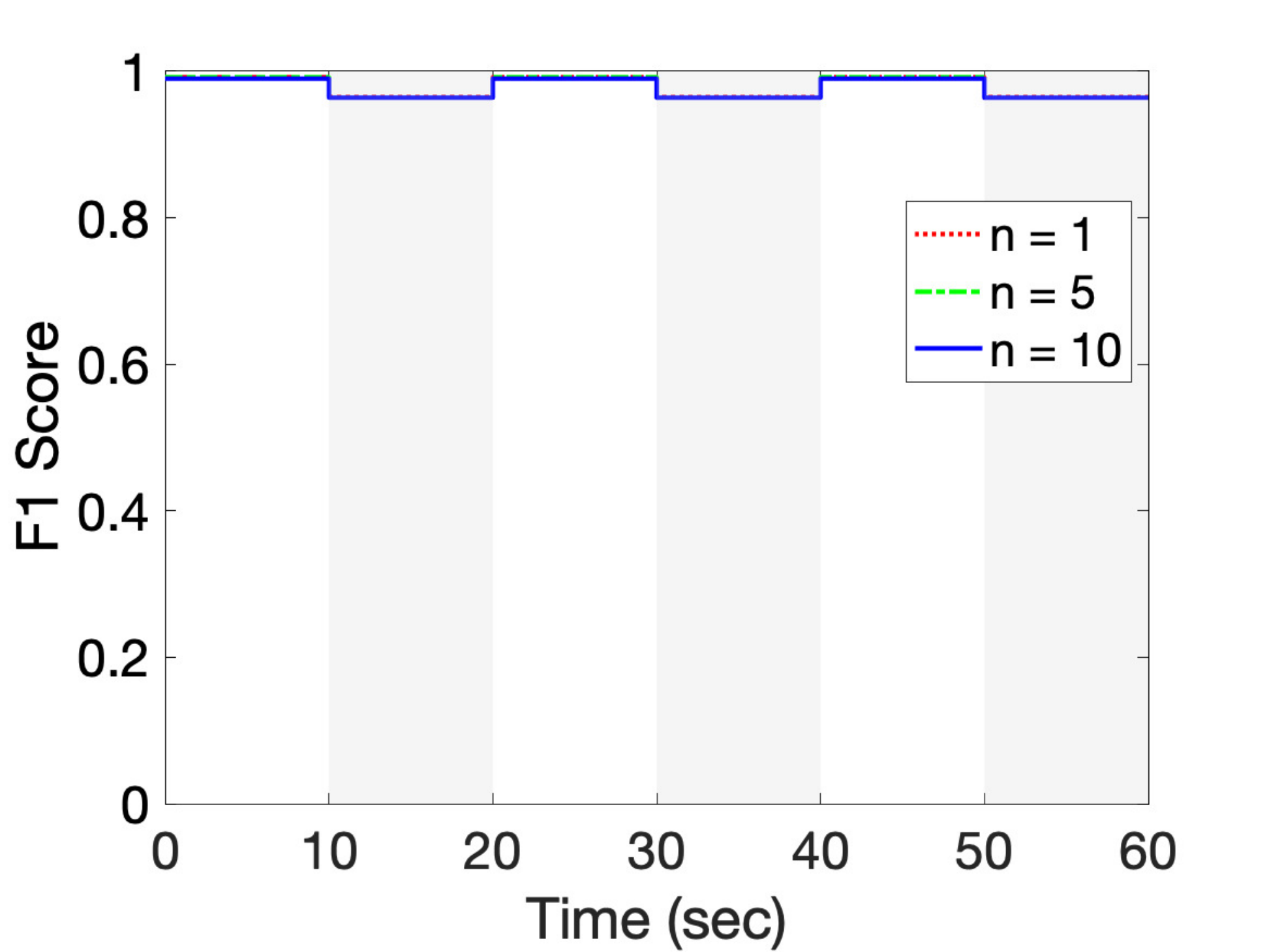}
 }
 \subfigure[$n = 5$]{
 \includegraphics[width=0.46\columnwidth]{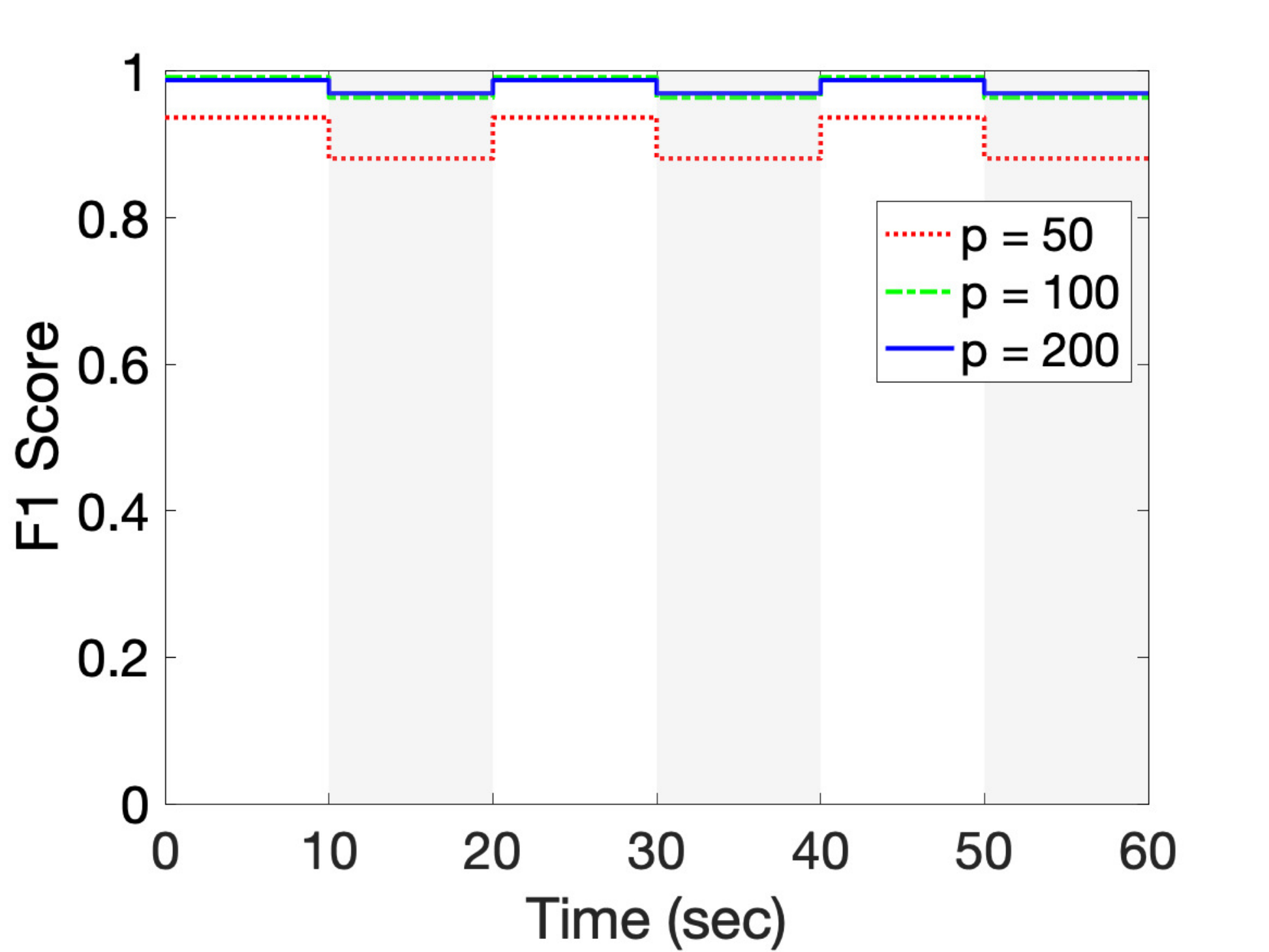}
 }
 \subfigure[$n = 5$, $p = 100$]{
 \includegraphics[width=0.46\columnwidth]{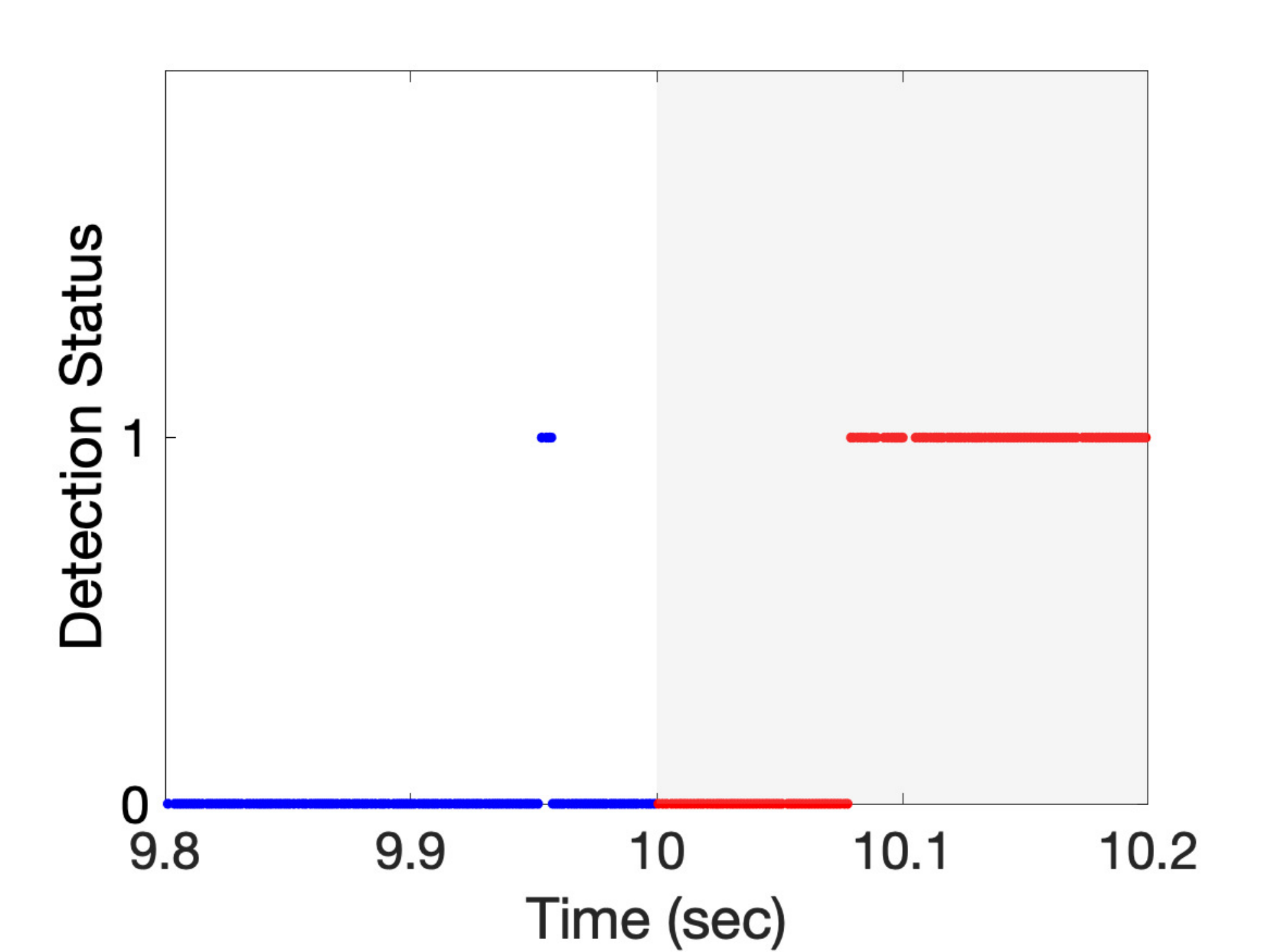}
 }
 \subfigure[$n = 5$, $p = 100$]{
 \includegraphics[width=0.46\columnwidth]{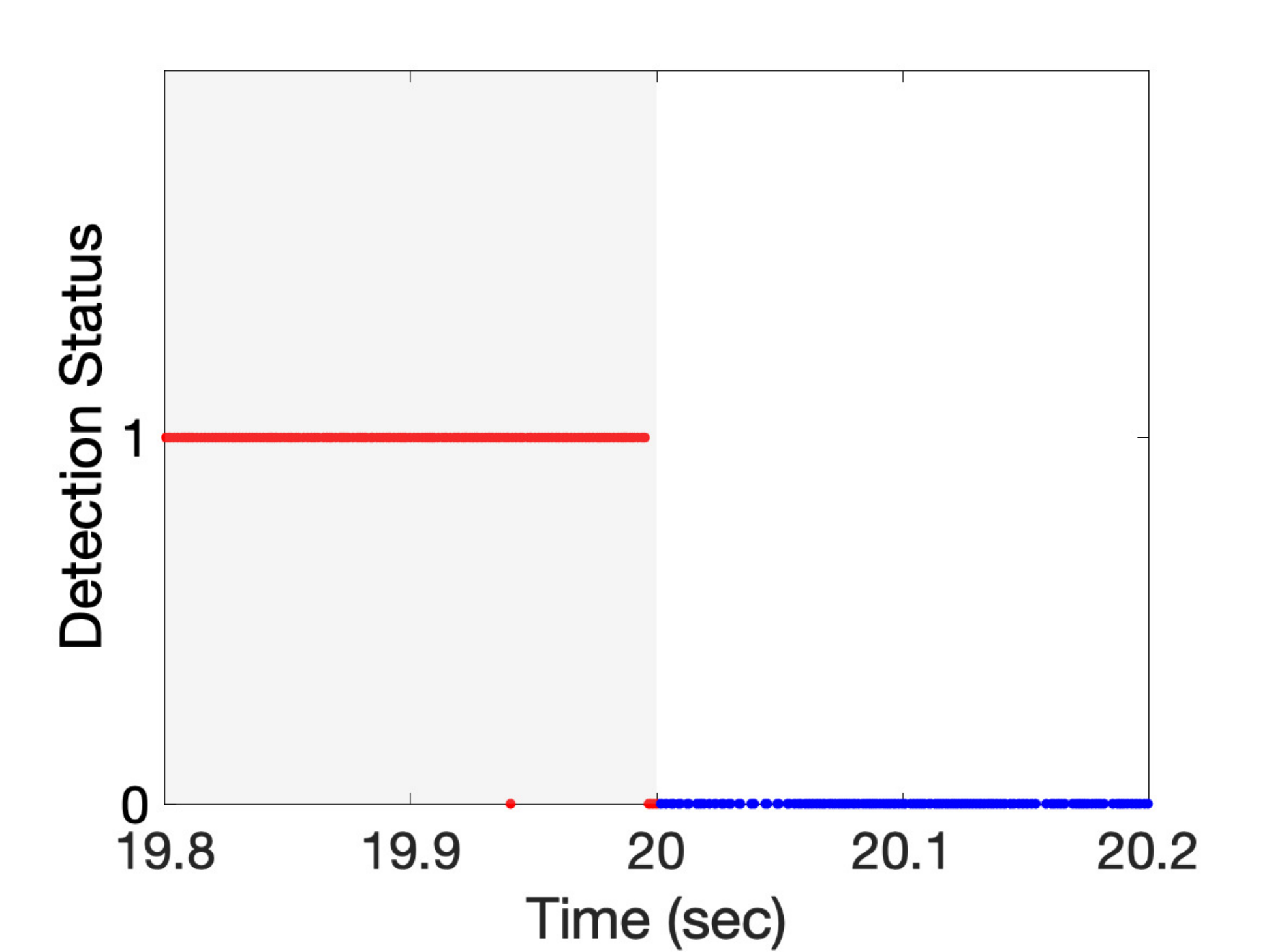}
 }
 \caption{Performance of logistic regression classifier as the bottleneck moves between WLAN and backhaul due to cross-traffic (gray shaded areas indicate when the bottleneck lies in the backhaul link i.e. when the cross-traffic is active).   Experimental data, setup in Section \ref{sec:classsetup}, AMDSU aggregation ($N_{max}=128$), gigabit ethernet backhaul.}\label{fig:fig20}
\end{figure}

We collect data for a setup where the backhaul link to the AP is gigabit ethernet.  When the WLAN is the bottleneck the send rate with a single client station is around 500Mbps, e.g. see Figure \ref{fig:fig23}.  We use iperf to generate UDP cross-traffic of 600Mbps to move the bottleneck from the WLAN to the backhaul link.    Figures \ref{fig:fig20}(a)-(b) shows time histories of the measured F1 score of the logistic regression bottleneck classifier as the cross-traffic switches on and off, so moving the bottleneck back and forth between backhaul and WLAN.   The gray shaded areas indicate when the bottleneck lies in the backhaul link i.e. when the cross-traffic is active.  Results are shown as the number $n$ of features used and the number of packets $p$ used to estimate the loss rate are both varied.   It can be seen that the performance is insensitive to the choice of $n$ and  close to $100\%$ when $p\ge 100$.  

Using $n=5$ and $p=100$ (the same as used in the previous section),  Figures \ref{fig:fig20}(c)-(d) show more detail of the performance time histories.  Figure \ref{fig:fig20}(c) shows the classifier output following a transition of the bottleneck from the WLAN to the backhaul, and Figure \ref{fig:fig20}(d) shows the output for a transition of the bottleneck from backhaul to WLAN.   It can be seen that the estimator detects the transitions within 100ms or less.  We observe similar performance under a range of network conditions, including for data from a production eduroam network, but omit it here since it adds little.


\subsection{Related Work}\label{sec:related}
In recent years there has been an upsurge in interest in userspace transports due to their flexibility and support for innovation combined with ease of rollout.  This has been greatly facilitated by the high efficiency possible in userspace with the support of modern kernels.  Notable examples of new transports developed in this way include Google QUIC \cite{quic}, UDT \cite{udt} and Coded TCP \cite{ctcp14,fec16,multipath17}.  ETSI has also recently set up a working group to study next generation protocols for 5G \cite{etsi}.   The use of performance enhancing proxies, including in the context of WLANs, is also not new e.g. RFC3135 \cite{RFC3135} provides an entry point into this literature.   However, none of these exploit the use of aggregation in WLANs to achieve high rate, low delay communication.      

Interest in using aggregation in WLANs pre-dates the development of the 802.11n standard in 2009 but has primarily focused on analysis and design for wireless efficiency, managing loss etc.  For a recent survey see for example \cite{aggsurvey}.   The literature on throughput modelling of WLANs is extensive but much of it focuses on so-called saturated operation, where transmitters always have a packet to send, see for example \cite{li09} for early work on saturated throughput modelling of 802.11n with aggregation.  When stations are not saturated (so-called finite-load operation) then for WLANs which use aggregation (802.11n and later) most studies resort to the use of simulations to evaluate performance due to the complex interaction between arrivals, queueing and aggregation with CSMA/CA service.   Notable exceptions include \cite{kuppa06,boris09} which adopt a bulk service queueing model that assumes a fixed, constant level of aggregation and \cite{kim08} which extends the finite load approach of \cite{malone07} for 802.11a/b/g but again assumes a fixed level of aggregation.

While measurements of round-trip time might be used to estimate the onset of queueing and adjust the send rate, it is known that this can be inaccurate when there is queueing in the reverse path~\cite{Pathak}.   Furthermore, using RTT to detect queueing is known to give inaccurate results in 802.11 networks~\cite{Malone}.   Accurately measuring one-way delay is also known in general to be challenging\footnote{The impact of  clock offset and skew between sender and receiver applies to all network paths.   In addition, when a wireless hop is the bottleneck then the transmission delay can also change significantly over time depending on the number of active stations e.g. if a single station is active and then a second station starts transmitting the time between transmission opportunities for the original station may double.}.   In contrast, the number of packets aggregated in a frame is relatively easy to measure accurately and reliably at the receiver, as already noted. 

\begin{figure}
\centering
\subfigure[]{
\includegraphics[width=0.46\columnwidth]{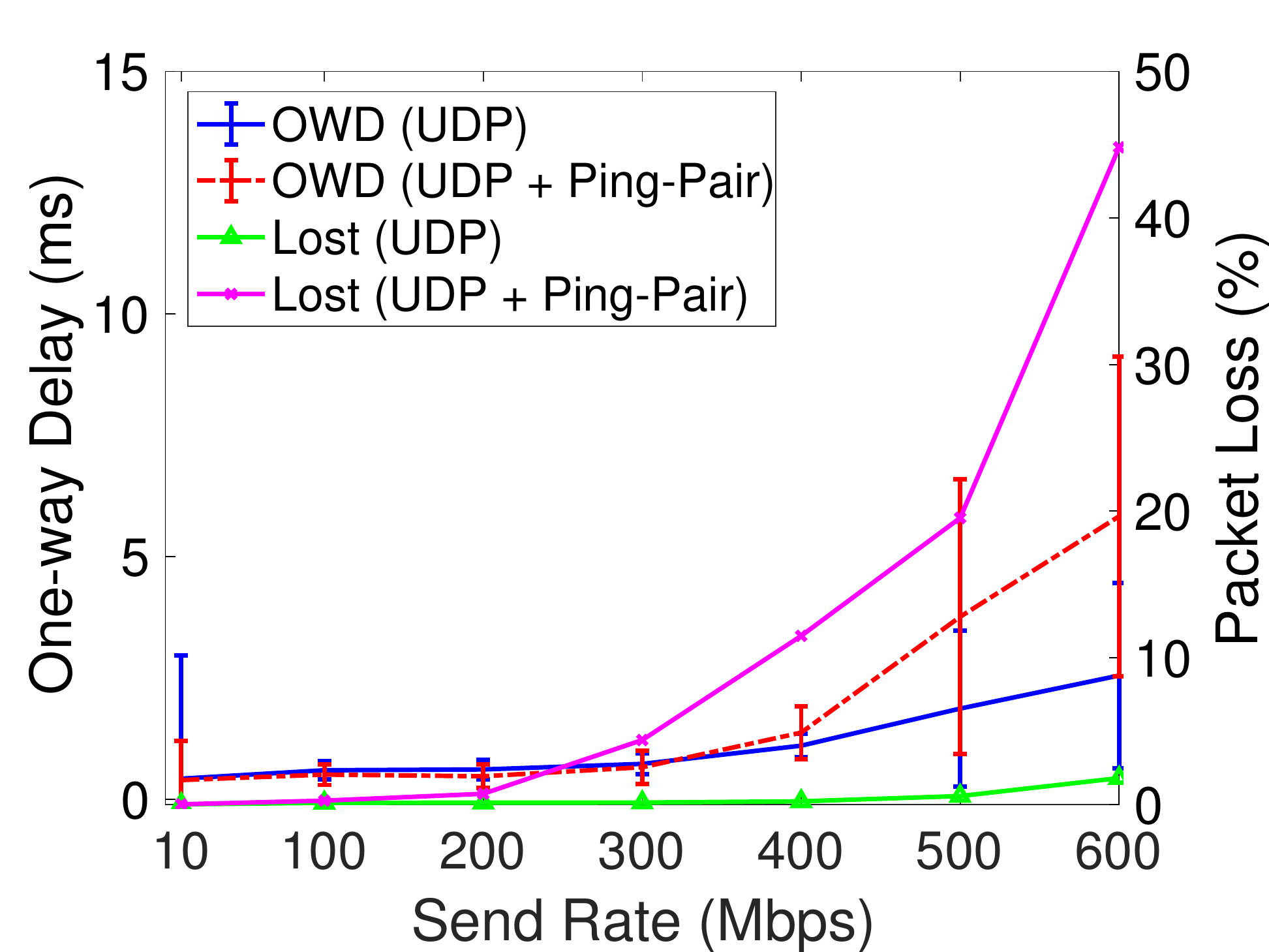}
}
\subfigure[]{
\includegraphics[width=0.46\columnwidth]{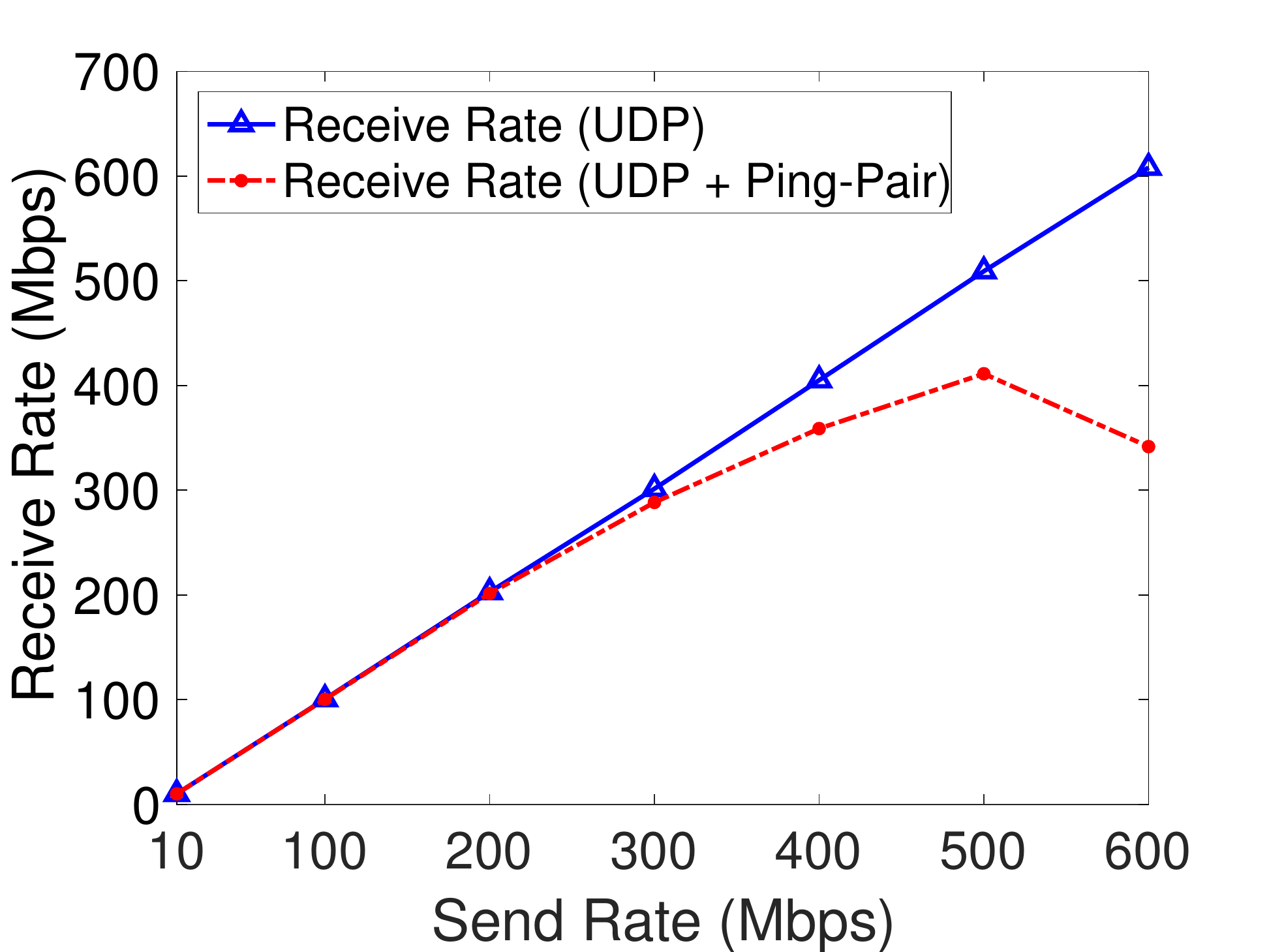}
}
\caption{Performance of the Ping-Pair algorithm~\cite{Pathak} in an 802.11ac WLAN as the send rate of a downlink UDP flow to a mobile handset is varied. Experimental data, setup in Appendix, AMDSU aggregation.}\label{fig:fig9}
\end{figure}

Recently, \cite{Das} proposes an elegant Ping-Pair method for detecting queue build up in 802.11ac APs.  In this approach, which is now used by Skype, a wireless client sends a pair of back-to-back ICMP echo requests to the AP with high and normal priorities specified by DSCP value, respectively.   The high priority packets are queued separately from the normal priority packets at the AP, and serviced more quickly.  Hence, the difference in RTTs between the two pings can provide an indication of queueing at the AP and \cite{Das} proposes thresholding this difference based on a predefined threshold inferred from a decision tree classifier in order to detect the queue build-up.  However, this active probing approach creates significant load on the network and so itself can cause queue buildup.   For example, Figure \ref{fig:fig9}(a) plots the measured one-way delay and loss with and without ping-pairs as the send rate at which UDP data packets are sent from server to client is varied.    Figure \ref{fig:fig9}(b) plots the corresponding goodput (the rate at which packets are received at the client, i.e. after queue overflow losses at the AP).   It can be seen that use of ping-pairs causes packet loss to start to occur for send rates above 200Mbps compared to send rates above 400Mbps without packet pairs, and similarly the one-way delay is roughly doubled with ping-pairs for send rates above 400Mbps.

TCP BBR~\cite{BBR} is currently being developed by Google and this also targets high rate and low latency, although not specifically in edge WLANs.  The BBR algorithm tries to estimate the bottleneck bandwidth and adapt the send rate accordingly to try to avoid queue buildup.  The delivery rate in BBR is defined as the ratio of the in-flight data when a packet departed the server to the elapsed time when its ACK is received.  This may be inappropriate, however, when the bottleneck is a WLAN hop since aggregation can mean that increases in rate need not correspond to increases in delay plus a small queue at the AP can be benificial for aggregation and so throughput.

\section{Summary \& Conclusions}\label{sec:conclusions}
In this paper we consider transport layer approaches for achieving high rate, low delay communication over edge paths where the bottleneck is an 802.11ac WLAN which can aggregate multiple packets into each frame.   We first show that regulating send rate so as to maintain a target aggregation level can be used to realise high rate, low latency communication over 802.11ac WLANs.  We then address two important practical issues arising in production networks, namely that (i) many client devices are non-rooted mobile handsets/tablets and (ii) the bottleneck may lie in the backhaul rather than the WLAN, or indeed vary between the two over time.  We show that both these issues can be resolved by use of simple and robust machine learning techniques.  We present a prototype transport layer implementation of our low latency rate allocation approach and use this to evaluate performance under real radio channel conditions.

\begin{appendices}
\section{Hardware \& Software Setup}\label{sec:setup}
\subsection{Experimental Testbed}
Our experimental testbed uses an Asus RT-AC86U Access Point (which uses a Broadcom 4366E chipset and {supports 802.11ac MIMO with up to four spatial streams}).   It is configured to use the 5GHz frequency band with 80MHz channel bandwidth.   This setup allows high spatial usage (we observe that almost always three spatial streams are used) and high data rates (up to MCS 11).  Note that we also carried out experiments with different chipsets at the AP (e.g., QCA chipsets) and did not observe any major differences.   By default aggregation supports AMSDU's and allows up to 128 packets to be aggregated in a frame (namely 64 AMSDUs each containing two packets).  In our tests in Section \ref{sec:low} we disabled AMSDU's to force AMPDU aggregation since this facilitates monitoring, in which case up to 64 packets can be aggregated in a frame.    

A Linux server connected to this AP via a gigabit switch uses iperf 2.0.5 to generate UDP downlink traffic to the WLAN clients.     Iperf inserts a sender-side timestamp into the packet payload and since the various machines are tightly synchronised over a LAN this can be used to estimate the one-way packet delay (the time between when a packet is passed into the socket in the sender and when it is received).  Note, however, that in production networks accurate measurement of one-way delay is typically not straightforward as it is difficult to maintain accurate synchronisation between server and client clocks (NTP typically only synchronises clocks to within a few tens of milliseconds).  

In Section \ref{sec:low}, where the clients use MAC timestamps to measure the aggregation level of received frames, the WLAN clients are Linux boxes running Debian Stretch and with Broadcom BCM4360 802.11ac NICs.    

In Section \ref{sec:noroot} a non-rooted Samsung Galaxy Tab S3 running Android Oreo is used as the client (we also carried out experiments using a non-rooted Google Pixel 2 running Android Pie and did not observe any significant differences).  Due to the lack of root privilege this client is restricted to using kernel timestamps to estimate the aggregation level of received frames.   A separate machine running Debian Stretch and equipped with a Broadcom BCM4360 802.11ac NIC is used in monitor mode to sniff network traffic and so provide ``ground truth'' since it can log MAC timestamps.    Iperf inserts a unique id number into the packet payload and this is used to synchronise measurements taken by the sniffer and by the mobile handset and in this way we can measure both the handset kernel timestamp and the packet MAC timestamp.   The antennas of the sniffer are placed in the path between AP and handset so as to improve reception with MIMO operation, and it is verified that all frames are captured.  

\subsection{Prototype Rate Allocation Implementation}\label{sec:proto}

\begin{figure}
\centering
\includegraphics[width=0.7\columnwidth]{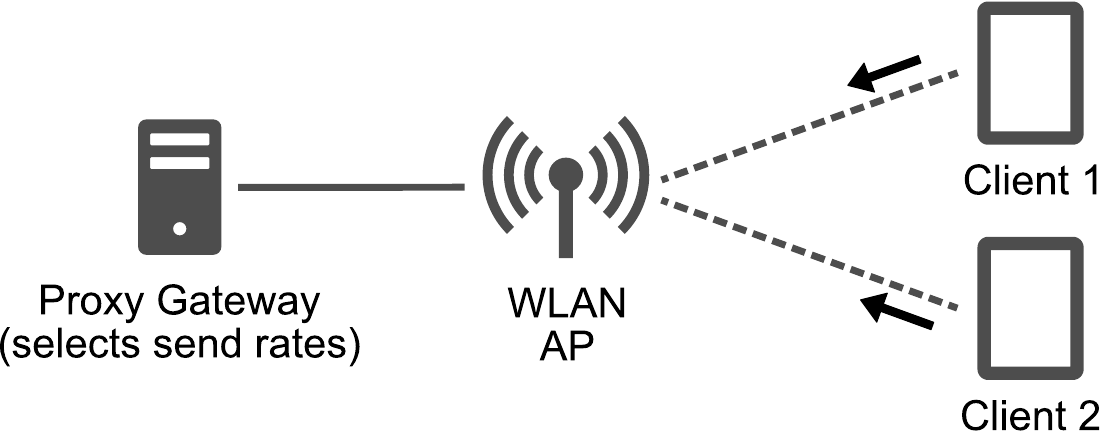} 
\caption{Schematic of scheduler architecture.  Clients send reports of observed aggregation level and MCS rate to proxy which then uses this information to adjust the downlink send rate to each station.}\label{fig:prototype}
\end{figure}

We implemented feedback of measured aggregation level from the clients to the sender using the software architecture illustrated in Figure \ref{fig:prototype}.  Clients measure/estimate the aggregation level of received frames and periodically report this data back to the sender at intervals of $\Delta$ seconds as the payload of a UDP packet.  The sender uses a modified version of iperf 2.0.5 where we implement the feedback collector and our rate control algorithm (see later for details).  Recall that we are considering next generation edge transports and so the sender would typically be located in the cloud close to the network edge.  While it may be located on the wireless access point this is not essential, and indeed we demonstrate this feature in all of our experiments by making use of a proprietary closed access point.   

\subsection{NS3 Simulator Implementation}
While we mainly use experimental measurements, to allow performance evaluation with larger numbers of client stations and with controlled changes to channel conditions we also implemented our approach in the NS3 simulator.  Based on the received feedbacks it periodically configures the sending rate of {\tt udp-client} applications colocated at a single node connected to an Access Point. Each wireless station receives a UDP traffic flow at a {\tt udp-server} application that we modified to collect frame aggregation statistics and periodically transmit these to the controller at intervals of $\Delta$ ms.  {To make the behaviour of the AP closer to reality we also introduced into its codebase a round-robin packet scheduler and per destination queues.} We configured 802.11ac to use a physical layer operating over an $80$MHz channel, VHT rates for data frames and legacy rates for control frames, PHY MCS=9 and with the number of spatial streams NSS = 2 i.e. similarly to the experimental setup.  As validation we reproduced a number of the simulation measurements in our experimental testbed and found them to be in good agreement. {The new NS3 code and the software that we used to perform experimental evaluations are available open-source}\footnote{Code can be obtained by contacting the corresponding author.}. 

\end{appendices}

\bibliographystyle{IEEEtran}
\bibliography{references.bib}

\begin{IEEEbiography}{Hamid Hassani} received B.S. and M.S. degrees with honors in electrical engineering from the University of Zanjan and K. N. Toosi University of Technology, Iran, in 2010 and 2013, respectively. He worked for three years as a RAN engineer and data analyst at MCI, Iran, where he twice received an outstanding employee award. He is currently pursuing a postgraduate degree at Trinity College Dublin, Ireland, under the supervision of Prof. Doug Leith. His research interest includes wireless networks, machine learning, and optimization.
\end{IEEEbiography}

\begin{IEEEbiography}{Francesco Gringoli} received  the  Laurea  degree  in  telecommunications  engineering  from the  University  of  Padua,  Italy,  in  1998  and  the PhD degree  in  information  engineering  from the  University  of  Brescia,  Italy,  in  2002.  Since 2018 he is Associate Professor  of Telecommunications at the Dept. of Information Engineering at the University of Brescia. His research interests include security assessment, performance evaluation and medium access control in Wireless LANs. He is a senior member of the IEEE.
\end{IEEEbiography}

\begin{IEEEbiography}{Doug Leith} graduated from the University of Glasgow in 1986 and was awarded his PhD, also from the University of Glasgow, in 1989. In 2001, Prof. Leith moved to the National University of Ireland, Maynooth and then in Dec 2014 to Trinity College Dublin to take up the Chair of Computer Systems in the School of Computer Science and Statistics.  His current research interests include wireless networks, network congestion control, distributed optimization and data privacy.  
\end{IEEEbiography}
\end{document}